\documentclass[10pt]{article}

\usepackage{epsfig,epsf}

\usepackage{amsmath}

\usepackage{amsfonts}
\usepackage{cite}
\usepackage{graphicx}
\renewcommand{\theequation}{\arabic{section}.\arabic{equation}}
\renewcommand{\thefootnote}{\fnsymbol{footnote}}
\renewcommand{\thetable}{\arabic{table}}

\renewcommand{\theequation}{\arabic{section}.\arabic{equation}}
\renewcommand{\thefootnote}{\fnsymbol{footnote}}
\renewcommand{\thetable}{\arabic{table}}

\def \tr {\mathop{\rm tr}\nolimits}

\newcommand{\as}{\ifmmode\alpha_{\rm s}\else{$\alpha_{\rm s}$}\fi}
\newcommand{\asbar}{\ifmmode\bar{\alpha}_{\rm s}\else{$\bar{\alpha}_{\rm s}$}\fi}

\newcommand \CCR {\mathrm{R}}
\newcommand \CCL {\mathrm{L}}

\font\cmss=cmss12 
\def\inbar{\,\vrule height1.5ex width.4pt depth0pt}
\def\IC{\relax\hbox{$\inbar\kern-.3em{\rm C}$}}
\def\IZ{\relax{\hbox{\cmss Z\kern-.4em Z}}}
\def\IR{{\hbox{{\rm I}\kern-.2em\hbox{\rm R}}}}

\def\IP{{\hbox{{\rm I}\kern-.2em\hbox{\rm P}}}}
\def\II{\hbox{{1}\kern-.25em\hbox{l}}}

\newbox\lett\newdimen\lheight\newdimen\lwidth
\def\ontop#1#2{\setbox\lett=\hbox{#2}\lheight\ht\lett
\multiply\lheight by 12 \divide\lheight by 10\relax%
\lwidth\wd\lett \multiply\lwidth by 8 \divide\lwidth by 10\relax #2\kern-\lwidth%
\raise\lheight\hbox{{$\scriptstyle #1$}}\kern.1ex}

\def\inbar{\,\vrule height1.5ex width.4pt depth0pt}

\begin{document}
\begin{titlepage}

\vspace*{1cm}

\begin{center}
{\Large \bf{Conformal algebra: R-matrix and star-triangle relation}}

 %\marginpar{\bf Pomenyal Title}

\vspace{1cm}

{\large \sf D. Chicherin$^{ca}$\footnote{\sc e-mail: chicherin@pdmi.ras.ru},
  S. Derkachov$^{a}$\footnote{\sc e-mail: derkach@pdmi.ras.ru} and  A.P.
Isaev$^b$\footnote{\sc e-mail: isaevap@theor.jinr.ru} \\
}

\vspace{0.5cm}

\begin{itemize}
\item[$^a$]
{\it St. Petersburg Department of Steklov Mathematical Institute
of Russian Academy of Sciences,
Fontanka 27, 191023 St. Petersburg, Russia}
\item[$^b$]
{\it  Bogoliubov Lab. of Theoretical Physics,
JINR, 141 980 Dubna, Moscow Reg.,  \\
and ITPM, M.V.Lomonosov Moscow State University, Russia}
\item[$^c$]
{\it Chebyshev Laboratory, St.-Petersburg State University,\\
14th Line, 29b, Saint-Petersburg, 199178 Russia}
\end{itemize}
\end{center}

\vspace{0.5cm}

\begin{abstract}
The main purpose of this paper is the construction of the
$\mathrm{R}$-operator which acts in the tensor product of two
 infinite-dimensional representations of the conformal
 algebra and solves Yang-Baxter equation.
We build the $\mathrm{R}$-operator as a product of more elementary operators
$\mathrm{S}_1 , \mathrm{S}_2$ and $\mathrm{S}_3$. Operators
$\mathrm{S}_1$ and $\mathrm{S}_3$ are identified with intertwining
operators of
 %\marginpar{\bf "equivalent" -- zamenil na "two"}
 two irreducible representations of the
conformal algebra and the operator $\mathrm{S}_2$ is obtained from the
intertwining operators $\mathrm{S}_1$ and $\mathrm{S}_3$ by a certain
duality transformation. There are star-triangle relations for the basic
building blocks $\mathrm{S}_1 , \mathrm{S}_2$ and $\mathrm{S}_3$ which produce all other relations for the general $\mathrm{R}$-operators.
In the
case of the conformal algebra of n-dimensional Euclidean space we
construct the $\mathrm{R}$-operator for the scalar (spin part is equal
to zero) representations and prove that the star-triangle
relation is a well known star-triangle relation for propagators of
scalar fields. In the special case of the conformal algebra
of the $4$-dimensional Euclidean space, the $\mathrm{R}$-operator is
obtained for more general class of infinite-dimensional (differential) representations with
nontrivial spin parts. As a result, for the case of the $4$-dimensional
Euclidean space, we generalize the scalar star-triangle relation to the
most general star-triangle relation for the propagators of particles
with arbitrary spins.
\end{abstract}

\vspace{4cm}

\end{titlepage}

{\small \tableofcontents}
\renewcommand{\refname}{References}
\renewcommand{\thefootnote}{\arabic{footnote}}
\setcounter{footnote}{0} \setcounter{equation}{0}

%%%%%%%%%%%%%%%%%%%%%%%%%%%%%%%%%%%%%%%%%%%%%%%%%%%%%%%%%%%%%%%%%%%%%%%%%%%%%%%%%%%%%%%%%%%%%%
\section{Introduction}
%%%%%%%%%%%%%%%%%%%%%%%%%%%%%%%%%%%%%%%%%%%%%%%%%%%%%%%%%%%%%%%%%%%%%%%%%%%%%%%%%%%%%%%%%%%%%%

\mbox{} Recently, the quantum integrable spin chains with
higher rank symmetry algebras have attracted much attention~\cite{SLN}.
However, the most part of the methods developed so far enable one to deal
only with $s\ell(N)$-symmetric models for which
finite-dimensional as well as infinite-dimensional representations
in the quantum space have been analyzed thoroughly.
The fundamental equations which underlie integrability are
the
 %\marginpar{\bf "general" zamenil na "universal" i redakt.
 %pravka + ssylki do konca stranicy}
 universal  Yang-Baxter $\mathrm{RRR}$-relation and its particular cases:
the $\mathrm{RLL}$-relation  with the general $\mathrm{R}$-operator acting in two
  quantum spaces and
the $\mathrm{RLL}$-relation  with $\mathrm{R}$-matrix acting in two
finite-dimensional auxiliary spaces (e.g., in spaces of the defining representations of $s\ell(N)$).
The last case of the $\mathrm{RLL}$-relation
is also obtained by means of the evaluation homomorphism of the
$s\ell(n)$-type Yangian: $Y(s\ell(N)) \to {\cal U}(s\ell(N))$.
For $s\ell(N)$-symmetric models the
general $\mathrm{R}$-operator acting in two infinite-dimensional
quantum spaces is known and it serves as a local building block in the construction of the Baxter $\mathrm{Q}$-operators \cite{DM}.

Much less is known about integrable quantum lattice models and spin chains with $so(N)$ symmetry
(see, however, \cite{Resh}, \cite{Okad}, \cite{KuSu}).
Substantially, it is related
to the fact that one of the basic algebraic object --  the Yangian $Y(so(N))$ \cite{Drin}
which can be defined by the $\mathrm{RLL}$-relation with $so(N)$-type $\mathrm{R}$-matrix acting in two
finite-dimensional auxiliary spaces (the spaces of $so(n)$ defining representations) does not possess
the evaluation homomorphism $Y(so(N)) \to {\cal U}(so(N))$. One can think that in this case the Yangian $Y(so(N))$
could be substituted by a more sophisticated Olshanskii twisted Yangian or
by the standard reflection equation algebra for which
the evaluation  homomorphism exists (see \cite{Mol} and \cite{IsOM}, respectively).
 %However for these type of algebras the formulation of the integrable lattice models is still unknown.
However these type of algebras are used only for the formulation of integrable open spin chain models
with nontrivial boundary conditions.

In this paper our aim is to adapt some methods developed for $s\ell(N)$-symmetric spin chains to spin chains with
$so(p+1,q+1)$ symmetry which is interpreted as the conformal symmetry in $\mathbb{R}^{p,q}$.
Here we make the first step in this direction.

The plan of the paper is the following.

In Section 2 we recall
basic facts about conformal algebra ${\sf conf}(\mathbb{R}^{p,q})=so(p+1,q+1)$
and its representations~\cite{MackSalam,Koller,Mack,DMPPT,TMP,DP}.
We construct representation of the
 conformal algebra $so(p+1,q+1)$ in the space of tensor fields
 by the method of induced representations.
This material is more or less standard~\cite{MackSalam,Koller,Mack,DMPPT,TMP,DP}.
Our approach is slightly different and appropriate for our own purposes
and we include it for the completeness.
There is also alternative approach --
the so called embedding formalism \cite{FGG,Siegel,Weinberg,Shadow,CPPR,Kuz}.

In Section 3 we collect some basic facts about $\mathrm{L}$-operators.
The $\mathrm{L}$-operator for $s\ell(N)$-symmetric quantum spin chain can
be constructed from the Yangian $Y(s\ell(N))$ by means of the
evaluation homomorphism $Y(s\ell(N)) \to {\cal U}(s\ell(N))$ and
is expressed as a polarized Casimir operator for $g\ell(N)$:
 $$
 \mathrm{L}(u) = u \cdot 1 + T(E_{ij}) \otimes T'(E_{ij}) \; ,
 $$
 where $u$ is a parameter, $E_{ij}$ are generators of $g\ell(N)$ and in the first
space (auxiliary space) of the tensor product the fundamental (defining) representation $T$ is taken and
in the second space (quantum space) one can choose an arbitrary representation $T'$.
Such $\mathrm{L}$-operators were considered in \cite{KS,KRS,KR,FRT}.
If we fix $T'$ as a differential (induced) representation~\cite{GN}, then
the $\mathrm{L}$-operator exhibits a remarkable factorization property~\cite{DM}
and respects $\mathrm{RLL}$-relation with Yang's $\mathrm{R}$-matrix.

 Henceforth, we define the {\it $\mathrm{L}$-operator} as an operator $\mathrm{L}$ acting in the tensor product
 of some finite-dimensional auxiliary space and arbitrary quantum space (generically infinite-dimensional) and furthermore $\mathrm{L}$ respects $\mathrm{RLL}$-relation
 with a certain numerical $\mathrm{R}$-matrix acting in the auxiliary spaces.
 For the
conformal algebra $so(p+1,q+1)$ of the pseudo-Euclidean space $\mathbb{R}^{p+1,q+1}$,
we consider the operator $\mathrm{L}$  which is constructed
from the $so(p+1,q+1)$ polarized Casimir operator acting  in the tensor product
of two spaces: the first one (the auxiliary space) is
the space of a spinor representation (instead of fundamental one) and  the second
quantum space is the space of the differential representation of the
conformal algebra $so(p+1,q+1)$.
 It happens that in general this operator
respects $\mathrm{RLL}$-relation only if we choose special (scalar) differential
representation of the conformal algebra in the quantum space when spin part $S_{\mu\nu}$ of
 the Lorentz generators is equal to zero. Corresponding numerical $\mathrm{R}$-matrix
 (acting in the spaces of spinor representation) is rather
nontrivial. For the first time it appeared in~\cite{Witten}
(see also \cite{Resh,Okad}), where the
$\mathrm{RLL}$-relation for the $so(N)$-invariant
$\mathrm{L}$-operator with fundamental (defining) representation in the
quantum space was established. Thus, we generalize this result. Namely we
prove that in the $\mathrm{L}$-operator
one can replace (in the quantum space) the defining representation to the infinite-dimensional
scalar representation parameterized by the conformal dimension $\Delta$. This new
conformal $\mathrm{L}$-operator can be factorized as well
 (similar to the $s\ell(N)$ case) and as we will see it
corresponds to a certain integrable systems~\cite{Lipat,Zam1}.

In Section 4 we specify formulae of the previous section to
the case of the conformal algebra $so(2,4)$ in 4-dimensional
Minkowski space $\mathbb{R}^{1,3}$ (actually these formulas can be easily
generalized to the case of any conformal algebra $so(p+1,5-p)$ in 4-dimensional
space $\mathbb{R}^{p,4-p}$, where $p=0,1,2$).
 This case is a special one since
$so(2,4)$ is isomorphic to $su(2,2)$ (for complexifications
we have $so(6,\mathbb{C})=s\ell(4,\mathbb{C})$) and consequently we can establish connection
with the known construction \cite{DM} developed for $s\ell(N,\mathbb{C})$.
Indeed, as we show the $\mathrm{L}$-operators for these two algebras are related
 by an appropriate change of variables.
The numerical $\mathrm{R}$-matrix for both
$\mathrm{L}$-operators is the Yang's one (for the $so(2,4)$ case it is shown
in Subsection 3.2) and in the quantum space, for the conformal
 $\mathrm{L}$-operator, we
 obtain  the general differential representation of the conformal algebra with nontrivial spin part $S_{\mu\nu}$, i.e.
we deal with representation $\rho_{\Delta,\ell,\dot{\ell}}$ of
$so(2,4)$ parameterized by conformal dimension $\Delta$ and two spin
variables $\ell,\dot{\ell}$.

 %\add{Frankly speaking we have some problems with next Subsection 4.2.
 %Problems are of the same type as for Minkowski space -- the absence of explicit
 %regularization of integral operators.\\
 %In \cite{DM} we have considered complex group $SL(n,\mathbb{C})$
 %so that we have $\frac{n(n-1)}{2}$ complex variables $z_{ik}$ and $\frac{n(n-1)}{2}$
 %complex conjugate variables $\bar{z}_{ik}$.
 %In our case $SL(4,\mathbb{C})$ we have 6 complex variables and 6
 %complex conjugate variables. In Subsection 4.1 everything is OK because we woork
 %with differential operators and one cane restrict everything to complex variables
 %and forget about complex conjugated variables -- the holomorphic and antiholomorphic
 %sectors cane be separated. In Subsection 4.2 the situation is different because
 %operators like $\partial_z^{\alpha}$ for noninteger
 %$\alpha$ needs antiholomorphic part $\partial_{\bar{z}}^{\bar{\alpha}}$ so that
 %only the operator $\partial_z^{\alpha}\cdot\partial_{\bar{z}}^{\bar{\alpha}}$
 %can be defined as usual integral operator acting on the function $f(z,\bar{z})$ defined on
 %usual $\mathbb{R}^2$. We omit the antiholomorphic part everywhere in Subsection 4.2 so
 %that intertwining operators are not properly defined and have only formal and illustrative meaning.}

In Subsection 4.2, following the approach
\cite{DM} which was developed for the $s\ell(N,\mathbb{C})$ case, we reproduce
intertwining operators for the product of two $so(6,\mathbb{C})$-type
$\mathrm{L}$-operators. These operators are building blocks in
construction of $\mathrm{R}$-operator which acts
in the tensor product of two infinite-dimensional representations. As we indicate
at the end of Subsection 4.2 the
form of these intertwining operators is not manifestly Lorentz covariant. Moreover
these operators are not properly defined can be treated only formally.
That is why, in the next Section 5, the same intertwining operators and
$\mathrm{R}$ operator are constructed directly without using the
isomorphism $so(6,\mathbb{C})=s\ell(4,\mathbb{C})$.

 %\add{I did not corrected the previous absatz because mentioned problems and
 %it is not clear for me what we should to do with Subsection 4.2}

The main purpose of Section 5 is the construction of the general
$\mathrm{R}$-operator which acts in the tensor product $\rho_1 \otimes \rho_2$ of two
infinite-dimensional representations of the conformal algebra $so(n+1,1)={\sf conf}(\mathbb{R}^n)$
  and solves $\mathrm{RLL}$-relation with conformal
$\mathrm{L}$-operators.
For the simplicity we restrict ourselves to the case of Euclidean
space $\mathbb{R}^n$ because in this situation all integral operators are
well defined in generic situation.
In the case of the conformal algebra $so(n+1,1)={\sf conf}(\mathbb{R}^n)$
 %of the n-dimensional Euclidean space
 we construct $\mathrm{R}$-operator for
 the scalar ($S_{\mu\nu}=0$) representations
  $\rho_{\Delta_1} \otimes \rho_{\Delta_2}$ and in
the special case of $so(5,1)$, i.e. the conformal algebra of the
$4$-dimensional Euclidean space, the $\mathrm{R}$-operator is
constructed for a rather general class of representations
$\rho_{\Delta_1,\ell_1,\dot{\ell}_1} \otimes
\rho_{\Delta_2,\ell_2,\dot{\ell}_2}$ with nontrivial spin parts.

We build the general $\mathrm{R}$-operator as a product of simpler
operators $\mathrm{S}_1 , \mathrm{S}_2 , \mathrm{S}_3$ which respect
relations of $\mathrm{RLL}$-type: $\mathrm{S LL'}=
\mathrm{L^{\prime\prime}L^{\prime\prime\prime}S}$. Each $\mathrm{L}$-operator depends on
the set of four parameters $(u,\Delta,\ell,\dot{\ell})$ and the
$\mathrm{RLL}$-relation implies that the $\mathrm{R}$-operator,
 intertwining the product $(\mathrm{L}_1 \cdot \mathrm{L}_2)$ of
two $\mathrm{L}$-operators,
 interchanges the sets of their parameters:
$(u,\Delta_1,\ell_1,\dot{\ell}_1) \leftrightarrow
(v,\Delta_2,\ell_2,\dot{\ell}_2)$. Consequently it is reasonable to
implement this transposition in several steps and to consider operators
$\mathrm{S}_1,\mathrm{S}_2,\mathrm{S}_3$ which intertwine two
L-operators and transpose (or change) only a part of their parameters.
 The operators $\mathrm{S}_1$ and $\mathrm{S}_3$ separately
 change the parameters in the first and second factor of the product
 $\mathrm{L}_1 \cdot \mathrm{L}_2$.
 Actually the operators
$\mathrm{S}_1$ and $\mathrm{S}_3$ can be identified with the
 intertwining operators of
 two irreducible representations of the
conformal algebra~\cite{DMPPT,DP} so that our construction has a
 transparent representation theory meaning. The operator $\mathrm{S}_2$
 interchanges parameters between the factors $\mathrm{L}_1$ and
 $\mathrm{L}_2$ and the form of $\mathrm{S}_2$
is obtained from the intertwining operators $\mathrm{S}_1$ and
$\mathrm{S}_3$ by some kind of the duality transformation. This duality
transformation is very similar to the one obtained in the spin chain
model~\cite{Lip} (see also~\cite{Der}). It also resembles the dual
conformal transformation for the Feynman diagrams~\cite{Br,dual0}
and for the scattering amplitudes in maximally supersymmetric $N=4$ super
Yang-Mills theory~\cite{dual1,dual2}.\\
Finally, the general $\mathrm{R}$-operator that implements
  a special transposition
in the set of parameters can be factorized in a product of operators
 $\mathrm{S}_1,\mathrm{S}_2,\mathrm{S}_3$ which represent basic
elementary transpositions. There are certain relations for the basic
building blocks which produce all other relations for
$\mathrm{R}$-operators. Indeed, the Coxeter (braid) three-term
relations in the symmetric group are represented as follows
$\mathrm{S}_1 \mathrm{S}_2 \mathrm{S}_1 = \mathrm{S}_2 \mathrm{S}_1
\mathrm{S}_2$ and $\mathrm{S}_3 \mathrm{S}_2 \mathrm{S}_3 =
\mathrm{S}_2 \mathrm{S}_3 \mathrm{S}_2$.
  %\marginpar{\bf redaktorskaya pravka do konca Vvedeniya}
These Coxeter three-term relations can be
interpreted as star-triangle relations and play the important role
in the construction of the general $\mathrm{R}$-operator.
In the n-dimensional scalar case these relations are well known star-triangle relations
\cite{FrP}, \cite{Vas} for propagators of scalar fields.
 In the case of the conformal algebra $so(5,1)$ of $4$-dimensional Euclidean
space we prove a new star-triangle relation for generic representations of
the type $\rho_{\Delta,\ell,\dot{\ell}}$ included spin degrees of
 freedom, i.e. we generalize the scalar star-triangle
relation to the star-triangle relation for the propagators of the spin
particles.

Recall that the star-triangle relations happen to be
a corner stone in the integrability of many lattice models of
statistical mechanics \cite{Bax} (see also papers \cite{BazSer} and references therein).
At the end of Subsection 5.1 we show that
the scalar star-triangle relations can be used for the formulation
of the n-dimensional variant of the integrable lattice
model proposed by Lipatov \cite{Lipat} 
 %to describe the high energy scattering of hadrons in multicolor QCD
(see also \cite{Zam1} where
another integrable lattice models were constructed and investigated by
means of the scalar star-triangle relations). To our knowledge 
integrable models 
 %of the types \cite{Lipat} and \cite{Zam1} 
 related to
the new spinorial star-triangle relation of $so(5,1)$ type are still unknown.

In Appendix A,  we prove this new star-triangle
 identity directly evaluating corresponding integrals.
In Appendix B, we
sketch a useful technique~\cite{VDK} for calculations with the algebra
of gamma-matrices needed to prove $\mathrm{RLL}$-relation
 for spinorial $\mathrm{R}$-matrix in Section 3.

%%%%%%%%%%%%%%%%%%%%%%%%%%%%%%%%%%%%%%%%%%%%%%%%%%%%%%%%%%%%%%%%%%%%%%%%%%%%%%%%%%%%%%%%%%%%%
%\subsection{$\mathrm{L}$-operator from conformal algebra}
\section{Conformal algebra in $\mathbb{R}^{p,q}$ }
%%%%%%%%%%%%%%%%%%%%%%%%%%%%%%%%%%%%%%%%%%%%%%%%%%%%%%%%%%%%%%%%%%%%%%%%%%%%%%%%%%%%%%%%%%%%%

\mbox{} In this Section we summarize some facts about conformal Lie algebras; these facts are needed
in subsequent Sections.
Let $\mathbb{R}^{p,q}$ be a pseudoeuclidean space with the metric
$$
g_{\mu \nu} = {\rm diag}(\underbrace{1,\dots,1}_p , \underbrace{-1,\dots,-1}_q) \; .
$$
Denote by $\textsf{conf}(\mathbb{R}^{p,q})$ a Lie algebra
of the conformal group in $\mathbb{R}^{p,q}$ with basis elements
$\{L_{\mu \nu} , P_{\mu},K_{\mu},D \}$
$(\mu,\nu = 0,1,\dots,p+q-1)$ and commutation relations:
$$
[ D\, ,\, P_{\mu} ] = i\, P_{ \mu }\ \,, \ \ \  [ D\, ,\, K_{\mu} ]  = - i\, K_{\mu}\ \,,
\ \ \  [ L_{\mu \nu}\, ,\, L_{\rho \sigma} ]  = i\,( g_{\nu \rho}\, L_{\mu
\sigma} + g_{\mu \sigma}\, L_{\nu \rho} - g_{\mu \rho}\, L_{\nu
\sigma} - g_{\nu \sigma}\, L_{\mu \rho} )
$$
 \begin{equation}
 \label{cnfA}
[ K_{\rho}\, ,\, L_{\mu \nu} ] = i\,( g_{\rho \mu}\, K_{\nu} - g_{\rho
\nu}\, K_{\mu} ) \; , \;\;\;  [ P_{\rho}\, ,\, L_{\mu \nu} ]  = i\, ( g_{\rho
\mu}\, P_{\nu} - g_{\rho \nu}\, P_{\mu} ) \; ,
 \end{equation}
$$
[ K_{\mu}\, ,\, P_{\nu} ] = 2 i\, ( g_{\mu \nu}\, D - L_{\mu \nu} ) \; , \;\;\;
[ P_{\mu}\, ,\, P_{\nu} ] = 0 \; , \;\;\; [ K_{\mu}\, ,\, K_{\nu} ] = 0 \; , \;\;\;
[ L_{\mu \nu}\, ,\, D ] = 0 \; .
$$
Note that elements $L_{\mu \nu}$ generate the Lie algebra $so(p,q)$ of the rotation
group $SO(p,q)$ in $\mathbb{R}^{p,q}$.

It is known~\cite{Dirac} that conformal Lie
algebra (\ref{cnfA}) is isomorphic to the algebra $so(p+1,q+1)$ with generators $M_{ab}$
$(a,b = 0,1,\dots, p+q+1)$ subject relations
\begin{equation}
 \label{cnfA5}
[ M_{ab} , M_{dc} ] = i( g_{bd} M_{ac}
+ g_{ac} M_{bd} - g_{ad} M_{bc} -
g_{bc} M_{ad} )  \; ,
 \end{equation}
where $g_{ab}$ is the metric for $\mathbb{R}^{p+1,q+1}$:
\begin{equation}
 \label{cnfA5a}
 g_{ab} = {\rm diag}(\underbrace{1,\dots,1}_p , \underbrace{-1,\dots,-1}_q , 1,-1) \; .
  \end{equation}
The isomorphism of Lie algebras $so(p+1,q+1)$ and $\textsf{conf}(\mathbb{R}^{p,q})$
is established by the relations (see, e.g., \cite{MackSalam}):
 \begin{equation}
\label{cnfA2}
\begin{array}{c}
L_{\mu\nu}  = M_{\mu\nu}  \; , \;\;\; K_\mu = M_{n,\mu} - M_{n+1,\mu} \; , \\ [0.2cm]
 \;\;\; P_\mu = M_{n,\mu} + M_{n+1,\mu} \; , \;\;\; D = - M_{n,n+1} \; , \;\;\;\; (n=p+q) \; .
 \end{array}
 \end{equation}
Using these formulas one can write relations (\ref{cnfA}) in the concise form (\ref{cnfA5}).
Define a quadratic Casimir operator
 \begin{equation}
\label{kaz}
 C_2 =  \frac{1}{2} M_{ab} \, M^{ab} \; ,
 \end{equation}
 which is the center element in the enveloping algebra of $so(p+1,q+1)=\textsf{conf}(\mathbb{R}^{p,q})$.
 In terms of generators of $\textsf{conf}(\mathbb{R}^{p,q})$
 the operator $C_2$ (\ref{kaz}) is written as
 \begin{equation}
\label{cnfA2b}
C_2  = \frac{1}{2}\left( L_{\mu \nu} L^{\mu \nu} +  P_{\mu}
K^{\mu} + K_{\mu} P^{\mu} \right) - D^2  \; ,
\end{equation}
where the identification (\ref{cnfA2}) has been used.

Now we describe matrix representations
for the conformal algebra $\textsf{conf}(\mathbb{R}^{p,q})=so(p+1,q+1)$ which we call
 \underline{spinor representations}.
 We consider only the case of even dimensions $n=p+q$ (the generalization to
the odd dimensional case is straightforward).
Let $\gamma_\mu$ $(\mu=0,\dots,n-1)$ be $2^{n\over 2}$-dimensional
gamma-matrices in $\mathbb{R}^{p,q}$:
 \begin{equation}
 \label{cnfA4a}
 \gamma_\mu \, \gamma_\nu  + \gamma_\nu \, \gamma_\mu = 2 \, g_{\mu\nu} \, I \; , \;\;\;
   \end{equation}
 \begin{equation}
 \label{cnfA4b}
 \gamma_{n+1} \equiv \alpha \; \gamma_0 \cdot \gamma_1 \cdots \gamma_{n-1}  \; ,
 \;\;\;\; \alpha^2 = (-1)^{q+n(n-1)/2} \; ,
 \end{equation}
 where  $I$ is the unit matrix and
 the constant $\alpha$ is such that $\gamma_{n+1}^2=I$.
  %, i.e. $\alpha^2 = (-1)^{q+n(n-1)/2}$.
  Using gamma-matrices (\ref{cnfA4a}) one can construct
 gamma-matrices $\Gamma_a$ in the space $\mathbb{R}^{p+1,q+1}$ with the metric $g_{ab}$ (\ref{cnfA5a}):
 %\begin{equation}\label{cnfA4}
 %\Gamma_\mu = \sigma_2 \otimes \gamma_\mu \;\;\;\;\; (\mu=0,\dots,n-1) \; , \;\;\;
 %\Gamma_{n} = \sigma_1 \otimes I \; , \;\;\;
 %\Gamma_{n+1} = i \, \sigma_2 \otimes \gamma_{n+1}   \; .\end{equation}
 \begin{equation}
 \label{cnfA4}
 \begin{array}{c}
 \Gamma_\mu = \sigma_2 \otimes \gamma_\mu =
 \begin{pmatrix} O & -i \gamma_\mu \\ i \gamma_\mu & O \end{pmatrix} \;\;\;\;\; (\mu=0,\dots,n-1) \; , \;\;\;
  \\ [0.3cm]
  \Gamma_{n} = \sigma_1 \otimes I = \begin{pmatrix} O & I \\ I & O \end{pmatrix} \; , \;\;\;
 \Gamma_{n+1} = i \, \sigma_2 \otimes \gamma_{n+1} =
  \begin{pmatrix} O & \gamma_{n+1} \\ -\gamma_{n+1} & O \end{pmatrix}  \; , \\ [0.3cm]
  \Gamma_{n+3} = -\alpha \, \Gamma_0 \cdot \Gamma_1 \cdots \Gamma_{n+1} =
  \begin{pmatrix} I & \; O \\ O & -I \end{pmatrix} \; ,
 \end{array}
 \end{equation}
 where $O$ is the $2^{n\over 2}$-dimensional zero matrix.
Here and below we use the standard Pauli matrices
\begin{equation}
 \label{pauli}
\sigma_1 = \begin{pmatrix} 0 & 1 \\ 1 & 0 \end{pmatrix} \; , \;\;\;
\sigma_2 = \begin{pmatrix} 0 & -i \\ i & 0 \end{pmatrix} \; , \;\;\;
\sigma_3 = \begin{pmatrix} 1 & 0 \\ 0 & -1 \end{pmatrix} \; .
 \end{equation}
 Matrices (\ref{cnfA4}), as it follows from (\ref{cnfA4a}), indeed satisfy
 Clifford relations for gamma-matrices in $\mathbb{R}^{p+1,q+1}$:
 \begin{equation}
 \label{cnfA40}
 \Gamma_a \, \Gamma_b  + \Gamma_b \, \Gamma_a = 2 \, g_{ab} \, {\bf I} \; , \;\;\;\;
 {\bf I} \equiv I_2 \otimes I \; ,
 \end{equation}
 where $I_2$ is the $2 \times 2$ unit matrix. Now the standard spinor representation $T$ for the
 generators $M_{ab}$ of the Lie algebra $so(p+1,q+1)$ (\ref{cnfA5}) is
 \begin{equation}
 \label{cnfA6}
 T(M_{ab}) = \frac{i}{4} \left( \Gamma_a \, \Gamma_b  - \Gamma_b \, \Gamma_a \right) \; .
 \end{equation}
 Substitution of (\ref{cnfA4}) into (\ref{cnfA6}) and using (\ref{cnfA2}) gives the
 spinor representation for $\textsf{conf}(\mathbb{R}^{p,q})$
 \begin{equation}
\label{cnfA8}
\begin{array}{c}
T(L_{\mu\nu}) = \frac{i}{4} \, I_2 \otimes \left[ \gamma_\mu ,  \, \gamma_\nu \right] \; , \;\;\;
 T(K_\mu) = \frac{1}{2} (I_2 \otimes  \gamma_{n+1} \, \gamma_\mu -
  \sigma_3 \otimes \gamma_\mu )  \; , \\ [0.3cm]
 \;\;\; T(P_\mu) = -\frac{1}{2} (I_2 \otimes  \gamma_{n+1} \, \gamma_\mu +
  \sigma_3 \otimes \gamma_\mu ) \; , \;\;\;
  T(D) =  \frac{i}{2} \, \sigma_3 \otimes \gamma_{n+1}  \; , \\ [0.3cm]
 \mu,\nu=0,1, \dots,n-1  \; , \;\;\;\; n=p+q   \; .
 \end{array}
 \end{equation}
This representation is reducible since all matrices (\ref{cnfA8}) have the block diagonal form
 \begin{equation}
\label{cnfA7}
\begin{array}{c}
T(L_{\mu\nu}) =
\left( \begin{array}{cc}
\frac{i}{4} \, \left[ \gamma_\mu ,  \, \gamma_\nu \right] & \mathbf{0} \\
\mathbf{0} & \frac{i}{4} \, \left[ \gamma_\mu ,  \, \gamma_\nu \right]
\end{array}
\right) \; , \;\;\;
 T(K_\mu) =
 \left( \begin{array}{cc}
 -\frac{1}{2} (1 -\gamma_{n+1}) \, \gamma_\mu  & \mathbf{0} \\
\mathbf{0} &   \frac{1}{2} (1+ \gamma_{n+1}) \, \gamma_\mu
 \end{array}
 \right)  \; , \\ [0.2cm]
 \;\;\; T(P_\mu) =
  \left( \begin{array}{cc}
 -\frac{1}{2} (1+\gamma_{n+1}) \, \gamma_\mu & \mathbf{0} \\
  \mathbf{0} &   \frac{1}{2} (1- \gamma_{n+1}) \, \gamma_\mu
 \end{array}
 \right) \; , \;\;\;
  T(D) =   \left( \begin{array}{cc}
  \frac{i}{2} \,  \gamma_{n+1}  & \mathbf{0} \\
  \mathbf{0} &   -\frac{i}{2} \, \gamma_{n+1}
 \end{array}
 \right)    \; .
 \end{array}
 \end{equation}
 Thus, the representation (\ref{cnfA8}),
 (\ref{cnfA7}) can be decomposed into the sum of two $2^{n \over 2}$-dimensional
 representations of $\textsf{conf}(\mathbb{R}^{p,q})$. In fact these two representations are related
 to each other by the obvious automorphism  of the conformal algebra (\ref{cnfA}):
 $$
 L_{\mu\nu} \to L_{\mu\nu} \; , \;\;\;  P_{\mu} \rightarrow - K_{\mu}
 \; , \;\;\;  K_{\mu} \rightarrow - P_{\mu} \; , \;\;\; D \to - D \; .
 $$
One of these  representations, after applying the
 commutation relations for gamma-matrices, can be written in the form
 \begin{equation}
\label{cnfA10}
\begin{array}{c}
 T_1(L_{\mu\nu}) = \frac{i}{4} \, \left[ \gamma_\mu ,  \, \gamma_\nu \right]
 \equiv \ell_{\mu \nu} \, , \;\;\;\;
  T_1(K_\mu) =  \gamma_\mu \, \frac{(1 - \gamma_{n+1})}{2} \equiv k_{\mu}  \, , \;\; \\ [0.2cm]
    T_1(P_\mu) =  \gamma_\mu \, \frac{(1 +\gamma_{n+1})}{2}  \equiv p_{\mu} \,  , \;\;\;\;
    T_1(D) = - \frac{i}{2} \, \gamma_{n+1} \equiv d \, ,
    \end{array}
\end{equation}
and it is not hard to check directly that the operators (\ref{cnfA10})
 possess needed commutation relations (\ref{cnfA}).

 \vspace{0.2cm}

 Further we use the common
representation (see, e.g., recurrence (\ref{cnfA4})) for the gamma-matrices (\ref{cnfA4a}):
 \begin{equation}
 \label{cnfA11s}
\gamma_{\mu} = \left(
  \begin{array}{cc}
  \mathbf{0} & \boldsymbol{\sigma}_{\mu} \\
  \overline{ \boldsymbol{\sigma} }_{\mu} & \mathbf{0}
  \end{array}
\right) \, , \;\; \gamma_{n+1} =  \left(
  \begin{array}{cc}
  \boldsymbol{1} & \mathbf{0} \\
  \mathbf{0} & - \boldsymbol{1}
  \end{array}
\right) \;\;\; \Longrightarrow \;\;\; \frac{I + \gamma_{n+1} }{2} = \left(
  \begin{array}{cc}
  \boldsymbol{1} & \mathbf{0} \\
  \mathbf{0} & \mathbf{0}
  \end{array}
\right) \, , \;\; \frac{I - \gamma_{n+1} }{2} = \left(
  \begin{array}{cc}
  \mathbf{0} & \mathbf{0} \\
  \mathbf{0} & \boldsymbol{1}
  \end{array}
\right) \; ,
 \end{equation}
 where $\boldsymbol{1}$, $\boldsymbol{\sigma}_{\mu}= ||(\boldsymbol{\sigma}_{\mu})_{\alpha\dot{\alpha}}||$
  and $\overline{ \boldsymbol{\sigma} }_{\mu} = ||(\overline{ \boldsymbol{\sigma} }_{\mu})^{\dot{\alpha}\alpha}||$
are $2^{{n \over 2}-1}$ - dimensional matrices;
 $\boldsymbol{1}$ is unit  matrix and $\boldsymbol{\sigma}_{\mu}$, $\overline{ \boldsymbol{\sigma} }_{\mu}$ satisfy
\begin{equation}
\label{idsi}
\boldsymbol{\sigma}_{\mu} \overline{ \boldsymbol{\sigma} }_{\nu} +
  \boldsymbol{\sigma}_{\nu} \overline{ \boldsymbol{\sigma} }_{\mu} = 2 \, g_{\mu \nu} \boldsymbol{1} \; , \;\;\;
\overline{ \boldsymbol{\sigma} }_{\mu} \boldsymbol{\sigma}_{\nu} +
   \overline{ \boldsymbol{\sigma} }_{\nu} \boldsymbol{\sigma}_{\mu} = 2 \, g_{\mu \nu} \boldsymbol{1}  \; ,
   %\;\;\;   \boldsymbol{\sigma}_{\mu}^{\dagger} = \boldsymbol{\sigma}_{\mu} \; , \;\;\; %\overline{\boldsymbol{\sigma}}_{\mu}^\dagger = \overline{\boldsymbol{\sigma}}_{\mu}  \; .
\end{equation}
Equations (\ref{idsi}) follow from identities (\ref{cnfA4a}) and representation (\ref{cnfA11s})
 for $\gamma_{\mu}$.
In terms of gamma-matrices (\ref{cnfA11s}) conformal generators (\ref{cnfA10}) can be represented as
 \begin{equation}
 \label{spinR1}
 \ell_{\mu\nu} =  \frac{i}{4}[ \gamma_{\mu} , \gamma_{\nu} ] =
  \left(
  \begin{array}{cc}
  \frac{i}{4}(\boldsymbol{\sigma}_{\mu} \overline{ \boldsymbol{\sigma} }_{\nu} -
  \boldsymbol{\sigma}_{\nu} \overline{ \boldsymbol{\sigma} }_{\mu}) & \mathbf{0} \\
  \mathbf{0} & \frac{i}{4}(\overline{ \boldsymbol{\sigma} }_{\mu} \boldsymbol{\sigma}_{\nu} -
   \overline{ \boldsymbol{\sigma} }_{\nu} \boldsymbol{\sigma}_{\mu})
  \end{array}
\right)
= \left(
  \begin{array}{cc}
  \boldsymbol{\sigma}_{\mu\nu}  & \mathbf{0} \\
  \mathbf{0} & \overline{ \boldsymbol{\sigma} }_{\mu\nu}
  \end{array}
\right) \; ,
 \end{equation}
 $$
p^{\mu} = \left(
  \begin{array}{cc}
  \mathbf{0} & \mathbf{0} \\
  \overline{ \boldsymbol{\sigma} }^{\mu} & \mathbf{0}
  \end{array}
\right)\; , \;\;\; k^{\mu} = \left(
  \begin{array}{cc}
  \mathbf{0} & \boldsymbol{\sigma}^{\mu} \\
  \mathbf{0} & \mathbf{0}
  \end{array}
\right)\; , \;\;\;  d = -\frac{i}{2}\left(
  \begin{array}{cc}
  \boldsymbol{1} & \mathbf{0} \\
  \mathbf{0} & - \boldsymbol{1} \\
  \end{array}
\right) \; .
$$
Recall that the matrices $\ell_{\mu\nu}$ as well as their diagonal blocks
\begin{equation}
\label{spinR}
\boldsymbol{\sigma}_{\mu \nu} = \frac{i}{4}(\boldsymbol{\sigma}_{\mu} \overline{ \boldsymbol{\sigma} }_{\nu} -
  \boldsymbol{\sigma}_{\nu} \overline{ \boldsymbol{\sigma} }_{\mu})
  = ||(\boldsymbol{\sigma}_{\mu\nu})_{\alpha}^{\;\; \beta}||  \; , \;\;\;
 \overline{\boldsymbol{\sigma}}_{\mu \nu} =  \frac{i}{4}(\overline{ \boldsymbol{\sigma} }_{\mu} \boldsymbol{\sigma}_{\nu} -
   \overline{ \boldsymbol{\sigma} }_{\nu} \boldsymbol{\sigma}_{\mu})
   = ||( \overline{\boldsymbol{\sigma}}_{\mu \nu})^{\dot{\alpha}}_{\;\; \dot{\beta}}||  \; ,
   \end{equation}
 are different spinor representations of the basis elements $L_{\mu\nu}$ of the algebra $so(p,q)$.
 %form the basis of the algebra $so(p,q)$ in spinor representations.

 \vspace{0.2cm}

 \noindent
 {\bf Remark 1.} It is well known that
 any two $2^{n/2}$-dimensional representations of the Clifford algebra (\ref{cnfA4a}) are equivalent.
 Since the both sets of matrices $\{ \gamma_\mu \}$  and $\{ \gamma_\mu^\dagger \}$
 represent the same Clifford algebra (\ref{cnfA4a}) we have
\begin{equation}
\label{cnfA4d}
 \gamma_\mu^\dagger   = {\bf C} \cdot \gamma_\mu \cdot {\bf C}^{-1} \; , \;\;\;\; \mu =0,\dots,n-1 \; ,
 \end{equation}
 where matrix ${\bf C}\in {\sf Mat}(2^{n/2})$ can be fixed such that ${\bf C}^\dagger = {\bf C}$.
 For matrices $\ell_{\mu\nu}$ (\ref{spinR1})
 and corresponding group elements
\begin{equation}
 \label{cnfA4p}
\mathbf{U} = \exp(i \omega^{\mu\nu}\ell_{\mu\nu}) =
\begin{pmatrix} ||\mathbf{\Lambda}_{\alpha}^{\;\; \beta}|| & {\bf 0} \\ {\bf 0} &
||\overline{\mathbf{\Lambda}}^{\dot{\alpha}}_{\;\; \dot{\beta}}|| \end{pmatrix}
 \; , \;\;\;\;\;\; \det(\mathbf{U})=1 \; ,
 \end{equation}
where $\omega^{\mu\nu} \in \mathbb{R}$, relations (\ref{cnfA4d}) give
 \begin{equation}
\label{cnfA4t}
\ell_{\mu\nu}^\dagger = {\bf C} \ell_{\mu\nu} {\bf C}^{-1} \;\; \Rightarrow \;\;
 \mathbf{U}^\dagger {\bf C}  \mathbf{U} = {\bf C} \; .
 \end{equation}
 The last equation means that $\mathbf{U}$ are pseudo-unitary matrices and their upper-diagonal
 blocks $\mathbf{\Lambda}$ (as well as their low-diagonal blocks $\overline{\mathbf{\Lambda}}$)
 generate matrix Lie group which is denoted as
 ${\sf Spin}(p,q)$.
  Definition (\ref{cnfA4b}) of $\gamma_{n+1}$ and relations (\ref{cnfA4d}) yield
 \begin{equation}
\label{cnfA4f}
 \gamma_{n+1}^\dagger =
 \alpha^* \; {\bf C} \cdot \gamma_{n-1} \cdots \gamma_0 \cdot {\bf C}^{-1}
 %= \alpha^* \alpha^2 (-1)^q \; {\bf C} \cdot \gamma_0 \cdots \gamma_{n-1} \cdot {\bf C}^{-1}
  = (-1)^q \, {\bf C} \cdot \gamma_{n+1} \cdot {\bf C}^{-1} \; .
 \end{equation}
 Note that there is a freedom  in the definition of $\gamma$-matrices (\ref{cnfA11s}) and matrices $\boldsymbol{\sigma}_{\mu}$, $\overline{\boldsymbol{\sigma}}_{\mu}$ (\ref{idsi}):
  \begin{equation}
\label{freed}
 \gamma_\mu \to \begin{pmatrix} x & {\bf 0} \\ {\bf 0} & y \end{pmatrix} \gamma_\mu
  \begin{pmatrix} x^{-1} & {\bf 0} \\ {\bf 0} & y^{-1}  \end{pmatrix} \;\; \Rightarrow \;\;
  \boldsymbol{\sigma}_{\mu} \to x\cdot \boldsymbol{\sigma}_{\mu} \cdot y^{-1} \; , \;\;\;\;
\overline{\boldsymbol{\sigma}}_{\mu} \to y \cdot \overline{\boldsymbol{\sigma}}_{\mu} \cdot x^{-1} \; ,
 \end{equation}
where $x,y \in {\sf Mat}(2^{n/2-1})$.
 Then, applying this freedom\footnote{The matrix $\gamma_{n+1}$ does
 not changed under the transformations (\ref{freed}) and one can bring one of
  the matrices $\gamma_\mu$ (for $\mu$ such that $g_{\mu\mu}=+1$),
 say $\gamma_0$, to the standard form $\gamma_0 =
  \begin{pmatrix} {\bf 0} & {\bf 1} \\ {\bf 1} & {\bf 0}\end{pmatrix}$.} and using
  relations (\ref{cnfA4d}), (\ref{cnfA4f}) and explicit form (\ref{cnfA11s}) of matrix
 $\gamma_{n+1} =  \gamma_{n+1}^\dagger$,  we partially fix the matrix ${\bf C}$
 according to the cases:
 %derive two cases:
 \begin{equation}
\label{cnfA4g}
\begin{array}{l}
 1.) \;\;\; q \; - \; {\rm even} \;\;\; \Rightarrow \;\;\;
  {\bf C} = \begin{pmatrix} {\bf c} & {\bf 0} \\ {\bf 0} & {\bf c} \end{pmatrix}
  \, , \;\;\;\;
  {\bf c}^\dagger = {\bf c} \; \in \; {\sf Mat}(2^{n/2-1}) \; ; \\ [0.3cm]
 2.) \;\;\;  q \; - \; {\rm odd} \;\;\; \Rightarrow \;\;\;
  {\bf C} = \begin{pmatrix} {\bf 0} &  {\bf g} \\ {\bf g} & {\bf 0} \end{pmatrix} \, , \;\;\;
    {\bf g}^\dagger = {\bf g} \; \in \; {\sf Mat}(2^{n/2-1}) \; .
  \end{array}
\end{equation}
Finally, from (\ref{cnfA4g}) we deduce the following relations for diagonal blocks
$\mathbf{\Lambda}$ and $\overline{\mathbf{\Lambda}}$ of the matrices $\mathbf{U}$ (\ref{cnfA4p}), (\ref{cnfA4t}):
 \begin{equation}
\label{cnfA4r}
\begin{array}{l}
 1.)  \;\;\; q \; - \; {\rm even} \;\;\; \Rightarrow \;\;\;
  \mathbf{\Lambda}^\dagger \cdot {\bf c} \cdot \mathbf{\Lambda} = \mathbf{c} \; , \;\;\;
  \overline{\mathbf{\Lambda}}^\dagger \cdot {\bf c} \cdot \overline{\mathbf{\Lambda}} = \mathbf{c}  \; ; \\ [0.2cm]
  2.) \;\;\;  q \; - \; {\rm odd} \;\;\; \Rightarrow \;\;\;
   \overline{\mathbf{\Lambda}}  = {\bf g}^{-1} \cdot (\mathbf{\Lambda}^{-1})^\dagger
  \cdot {\bf g} \; .
  \end{array}
\end{equation}

\vspace{0.2cm}

 \noindent
 {\bf Remark 2.} For the complexification of the
 %\marginpar{\bf Novyj Remark 2. Vozmozhno tol'ko ego
 %nado ostavit' vmesto Remark 1.}
 group ${\sf Spin}(p,q)$
 when parameters $\omega^{\mu \nu}$ in (\ref{cnfA4p}) are complex numbers the second relation
 in (\ref{cnfA4t}) is not valid. Here we present another conditions for blocks  $\mathbf{\Lambda},\overline{\mathbf{\Lambda}} \in {\sf Spin}(p,q)$
 which are correct even for the complex case and
 which will be used below. Again the sets of matrices $\{ \gamma_\mu \}$  and $\{ \gamma_\mu^{T} \}$
 represent the same Clifford algebra (\ref{cnfA4a}) and therefore we have
\begin{equation}
\label{A4d}
 \gamma_\mu^T   = \mathrm{C} \cdot \gamma_\mu \cdot \mathrm{C}^{-1} \; , \;\;\;\; \mu =0,\dots,n-1 \; .
 \end{equation}
 For matrices $\ell_{\mu\nu}$ (\ref{spinR1})
 and corresponding group elements (\ref{cnfA4p})
relations (\ref{A4d}) give
 \begin{equation}
\label{A4t}
\ell_{\mu\nu}^T = - \mathrm{C} \cdot \ell_{\mu\nu} \cdot \mathrm{C}^{-1} \;\; \Rightarrow \;\;
 \mathbf{U}^T \cdot \mathrm{C} \cdot \mathbf{U} = \mathrm{C} \; .
 \end{equation}
  Definition (\ref{cnfA4b}) of $\gamma_{n+1}$ and relations (\ref{A4d}) yield
 \begin{equation}
\label{A4f}
 \gamma_{n+1}^T =
 \alpha \; (-1)^n \, \mathrm{C} \cdot \gamma_{n-1} \cdots \gamma_0 \cdot \mathrm{C}^{-1}
  = (-1)^{n(n-1)/2} \, \mathrm{C} \cdot \gamma_{n+1} \cdot \mathrm{C}^{-1} \; ,
 \end{equation}
 where we have used that $n$ is the even number.
 Then, applying the freedom (\ref{freed}) and using
  relations (\ref{A4d}), (\ref{A4f}) and explicit form (\ref{cnfA11s}) of the matrix
 $\gamma_{n+1} =  \gamma_{n+1}^T$,  we fix operator $\mathrm{C}$ in (\ref{A4d})
 according to the cases:
 \begin{equation}
\label{A4g}
\begin{array}{l}
 1.) \;\;\; \frac{n(n-1)}{2} \; - \; {\rm even} \;\;\; \Rightarrow \;\;\;
  \mathrm{C} = \begin{pmatrix} \mathrm{c} & {\bf 0} \\ {\bf 0} & \mathrm{c} \end{pmatrix}
  \, , \;\;\;\;
  \mathrm{c}^T = \mathrm{c} \; \in \; {\sf Mat}(2^{n/2-1}) \; ; \\ [0.3cm]
 2.) \;\;\;  \frac{n(n-1)}{2} \; - \; {\rm odd} \;\;\; \Rightarrow \;\;\;
  \mathrm{C} = \begin{pmatrix} {\bf 0} &  \mathrm{g} \\ \mathrm{g}& {\bf 0} \end{pmatrix} \, , \;\;\;
    \mathrm{g}^T = \mathrm{g} \; \in \; {\sf Mat}(2^{n/2-1}) \; .
  \end{array}
\end{equation}
Finally, from (\ref{A4g}) we deduce the following relations for diagonal blocks
$\mathbf{\Lambda}$ and $\overline{\mathbf{\Lambda}}$ of the matrices $\mathbf{U}$ (\ref{cnfA4p}), (\ref{A4t}):
 \begin{equation}
\label{A4r}
\begin{array}{l}
 1.)  \;\;\; \frac{n(n-1)}{2} \; - \; {\rm even} \;\;\; \Rightarrow \;\;\;
  \mathbf{\Lambda}^T \cdot \mathrm{c} \cdot \mathbf{\Lambda} = \mathrm{c} \; , \;\;\;
  \overline{\mathbf{\Lambda}}^T \cdot
  \mathrm{c} \cdot \overline{\mathbf{\Lambda}} = \mathrm{c}  \; ; \\ [0.2cm]
  2.) \;\;\;  \frac{n(n-1)}{2} \; - \; {\rm odd} \;\;\; \Rightarrow \;\;\;
  \mathbf{\Lambda}^T  \cdot  \mathrm{g} \cdot \overline{\mathbf{\Lambda}}  = \mathrm{g} \; .
  \end{array}
\end{equation}

    \subsection{Differential realization
    for conformal algebra and induced representations}

\mbox{} The standard differential representation $\rho$ of elements
$\{L_{\mu \nu} , P_{\mu},K_{\mu},D \}$ of the algebra (\ref{cnfA}) is \cite{MackSalam}:
 \begin{equation}
\label{cnfA1}
\begin{array}{cc}
 \rho(P_{\mu}) = - i \partial_{x_\mu}  \equiv \hat{p}_{\mu} \; , &
 \rho(D) = x^{\mu} \hat{p}_{\mu} - i \Delta \; , \\ [0.2cm]
 \rho(L_{\mu \nu}) = \hat{\ell}_{\mu \nu} + S_{\mu \nu}   \; , & \;\;\;
 \rho(K_{\mu}) = 2 \, x^{\nu} \, (\hat{\ell}_{\nu \mu} + S_{\nu \mu})
+ (x^{\nu} x_{\nu}) \hat{p}_{\mu}   - 2 i \Delta x_{\mu} \; ,
\end{array}
\end{equation}
 $$
\hat{\ell}_{\mu \nu} \equiv (x_{\nu} \hat{p}_{\mu} -  x_{\mu} \hat{p}_{\nu}) \; ,
$$
where
$x_\mu$ are coordinates in $\mathbb{R}^{p,q}$, $\Delta \in \mathbb{R}$ -- conformal parameter,
$S_{\mu \nu}=-S_{\nu \mu}$ are spin generators with the same
commutation relations as for generators $L_{\mu \nu}$ (see (\ref{cnfA})):
 \begin{equation}
\label{spin}
[ S_{\mu \nu} , S_{\rho \sigma} ] = i( g_{\nu \rho} S_{\mu \sigma}
+ g_{\mu \sigma} S_{\nu \rho} - g_{\mu \rho} S_{\nu \sigma} -
g_{\nu \sigma} S_{\mu \rho} )  \; ,
 \end{equation}
and $[ S_{\mu \nu} , x_\rho] = 0 = [ S_{\mu \nu} , \hat{p}_\rho]$. Note that
in the differential representation (\ref{cnfA1}) the quadratic Casimir operator
(\ref{cnfA2b}) acquires the form:
\begin{equation}
\label{cnfA2z}
\rho(C_2) =  \frac{1}{2} \left( S_{\mu \nu} \, S^{\mu \nu}
- \hat{\ell}_{\mu \nu} \, \hat{\ell}^{\mu \nu} \right) + \Delta(\Delta - n)  \; .
\end{equation}

In this Subsection we obtain the differential realization (\ref{cnfA1}) of the conformal algebra
by means of the method of induced representations. Our method is slightly different from the method
which was used in \cite{MackSalam}.

First we pack generators (\ref{cnfA1}) into $2^{n/2}$-dimensional matrix
  \begin{equation}
 \label{cnfA16}
 \begin{array}{c}
\frac{1}{2} T_1(M^{ab}) \cdot \rho(M_{ab}) =
  \frac{1}{2} [ \ell^{\mu \nu} \cdot \rho(L_{\mu \nu}) +
  p^{\mu} \cdot \rho(K_{\mu}) +
  k^{\mu} \cdot \rho(P_{\mu}) ] - d \cdot \rho(D) = \\ [0.4cm]
 = \left(\!
  \begin{array}{cc}
  \mathbf{L} + \mathbf{S} + \frac{i}{2} \rho(D) \, \boldsymbol{1}  & \mathbf{p} \\ [0.2cm]
  \overline{\mathbf{K}}  & \;  \overline{\mathbf{L}} + \overline{\mathbf{S}}
   - \frac{i}{2} \rho(D) \, \boldsymbol{1}
  \end{array}
\! \right) \, ,
\end{array}
 \end{equation}
 where the representations $T_1$ and  $\rho$ were defined in (\ref{cnfA10}) and (\ref{cnfA1}), respectively.
In eq. (\ref{cnfA16}) and below we use notations
\begin{equation}
\label{TSPK}
\begin{array}{c}
\mathbf{L} = \frac{1}{2} \, \boldsymbol{\sigma}^{\mu\nu} \, \hat{\ell}_{\mu\nu} \; , \;\;\;
\overline{\mathbf{L}} = \frac{1}{2} \,
\overline{\boldsymbol{\sigma}}^{\mu\nu} \, \hat{\ell}_{\mu\nu}   \; , \;\;\;
\mathbf{p} = \frac{1}{2} \, \boldsymbol{\sigma}^{\mu} \, \hat{p}_\mu =
 -\frac{i}{2} \, \boldsymbol{\sigma}^{\mu} \, \partial_{x_\mu} \; , \\ [0.3cm]
\mathbf{S} = \frac{1}{2} \, \boldsymbol{\sigma}^{\mu\nu} \, S_{\mu\nu} \; , \;\;\;
\overline{\mathbf{S}} = \frac{1}{2} \, \overline{\boldsymbol{\sigma}}^{\mu\nu} \, S_{\mu\nu} \; , \;\;\;
\overline{\mathbf{K}} = \frac{1}{2} \, \overline{\boldsymbol{\sigma}}^{\mu} \, \rho(K_\mu)
\; , \;\;\;  \mathbf{x} = - i \, \overline{\boldsymbol{\sigma}}^{\mu} \, x_\mu  \; .
\end{array}
\end{equation}
Then we need the following technical result.
 %\vspace{0.2cm} \noindent {\bf Lemma.} {\it Operators (\ref{TSPK}) satisfy identities}
Namely, operators (\ref{TSPK}) satisfy identities
\begin{equation}
\label{cnfA14-a}
\mathbf{L} =  -\mathbf{p} \cdot \mathbf{x} - \frac{i}{2} (\hat{p}_\mu  x^\mu) \, \boldsymbol{1} \; , \;\;\;
\overline{\mathbf{L}} = \mathbf{x}  \cdot \mathbf{p}  +
\frac{i}{2} (x^\mu \hat{p}_\mu)  \, \boldsymbol{1} \; ,
\end{equation}
\begin{equation}
\label{cnfA14-b}
\overline{\mathbf{K}} =  (\mathbf{x} \cdot \mathbf{S} - \overline{\mathbf{S}} \cdot \mathbf{x})
 -  \mathbf{x} \cdot \mathbf{p} \cdot \mathbf{x} +
  \left(\Delta -\frac{n}{2} \right) \, \mathbf{x} \; .
\end{equation}
 %where $\mathbf{x} = - i \, \overline{\boldsymbol{\sigma}}^{\mu} \, x_\mu$.}
 %\noindent {\bf Proof.}
Indeed, the first identity in (\ref{cnfA14-a}) can be deduced as following
$$
\mathbf{L} = \frac{i}{4} \,
\boldsymbol{\sigma}^{\mu} \overline{ \boldsymbol{\sigma} }^{\nu} (\hat{p}_\mu x_\nu   - \hat{p}_\nu x_\mu ) =
- \frac{1}{2} \, \mathbf{p} \cdot \mathbf{x}  - \frac{i}{4} \,
(2 g^{\mu\nu} \boldsymbol{1} - \boldsymbol{\sigma}^{\nu} \overline{ \boldsymbol{\sigma} }^{\mu}) \hat{p}_\nu x_\mu  =
- \mathbf{p} \cdot \mathbf{x} - \frac{i}{2} \, (\hat{p}^\mu x_\mu) \,  \boldsymbol{1} \; ,
$$
where we have applied (\ref{idsi}). Second identity in (\ref{cnfA14-a}) is obtained analogously.
To prove identity (\ref{cnfA14-b}) we note that
\begin{equation}
\label{cnfA15-a}
\begin{array}{c}
\mathbf{x} \cdot \mathbf{S} - \overline{\mathbf{S}} \cdot \mathbf{x} =
 %\frac{1}{4} (\overline{\boldsymbol{\sigma}}^{\rho} x_\rho)
 %\cdot (\boldsymbol{\sigma}^{\mu} \overline{\boldsymbol{\sigma}}^{\nu} \, S_{\mu\nu})
 %- \frac{1}{4} (\overline{\boldsymbol{\sigma}}^{\mu} \boldsymbol{\sigma}^{\nu} \, S_{\mu\nu}) \cdot
 %(\overline{\boldsymbol{\sigma}}^{\rho} x_\rho) = \\ [0.2cm]
 %= \frac{1}{4}  \left( \overline{\boldsymbol{\sigma}}^{\rho}
 %\, \boldsymbol{\sigma}^{\mu} \, \overline{\boldsymbol{\sigma}}^{\nu} -
 %\overline{\boldsymbol{\sigma}}^{\mu} \, \boldsymbol{\sigma}^{\nu} \,
 %\overline{\boldsymbol{\sigma}}^{\rho} \right) x_\rho \, S_{\mu\nu}
 %= \frac{1}{4}  \left( 2 \, g^{\rho \mu} \, \overline{\boldsymbol{\sigma}}^{\nu} - 2 \, g^{\rho \nu} \,
 %\overline{\boldsymbol{\sigma}}^{\mu}  \right) x_\rho \, S_{\mu\nu} =
\overline{\boldsymbol{\sigma}}^{\nu} \, x^\mu  \, S_{\mu\nu}  \; , \\ [0.2cm]
 %\end{array} \end{equation} and \begin{equation} \label{cnfA15-b} \begin{array}{c}
\mathbf{x} \cdot \mathbf{p} \cdot \mathbf{x} =
 %- \frac{1}{2} \overline{\boldsymbol{\sigma}}^{\nu}  \boldsymbol{\sigma}^{\rho} \,
 %\overline{\boldsymbol{\sigma}}^{\mu} \, x_\nu  \, \hat{p}_\rho  \, x_\mu =
 %- \frac{1}{4} \left( \overline{\boldsymbol{\sigma}}^{\nu}  \boldsymbol{\sigma}^{\rho} \,
 %\overline{\boldsymbol{\sigma}}^{\mu} +
 %\overline{\boldsymbol{\sigma}}^{\mu}  \boldsymbol{\sigma}^{\rho} \,
 %\overline{\boldsymbol{\sigma}}^{\nu} \right) x_\nu \, \hat{p}_\rho \, x_\mu  = \\ [0.2cm]
 %= - \frac{1}{4} \left( 2 g^{\nu \rho} \, \overline{\boldsymbol{\sigma}}^{\mu} +
 %2 g^{\mu \rho} \, \overline{\boldsymbol{\sigma}}^{\nu}  - 2 g^{\mu \nu} \,
 %\overline{\boldsymbol{\sigma}}^{\rho}
 %\right) x_\nu \, (x_\mu \, \hat{p}_\rho - i \, g_{\mu\rho} ) = \\ [0.2cm] =
 - (\overline{\boldsymbol{\sigma}}^{\mu} x_\mu)  \, (x^\nu \, \hat{p}_\nu) + \frac{1}{2}  \, x^2 \,
(\overline{\boldsymbol{\sigma}}^{\mu} \hat{p}_\mu) +
 \frac{n}{2} \, i \, (\overline{\boldsymbol{\sigma}}^{\mu} x_\mu) \; ,
 \end{array}
\end{equation}
where we again applied (\ref{idsi}).
Then (\ref{cnfA14-b}) follows from (\ref{cnfA1}) and (\ref{cnfA15-a}).
 %and (\ref{cnfA15-b}).
 %$$ \overline{\mathbf{K}} = \overline{\boldsymbol{\sigma}}^{\mu}
 %\left( \! x_\mu  x^{\nu} \hat{p}_\nu   +  x^{\nu} S_{\nu \mu} - \frac{1}{2} \,  (x^2) \hat{p}_{\mu}
 %- i \Delta x_{\mu} \! \right) = - \frac{n}{2} \, \mathbf{x} - \mathbf{x} \cdot \mathbf{p} \cdot \mathbf{x}
 %+ \mathbf{x} \cdot \mathbf{S} - \overline{\mathbf{S}} \cdot \mathbf{x}
 %+ \Delta \, \mathbf{x} \; . $$ \hfill $\bullet$ \vspace{0.5cm}
 Now we substitute (\ref{cnfA14-a}), (\ref{cnfA14-b}) into (\ref{cnfA16}). As a result
the matrix (\ref{cnfA16}) can be written in the form
\begin{equation}
\label{cnfA27}
\frac{1}{2} T_1(M^{ab}) \cdot \rho(M_{ab}) = \left(
  \begin{array}{cc}
    \frac{\Delta -n}{2} \cdot \boldsymbol{1} + \mathbf{S} - \mathbf{p} \cdot  \mathbf{x} \; , & \;
    \mathbf{p} \\ [0.2cm]
  \mathbf{x} \cdot \mathbf{S} - \overline{\mathbf{S}} \cdot \mathbf{x}
    - \mathbf{x} \cdot \mathbf{p} \cdot \mathbf{x} +
     (\Delta-\frac{n}{2}) \cdot \mathbf{x}  \; , & \;\;
      - \frac{\Delta}{2} \cdot \boldsymbol{1} + \overline{\mathbf{S}} + \mathbf{x} \cdot \mathbf{p}
  \end{array}
\right) \; ,
\end{equation}
and this form of (\ref{cnfA16}) will be extensively used below.

\vspace{0.2cm}

Now we consider the set of matrices
 \begin{equation}
 \label{s6s4}
A = i (\omega^{\mu \nu} \, \ell_{\mu \nu} + a^\mu \, p_\mu + b^\mu \, k_\mu + \beta \, d ) \; , \;\;\;
(\omega^{\mu \nu}, \,  a^\mu, \, b^\mu, \, \beta \in \mathbb{R}) \; ,
 \end{equation}
which are the linear combinations of the generators (\ref{spinR1}). These matrices form a matrix Lie algebra.
 %Consider a group $G$ which is generated by Lie algebra (\ref{s6s4}).
The corresponding matrix Lie group $G$ is isomorphic to the group
${\sf Spin}(p+1,q+1)$. The elements $g \in G$ (at least that which are closed to unity)
can be represented in the exponential form
$$
 g = \exp(i \omega^{\mu \nu} \, \ell_{\mu \nu} + i a^\mu \, p_\mu + i b^\mu \, k_\mu + i \beta \, d ) \; .
$$
We stress that elements $g \in {\sf Spin}(p+1,q+1)$ satisfy one of the equations in (\ref{cnfA4r})
depending on the case of $(q+1)$ is even or odd.
The group $G \simeq {\sf Spin}(p+1,q+1)$
has a subgroup $H \subset G$ which is generated by elements $\{ \ell_{\mu\nu}, \; k_\mu, \; d \}$:
 \begin{equation}
 \label{hofH}
h = \exp(i \omega^{\mu \nu} \, \ell_{\mu \nu} + i b^\mu \, k_\mu + i \beta \, d ) \; \in \; H \; .
 \end{equation}
This fact immediately follows from  the commutation relations (\ref{cnfA}).
In the representation (\ref{spinR1}) the elements (\ref{hofH}) can be written in the matrix form
 \begin{equation}
 \label{hhL}
h =\begin{pmatrix}
e^{\beta \over 2} \cdot \exp(i \omega^{\mu \nu} \boldsymbol{\sigma}_{\mu\nu })
 & e^{\beta \over 2}\mathbf{\Lambda}_{0} \\
0 & e^{-{\beta \over 2}} \cdot \exp(i \omega^{\mu \nu} \overline{\boldsymbol{\sigma}}_{\mu\nu })
\end{pmatrix}  =
\begin{pmatrix}
\delta \cdot \mathbf{1} & 0 \\
0 & \overline{\delta} \cdot \mathbf{1}
\end{pmatrix} \cdot
\begin{pmatrix}
\mathbf{\Lambda} & \mathbf{\Lambda}_{0} \\
0 & \overline{\mathbf{\Lambda}}
\end{pmatrix}  \; ,
 \end{equation}
where we denote $\delta = \overline{\delta}^{\,\,-1} = e^{\beta \over 2}$. We recall that
matrices $\mathbf{\Lambda}$, $\overline{\mathbf{\Lambda}}$ were defined in (\ref{cnfA4p}), (\ref{cnfA4r})
and they satisfy $\det(\mathbf{\Lambda})=\det(\overline{\mathbf{\Lambda}})=1$.
The coset space $G/H$ can be parameterized by the special elements of ${\sf Spin}(p+1,q+1)$
$$
Z = \exp(- i x^\mu \, p_\mu) =
\begin{pmatrix}
{\bf 1} & \; 0 \\
- i x^\mu  \overline{\boldsymbol{\sigma}}_{\mu} & \; {\bf 1}
\end{pmatrix} =
\begin{pmatrix}
{\bf 1} & \; 0 \\
\mathbf{x} & \; {\bf 1}
\end{pmatrix}  \; ,
$$
and any element $g \in G$ is uniquely represented as a product $g=Z \cdot h$, where
$Z \in G/H$ and $h \in H$.
The group $G \simeq {\sf Spin}(p+1,q+1)$ acts on the coset space $G/H$ as following
\begin{equation}
\label{actgh}
g^{-1} \cdot Z = Z' \cdot h^{-1}  \;  , \;\;\;\;\;  \forall g \in G
\;\; , \;\; \forall Z  \in G/H \; ,
\end{equation}
where $h \in H$ and $Z' \in G/H$ depends on $g$ and $Z$. We take $g^{-1}$ and $Z'$ in the block form
 \begin{equation}
 \label{hmgm}
g^{-1} =
\begin{pmatrix}
A & B \\
C & D
\end{pmatrix} \; , \;\;\;
Z' =
\begin{pmatrix}
{\bf 1} & \; 0 \\
\mathbf{x}' & \; {\bf 1}
\end{pmatrix}  \in G/H  \; .
 \end{equation}
and from (\ref{actgh}) we deduce expressions:
 \begin{equation}\label{hx}
\mathbf{x}' = (C+ D \,\mathbf{x} )( A + B \,\mathbf{x})^{-1} \;\; , \
 \end{equation}
  \begin{equation}\label{hx1}
h^{-1} =
\begin{pmatrix}
A  + B \, \mathbf{x} & B \\
0 & D - \mathbf{x}' \, B
\end{pmatrix} \,.
 \end{equation}

For the subgroup $H$ consisting of elements $h$ (\ref{hhL}) we define
the representation $T$ which acts in the space of tensors
$\Phi_{\alpha_1 \dots \alpha_{2\ell}}^{\dot{\alpha}_1 \dots \dot{\alpha}_{2\dot{\ell}}}$ of the type
$(\ell,\dot{\ell})$:
 \begin{equation}\label{repind}
[T( h ) \cdot \Phi]_{\underline{\alpha}}^{ \underline{\dot{\alpha}}} =
 \delta^{\Delta} \; \overline{\delta}^{\Delta}
 \; t(\mathbf{\Lambda})_{\underline{\alpha}}^{\,\,\, \underline{\beta}} \;\cdot \;
\overline{t}(\overline{\mathbf{\Lambda}})^{\underline{\dot{\alpha}}}_{\,\,\, \underline{\dot{\beta}}} \;\cdot \; \Phi_{\underline{\beta}}^{\underline{\dot{\beta}}} \; .
 \end{equation}
Here we assume that parameters $\delta$, $\overline{\delta}$ of $h$ are independent
 and $\underline{\alpha}$ and $\underline{\dot{\alpha}}$ denote collections of indexes
$(\alpha_1 \dots \alpha_{2\ell})$ and $(\dot{\alpha}_1 \dots \dot{\alpha}_{2\dot{\ell}})$,
respectively. Matrices $t$ and $\overline{t}$ are two inequivalent representations of the subgroup
${\sf Spin}(p,q)\subset {\sf Spin}(p+1,q+1)$.
Matrix $t$ corresponds to the representation of the type $(\ell,0)$
 with undotted spinor indices while $\overline{t}$
corresponds to the representation of the type $(0,\dot{\ell})$ with dotted spinor indices.
In particular for $(1/2,0)$ and
$(0,1/2)$ type representations we have (see (\ref{cnfA4p})) $t(\mathbf{\Lambda})^{\alpha}_{\,\,\,\beta} = \mathbf{\Lambda}^{\alpha}_{\,\,\,\beta}$
and $\overline{t}(\overline{\mathbf{\Lambda}})_{\dot{\alpha}}^{\,\,\,\dot{\beta}}=
\overline{\mathbf{\Lambda}}_{\dot{\alpha}}^{\,\,\,\dot{\beta}}$, respectively.

Then we induce representation (\ref{repind}) of the subgroup $H$ to the representation $\rho$
of the whole group $G$. The representation $\rho$ acts in
 the space of tensor fields
  $\Phi_{\underline{\alpha}}^{ \underline{\dot{\alpha}}}(\mathbf{x})$ according to the rule
 \begin{equation}\label{repind1}
\rho(g) \cdot \Phi(\mathbf{x}) = [T(h) \cdot \Phi](\mathbf{x'}) \;\; ; \;\; h \in H \;\; ; \;\; g \in G \; ,
 \end{equation}
 where elements $g$, $h$ and parameters $\mathbf{x}$, $\mathbf{x'}$ are related
 by the formula (\ref{actgh}).

\vspace{0.5cm}

Our aim is to find the infinitesimal form of (\ref{repind1}). To do this
we first take the element $g^{-1}$ (\ref{hmgm}) in the infinitesimal form
 \begin{equation}
 \label{g-min}
g^{-1} =
\begin{pmatrix}
{\bf 1} - \varepsilon_{11} & -\varepsilon_{12} \\
- \varepsilon_{21} & {\bf 1} - \varepsilon_{22}
\end{pmatrix} = I - ||\varepsilon_{ij}||  \; ,
 \end{equation}
where the $2\times 2$ block matrix $||\varepsilon_{ij}||$ can be represented as linear combination (\ref{s6s4})
of $Spin(p+1,q+1)$ generators
and in particular we have $\tr (\varepsilon_{11}) = - \tr (\varepsilon_{22}) \in \mathbb{R}$.
It is easy to find from (\ref{hx}) that
$$
\mathbf{x'} = \mathbf{x} +
\left( - \varepsilon_{21} - \varepsilon_{22} \cdot \mathbf{x} + \mathbf{x} \cdot  \varepsilon_{11} +
\mathbf{x} \cdot  \varepsilon_{12} \cdot  \mathbf{x} \right)  \; ,
$$
and for the parameters $\delta$, $\overline{\delta}$
and diagonal blocks of matrix $h$ (\ref{hhL}) we have:
$$
\delta = 1 +   \mathrm{tr}[ \varepsilon_{11} +  \varepsilon_{12} \cdot \mathbf{x}] \; , \qquad
\overline{\delta} = 1 +  \mathrm{tr}[ \varepsilon_{22} - \varepsilon_{12} \cdot \mathbf{x}] \; ,
$$
$$
\mathbf{\Lambda}=
{\bf 1} + \left( \varepsilon_{11} + \varepsilon_{12} \cdot \mathbf{x} -
 \mathrm{tr}[ \varepsilon_{11} +\varepsilon_{12} \cdot \mathbf{x}] \cdot {\bf 1}  \right) \equiv
{\bf 1} + \varepsilon(\mathbf{x})
\; ,
$$
$$
\overline{\mathbf{\Lambda}}=
{\bf 1} + \left( \varepsilon_{22} - \mathbf{x} \cdot \varepsilon_{12} -
 \mathrm{tr}[ \varepsilon_{22} - \varepsilon_{12} \cdot \mathbf{x}] \cdot {\bf 1} \right) \equiv
{\bf 1} + \overline{\varepsilon}(\mathbf{x}) \; ,
$$
where to simplify formulas we normalize the trace such that $\mathrm{tr}({\bf 1})=1$. In particular this
normalization yields
 $$
 \mathrm{tr}[\boldsymbol{\sigma}_{\mu} \overline{\boldsymbol{\sigma}}_{\nu}]= g_{\mu\nu} \; , \;\;\;
  \mathrm{tr}[\boldsymbol{\sigma}_{\mu} \overline{\boldsymbol{\sigma}}_{\nu}
  \boldsymbol{\sigma}_{\lambda} \overline{\boldsymbol{\sigma}}_{\rho}]= 2
  \left( g_{\mu\nu}g_{\lambda\rho} - g_{\mu\lambda}g_{\nu\rho} + g_{\mu\rho}g_{\nu\lambda} \right) \; , \;\;\;
  \dots \; .
 $$
Further we assume that generators $S_{\mu \nu}$ (\ref{spin}) of infinitesimal Lorentz transformations
 are related to
matrix representations $t$ and $\overline{t}$ (\ref{repind}) of the Lorentz subgroup by means of the formulae
$$
t_{\underline{\alpha}}^{\,\,\, \underline{\beta}}({\bf 1} + \varepsilon(\mathbf{x}) ) =
\delta_{\underline{\alpha}}^{\;\, \underline{\beta}} +  2 \, \mathrm{tr} [ \varepsilon(\mathbf{x}) \, \mathbf{S}_{\underline{\alpha}}^{\,\,\, \underline{\beta}} ]
\; , \qquad
\overline{t}^{\underline{\dot{\alpha}}}_{\,\,\, \underline{\dot{\beta}}}
({\bf 1} + \overline{\varepsilon}(\mathbf{x})) =
\delta^{\underline{\dot{\alpha}}}_{\;\, \underline{\dot{\beta}}}
 + 2 \, \mathrm{tr} [ \overline{\varepsilon}(\mathbf{x}) \cdot \overline{\mathbf{S}}^{\underline{\dot{\alpha}}}_{\,\,\, \underline{\dot{\beta}}} ]  \; ,
$$
where $\mathbf{S}_{\underline{\alpha}}^{\,\,\, \underline{\beta}} = \frac{1}{2}
 (S^{\mu \nu})_{\underline{\alpha}}^{\,\,\, \underline{\beta}}
\cdot \boldsymbol{\sigma}_{\mu \nu}$ and
$\overline{\mathbf{S}}^{\underline{\dot{\alpha}}}_{\,\,\, \underline{\dot{\beta}}} =
 \frac{1}{2} (S^{\mu \nu})^{\underline{\dot{\alpha}}}_{\,\,\, \underline{\dot{\beta}}} \; \cdot \overline{\boldsymbol{\sigma}}_{\mu \nu}$ (see (\ref{TSPK})).
 Operators $(S^{\mu \nu})_{\underline{\alpha}}^{\,\,\, \underline{\beta}}$ and
 $(S^{\mu \nu})^{\underline{\dot{\alpha}}}_{\,\,\, \underline{\dot{\beta}}}$ define the action
 of generators $S_{\mu\nu}$ on the tensor fields of $(\ell,0)$ and $(0,\dot{\ell})$ types
 \begin{equation}
 \label{act-Phi1}
 \begin{array}{c}
(S^{\mu \nu})_{\underline{\alpha}}^{\,\,\, \underline{\beta}}
 \Phi_{\underline{\beta}} \; = \;
\left(\boldsymbol{\sigma}_{\mu\nu}\right)^{\ \ \alpha}_{\alpha_1}
\Phi_{\alpha \alpha_{2}\cdots \alpha_{2\ell}} + \cdots+
 \left(\boldsymbol{\sigma}_{\mu\nu}\right)^{\ \ \alpha}_{\alpha_{2\ell}}
\Phi_{\alpha_1 \cdots \alpha_{2\ell -1}\alpha} \; ,
 \\ [0.2cm]
 (S^{\mu \nu})^{\underline{\dot{\alpha}}}_{\,\,\, \underline{\dot{\beta}}}
 \Phi^{ \underline{\dot{\beta}}} \; = \;
 \left(\bar{\boldsymbol{\sigma}}_{\mu\nu}\right)^{\dot{\alpha}_1}_{\ \ \dot{\alpha}}
\Phi^{\dot{\alpha}\dot{\alpha}_2 \cdots \dot{\alpha}_{2\dot{\ell}}} + \cdots+ \left(\bar{\boldsymbol{\sigma}}_{\mu\nu}\right)^{\dot{\alpha}_{2\ell}}_{\ \ \dot{\alpha}}
\Phi^{\dot{\alpha}_1 \cdots \dot{\alpha}_{2\ell-1}\dot{\alpha}} \; .
 \end{array}
 \end{equation}
 Thus, for (\ref{repind1}) we have
$$
\rho(g) \, \Phi(\mathbf{x}) =  \left( 1 +
\left( \Delta \, \mathrm{tr}[\varepsilon_{11} + \varepsilon_{12} \, \mathbf{x} ]
 - \Delta \, \mathrm{tr}[\varepsilon_{22} - \varepsilon_{12} \, \mathbf{x} ] +
  2 \, \mathrm{tr} \left[ \left( \varepsilon_{11}+ \varepsilon_{12} \, \mathbf{x} \right) \mathbf{S} \right]
  +2 \, \mathrm{tr} \left[ \left( \varepsilon_{22} - \mathbf{x} \, \varepsilon_{12} \right) \overline{\mathbf{S}} \right] \, \right)  \right) \cdot
  $$
  \begin{equation}\label{phi}
 \cdot \, \Phi\left(\mathbf{x}+  \left(- \varepsilon_{21} - \varepsilon_{22} \, \mathbf{x}
  + \mathbf{x} \, \varepsilon_{11} + \mathbf{x} \, \varepsilon_{12} \, \mathbf{x} \right) \right) \; .
\end{equation}
According to (\ref{g-min})
we denote the infinitesimal part of $\rho(g)$ as $\rho(||\varepsilon_{ij}||)$
and write the l.h.s. of (\ref{phi}) in the form $\rho(g) \, \Phi(\mathbf{x}) =
\Phi(\mathbf{x}) + \rho(||\varepsilon_{ij}||) \cdot \Phi(\mathbf{x})$.
Next we transform infinitesimal part of r.h.s. of (\ref{phi}) in the form of the trace by using expansion
 \begin{equation}\label{phi1}
\Phi\left(\mathbf{x} - \epsilon(\mathbf{x}) \right) = \left( 1 + 2 \, \mathrm{tr} \left[ \epsilon(\mathbf{x}) \cdot \mathbf{p} \right] \right) \Phi(\mathbf{x}) \; , \;\;\;\;\; \mathbf{p} =
 - \frac{i}{2} \, \boldsymbol{\sigma}^\mu \, \partial_{x^\mu} \; ,
 \end{equation}
and cyclicity of the trace taking into account the noncommutativity of operators
 $\mathbf{x}$ and $\mathbf{p}$, e.g.,
$\mathrm{tr} \left[\mathbf{x} \cdot \varepsilon_{11} \cdot \mathbf{p}\right] =
\mathrm{tr} \left[\varepsilon_{11}(\mathbf{p} \cdot \mathbf{x}+\frac{n}{2})\right]$, etc.
 Note that the operator $\mathbf{p}$ is the same as in (\ref{TSPK}). As a result we
write (\ref{phi}) in the form
$$
\begin{array}{ccl}
\rho(||\varepsilon_{ij}||) \, \Phi(\mathbf{x}) & = &
    2 \, \mathrm{tr}
    \left[
    \varepsilon_{11} \cdot \left( \frac{\Delta-n}{2} + \mathbf{S} - \mathbf{p} \, \mathbf{x} \right) +
      \varepsilon_{12} \cdot \left( (\Delta - \frac{n}{2}) \, \mathbf{x} +  \mathbf{x} \, \mathbf{S}
        -  \overline{\mathbf{S}} \, \mathbf{x} - \mathbf{x} \, \mathbf{p} \, \mathbf{x} \right) + \right. \\ [0.2cm]
     & +  & \left. \varepsilon_{21} \cdot \mathbf{p}
      + \varepsilon_{22} \cdot \left( - \frac{\Delta}{2} + \overline{\mathbf{S}} + \mathbf{x} \, \mathbf{p} \right)
    \right] \Phi(\mathbf{x}) =
    %\Phi(\mathbf{x}) +
   \mathrm{Tr} \left[
    \begin{pmatrix} \varepsilon_{11} & \varepsilon_{12} \\ \varepsilon_{21} & \varepsilon_{22}
    \end{pmatrix} \, \left( T_1(M^{ab}) \otimes \rho(M_{ab}) \right) \right] \Phi(\mathbf{x}) \; .
    \end{array}
$$
From this formula we immediately recover  generators (\ref{cnfA1})
 of the conformal algebra collected in the blocks as they
appear in the matrix (\ref{cnfA27}).

\vspace{0.5cm}

At the end of this Section we list global forms (\ref{hx}) of four basic
conformal group transformations and give corresponding elements $h \in H$, $g \in G$
which are used in (\ref{repind1}).
\begin{itemize}
\item Translations

\begin{equation}
\label{tran}
g = e^{i a^{\mu} p_{\mu}} =
\begin{pmatrix} {\bf 1} & 0\\
 \mathbf{a} & {\bf 1} \end{pmatrix}
\; , \;\;\;\;
\mathbf{x}' = \mathbf{x} - \mathbf{a} \; , \;\;\;\;
h = \begin{pmatrix} {\bf 1} & 0\\
 0 & {\bf 1} \end{pmatrix}  \; , \;\;\;\; \mathbf{a} := i a^{\mu} \overline{\boldsymbol{\sigma}}_{\mu} \; .
\end{equation}

\item Lorentz rotations

\begin{equation}
\label{Lor}
g = e^{i \omega^{\mu\nu} \ell_{\mu\nu}} =
\begin{pmatrix} \mathbf{\Lambda} & 0\\
 0 & \overline{\mathbf{\Lambda}} \end{pmatrix}  \; , \;\;\;\;
\mathbf{x}' = \overline{\mathbf{\Lambda}}^{\, -1} \cdot \mathbf{x} \cdot \mathbf{\Lambda} \; , \;\;\;\;
h=\begin{pmatrix}
\mathbf{\Lambda} & 0\\
0 & \overline{\mathbf{\Lambda}}
\end{pmatrix} \; .
 \end{equation}

\item Dilatation

\begin{equation}
\label{Dilat}
g = e^{i \beta d} =
\begin{pmatrix}
e^{\frac{\beta}{2}} & 0\\ 0 & e^{-\frac{\beta}{2}}
\end{pmatrix} \; , \;\;\;\;
\mathbf{x}' = e^{\beta} \, \mathbf{x} \; , \;\;\;\;
h=\begin{pmatrix}
e^{\frac{\beta}{2}} & 0\\ 0 & e^{-\frac{\beta}{2}}
\end{pmatrix} \; .
\end{equation}

\item Special conformal transformations

\begin{equation}
\label{sp-con}
g = e^{i b^{\mu} k_{\mu}}
=\begin{pmatrix}
\mathbf{1} & \boldsymbol{b} \\
0 & \mathbf{1}
\end{pmatrix} \; , \;\;\;\; \mathbf{x}' = \mathbf{x} \cdot
(\mathbf{1} - \boldsymbol{b} \cdot\mathbf{x})^{-1} \; ,
\end{equation}
$$
h = \begin{pmatrix}
(\mathbf{1} - \boldsymbol{b} \cdot\mathbf{x})^{-1} \; , \; & \;\;
 (\mathbf{1} - \boldsymbol{b} \cdot\mathbf{x})^{-1} \cdot \boldsymbol{b}
  \cdot (1+ \mathbf{x}' \cdot \mathbf{b})^{-1} \\[0.2cm]
0 \; , & (1+ \mathbf{x}' \cdot \mathbf{b})^{-1}
\end{pmatrix} \; , \;\;\;\; \boldsymbol{b} := i b^{\mu} \boldsymbol{\sigma}_{\mu} \; .
$$

\end{itemize}
At the end of this Subsection we stress that all formulas (\ref{tran}) -- (\ref{sp-con}) are
written for the case when dimension $n=p+q$ is even.

\subsection{Spin operators $\mathbf{S}$ and $\overline{\mathbf{S}}$\label{spinSS}}

\mbox{} According to previous Subsection
we consider the representation of the conformal algebra in the space of tensor fields
$\Phi_{\alpha_1 \cdots \alpha_{2\ell}}^
{\dot{\alpha}_1\cdots \dot{\alpha}_{2\dot{\ell}}}(x)$
 (see (\ref{repind}) and (\ref{repind1})) of the type
$(\ell , \dot{\ell})$.
 Here and below in the capacity of the argument of fields $\Phi$ we use
 the point $x \in \mathbb{R}^{p,q}$ with coordinates $x_\mu$ instead of corresponding matrix
  $\mathbf{x}$ (\ref{TSPK}).
The generators $S_{\mu \nu}$ act on the tensor field
of the type $(\ell , \dot{\ell})$ according to the formulas (\ref{act-Phi1}):
 \begin{equation}
 \label{act-Phi}
 \begin{array}{c}
\left[S_{\mu \nu}\Phi\right]_{\alpha_1 \cdots \alpha_{2\ell}}^{\dot{\alpha}_1 \cdots \dot{\alpha}_{2\dot{\ell}}} =
\left(\boldsymbol{\sigma}_{\mu\nu}\right)^{\ \ \alpha}_{\alpha_1}
\Phi_{\alpha \alpha_{2}\cdots \alpha_{2\ell}}^{\dot{\alpha}_1 \cdots \dot{\alpha}_{2\dot{\ell}}} + \cdots+ \left(\boldsymbol{\sigma}_{\mu\nu}\right)^{\ \ \alpha}_{\alpha_{2\ell}}
\Phi_{\alpha_1 \cdots \alpha_{2\ell -1}\alpha}^{\dot{\alpha}_1 \cdots \dot{\alpha}_{2\dot{\ell}}} \; +
 \\ [0.2cm]
+ \left(\bar{\boldsymbol{\sigma}}_{\mu\nu}\right)^{\dot{\alpha}_1}_{\ \ \dot{\alpha}}
\Phi_{\alpha_1 \cdots \alpha_{2\ell}}^{\dot{\alpha}\dot{\alpha}_2 \cdots \dot{\alpha}_{2\dot{\ell}}} + \cdots+ \left(\bar{\boldsymbol{\sigma}}_{\mu\nu}\right)^{\dot{\alpha}_{2\ell}}_{\ \ \dot{\alpha}}
\Phi_{\alpha_1 \cdots \alpha_{2\ell}}^{\dot{\alpha}_1 \cdots \dot{\alpha}_{2\ell-1}\dot{\alpha}}\,.
 \end{array}
 \end{equation}

  First we discuss the very special case of representations of $so(p+1,q+1)$ when
  tensor fields
  $\Phi_{\alpha_1 \cdots \alpha_{2\ell}}^{\dot{\alpha}_1\cdots \dot{\alpha}_{2\dot{\ell}}}(x)$
   are such that
dotted and undotted indexes compose symmetric sets separately.
In this situation it is convenient to work with the generating functions
 \begin{equation}
 \label{gen-Phi}
\Phi(x , \lambda , \tilde{\lambda}) =
\Phi_{\alpha_1 \cdots \alpha_{2\ell}}^
{\dot{\alpha}_1 \cdots \dot{\alpha}_{2\dot{\ell}}}(x)
\, \lambda^{\alpha_1} \cdots \lambda^{\alpha_{2\ell}}
\, \tilde{\lambda}_{\dot{\alpha}_1} \cdots
\tilde{\lambda}_{\dot{\alpha}_{2\dot{\ell}}} \; ,
 \end{equation}
where $\lambda$ and $\tilde{\lambda}$ are auxiliary spinors.
Using these generating
functions the action (\ref{act-Phi}) of generators $S_{\mu\nu}$
can be written in a compact form
 \begin{equation}
 \label{difr}
\left[S_{\mu \nu}\Phi\right](x , \lambda , \tilde{\lambda})=
\left[\lambda\,\boldsymbol{\sigma}_{\mu\nu}\partial_{\lambda}+ \tilde{\lambda}\,\bar{\boldsymbol{\sigma}}_{\mu\nu}\partial_{\tilde{\lambda}}\right]
\Phi(x , \lambda , \tilde{\lambda}) \; ,
 \end{equation}
where
$$
\lambda\,\boldsymbol{\sigma}_{\mu\nu}\partial_{\lambda} = \lambda_{\alpha}\left(\boldsymbol{\sigma}_{\mu\nu}\right)^{\alpha}_{\,\,\, \beta}
\partial_{\lambda_{\beta}}\ \ \ ;\ \ \
 \tilde{\lambda}\,\bar{\boldsymbol{\sigma}}_{\mu\nu}\partial_{\tilde{\lambda}} = \tilde{\lambda}^{\dot{\alpha}}
\left(\bar{\boldsymbol{\sigma}}_{\mu\nu}\right)_{\dot{\alpha}}^{\,\,\, \dot{\beta}}
\partial_{\tilde{\lambda}^{\dot{\beta}}} \; .
$$
In accordance with (\ref{difr}) we obtain the
realization of the spin generators $S_{\mu\nu}$ as
differential operators over spinor variables
\begin{equation}
\label{SigmaMN}
S_{\mu \nu} =  \lambda \, \boldsymbol{\sigma}_{\mu \nu} \partial_{\lambda} +
\tilde{\lambda} \, \overline{\boldsymbol{\sigma}}_{\mu \nu} \partial_{\tilde{\lambda}}\,.
\end{equation}
One can easily show that operators $S_{\mu \nu}$ defined in (\ref{SigmaMN}) respect commutation relations
(\ref{spin}) for the algebra $so(p,q)$.

 Now we consider the 4-dimensional Minkowski case $n=4$, i.e. $\mathbb{R}^{p,q}=\mathbb{R}^{1,3}$.
  In this case the dimension of the spinor spaces is equal to $2^{n/2}=2$ and
  tensor fields $\Phi_{\alpha_1 \cdots \alpha_{2\ell}}^
{\dot{\alpha}_1\cdots \dot{\alpha}_{2\dot{\ell}}}(x)$ are automatically symmetric under permutations
of dotted and undotted indexes separately.
 For the Minkowski space $\mathbb{R}^{1,3}$, in the expressions for gamma-matrices (\ref{cnfA11s}),
  we choose
 \begin{equation}
 \label{sigmaM}
\boldsymbol{\sigma}_{\mu} = ( \sigma_0 , \sigma_1, \sigma_2, \sigma_3 ) \; ,
 %( \hat{\sigma}_0 , \vec{ \boldsymbol{\sigma} } ) \; ,
\qquad \overline{ \boldsymbol{\sigma} }_{\mu} =
 ( \sigma_0 , - \sigma_1, - \sigma_2, - \sigma_3 ) \; ,
 %( \hat{\sigma}_0 , - \vec{\boldsymbol{\sigma} } )
 \end{equation}
where $\sigma_0 = \mathrm{I}_2$ and $\sigma_1,\sigma_2,\sigma_3$
are standard Pauli matrices (\ref{pauli}). One can check that $\boldsymbol{\sigma}_{\mu}$,
$\overline{\boldsymbol{\sigma}}_{\mu}$ satisfy identities (\ref{idsi}) with $||g_{\mu\nu}|| =
{\rm diag}(+1,-1,-1,-1)$.
 To proceed further we note that
$$
\boldsymbol{\sigma}_{\mu \nu} \otimes \boldsymbol{\sigma}^{\mu \nu} =
\sigma_i \otimes \sigma_i \;\;\;, \;\;\;
\overline{\boldsymbol{\sigma}}_{\mu \nu} \otimes \overline{\boldsymbol{\sigma}}^{\mu \nu} =
\sigma_i \otimes \sigma_i
\;\;\;, \;\;\;
\boldsymbol{\sigma}_{\mu \nu} \otimes \overline{\boldsymbol{\sigma}}^{\mu \nu} =0 \; ,
$$
(sum over $i=1,2,3$ is implied) consequently by using (\ref{SigmaMN}) we get
for the self-dual components of $S_{\mu\nu}$
\begin{equation}
\label{spinorS}
\mathbf{S} = \frac{1}{2} \, \boldsymbol{\sigma}^{\mu\nu} \, S_{\mu\nu} =
\frac{1}{2} \, \sigma_i \cdot \left( \lambda \, \sigma_i \, \partial_{\lambda} \right)  =
\begin{pmatrix}
\frac{1}{2} \, \lambda_1 \partial_{\lambda_1} - \frac{1}{2} \, \lambda_2 \partial_{\lambda_2} &  \lambda_2 \partial_{\lambda_1} \\
 \lambda_1 \partial_{\lambda_2} &  - \frac{1}{2} \, \lambda_1 \partial_{\lambda_1} + \frac{1}{2} \, \lambda_2 \partial_{\lambda_2}
\end{pmatrix}
\end{equation}
and for anti-self-dual components of $S_{\mu\nu}$
\begin{equation}
\label{spinorSbar}
\overline{\mathbf{S}} = \frac{1}{2} \, \overline{\boldsymbol{\sigma}}^{\mu\nu} \, S_{\mu\nu} =
\frac{1}{2} \, \sigma_i \cdot \left( \tilde{\lambda} \, \sigma_i \, \partial_{\tilde{\lambda}} \right) =
\begin{pmatrix}
 \frac{1}{2} \, \tilde{\lambda}^{\dot{1}} \partial_{\tilde{\lambda}^{\dot{1}}} -
 \frac{1}{2} \, \tilde{\lambda}^{\dot{2}} \partial_{\tilde{\lambda}^{\dot{2}}} &
 \tilde{\lambda}^{\dot{2}} \partial_{\tilde{\lambda}^{\dot{1}}}  \\
 \tilde{\lambda}^{\dot{1}} \partial_{\tilde{\lambda}^{\dot{2}}} &
- \frac{1}{2} \, \tilde{\lambda}^{\dot{1}} \partial_{\tilde{\lambda}^{\dot{1}}}
+ \frac{1}{2} \, \tilde{\lambda}^{\dot{2}} \partial_{\tilde{\lambda}^{\dot{2}}}
\end{pmatrix}
\end{equation}
In fact the operator $\mathbf{S}$ is restricted to the space of
homogeneous polynomials in components of the spinor $\lambda$ of degree $2\ell$
(see (\ref{act-Phi}) and (\ref{gen-Phi})) so that one can choose new
 variables $\chi_1 = -\frac{\lambda_1}{\lambda_2}$, $t=-\lambda_2$ and obtain that $\mathbf{S}$ coincides with the following matrix $\mathbf{S}^{(\ell)}$ which contains parameter $\ell$ (the eigenvalue of the
 operator $\frac{1}{2}t\partial_t$):
\begin{equation}
 \label{sss}
 \begin{array}{c}
\mathbf{S}^{(\ell)}  = \begin{pmatrix}
\chi_1 \partial_{\chi_1} - \ell \; , & \; - \partial_{\chi_1} \\
\chi_1^2 \, \partial_{\chi_1} - 2 \ell \, \chi_1 \; , &
 \; - \chi_1 \, \partial_{\chi_1}  + \ell
\end{pmatrix} \equiv
\begin{pmatrix}
S_3 \; , & \; S_- \\
S_+ \; , & - S_3
 \end{pmatrix}
\; ,
\end{array}
\end{equation}
Similarly the operator $\bar{\mathbf{S}}$ is restricted to the space of
homogeneous polynomials in components of the spinor $\tilde{\lambda}$ of degree $2\dot{\ell}$ so that for the the choice
$\chi_2 = - \frac{\tilde{\lambda}^{\dot{1}}}{\tilde{\lambda}^{\dot{2}}}$ one obtains $\overline{\mathbf{S}} = \overline{\mathbf{S}}^{(\dot{\ell})}$, where
\begin{equation}
 \label{barsss}
 \begin{array}{c}
\overline{\mathbf{S}}^{(\dot{\ell})}  = \begin{pmatrix}
\chi_2 \partial_{\chi_{_2}} - \dot{\ell} \; , & \; - \partial_{\chi_{_2}} \\
\chi_2^2 \, \partial_{\chi_{_2}} - 2 \dot{\ell} \, \chi_2 \; , &
 \; - \chi_2 \, \partial_{\chi_{_2}}  + \dot{\ell}
\end{pmatrix} \equiv
\begin{pmatrix}
\overline{S}_3 \; , & \; \overline{S}_- \\
\overline{S}_+ \; , & - \overline{S}_3
 \end{pmatrix}
\; .
\end{array}
\end{equation}

For constructing general $\mathrm{R}$-operators in Section \ref{Lybe} we will need Euclidean
analogues of the previous formulas. For 4-dimaensional Euclidean space $\mathbb{R}^4$
we choose gamma-matrices (\ref{cnfA11s}) such that
$$
\boldsymbol{\sigma}_{\mu} = ( \sigma_0 , i \sigma_1, i \sigma_2, i \sigma_3 ) \; ,
 %( \hat{\sigma}_0 , \vec{ \boldsymbol{\sigma} } ) \; ,
\qquad \overline{ \boldsymbol{\sigma} }_{\mu} =
 ( \sigma_0 , - i \sigma_1, - i \sigma_2, - i \sigma_3 ) \; ,
 %( \hat{\sigma}_0 , - \vec{\boldsymbol{\sigma} } )
$$
and $\boldsymbol{\sigma}_{\mu}$, $\overline{ \boldsymbol{\sigma} }_{\mu}$
satisfy relations (\ref{idsi}) with $||g_{\mu\nu}|| =
{\rm diag}(+1,+1,+1,+1)$. Let us mention that
explicit experessions for
$\mathbf{S}^{(\ell)}$ , $\overline{\mathbf{S}}^{(\dot{\ell})}$
(\ref{sss}, \ref{barsss})
remains valid.

%%%%%%%%%%%%%%%%%%%%%%%%%%%%%%%%%%%%%%%%%%%%%%%%%%%%%%%%%%%%%%%%%%%%%%%%%%%%%%%%%%%%%%%%%%%%%
\section{$\mathrm{L}$-operators}
\label{Defs}\setcounter{equation}{0}
%%%%%%%%%%%%%%%%%%%%%%%%%%%%%%%%%%%%%%%%%%%%%%%%%%%%%%%%%%%%%%%%%%%%%%%%%%%%%%%%%%%%%%%%%%%%%

\mbox{} Let $V$  be a vector space and $I$ is the identity operator in $V$.
 Consider an operator $\mathrm{R}(u) \in {\rm End}(V \otimes V)$ which is a function
 of spectral parameter $u$ and satisfies
Yang-Baxter equation in the braid form
 \begin{equation}
 \label{Y-B1}
 \mathrm{R}_{12}(u - v) \, \mathrm{R}_{23}(u) \,  \mathrm{R}_{12}(v) =
  \mathrm{R}_{23}(v) \, \mathrm{R}_{12}(u)  \,  \mathrm{R}_{23}(u - v) \;\; \in  \;\; {\rm End}(V \otimes V \otimes V) \; .
 \end{equation}
 Here we use standard matrix notations of \cite{KS,FRT}, i.e.
 we denote by $\mathrm{R}_{23}(u)$ the operator $\mathrm{R}(u)$ which acts nontrivially
 in the second and third factors in $V \otimes V \otimes V$ and as identity $I$ on the first
 factor, then $\mathrm{R}_{12}(u) = \mathrm{R}(u) \otimes I$, etc.
Let $V'$  be another vector space and $I'$ is the identity operator in $V'$.
  We call operator
$\mathrm{L}(u) \in {\rm End}(V \otimes V')$ as \underline{$\mathrm{L}$-operator} in the spaces $V$ and $V'$
if it obeys intertwining relation
 \begin{equation}
 \label{kaz-d10}
 \mathrm{R}_{12}(u - v) \, \mathrm{L}_{13}(u) \,  \mathrm{L}_{23}(v) =
  \mathrm{L}_{13}(v) \, \mathrm{L}_{23}(u)  \,  \mathrm{R}_{12}(u - v)
   \;\; \in  \;\; {\rm End}(V \otimes V \otimes V') \; .
 \end{equation}
Here again indices $1,2,3$ indicate in which
factors of the space $V \otimes V \otimes V'$
the corresponding operators act nontrivially, e.g., $\mathrm{L}_{23}(v) = I \otimes \mathrm{L}(v)$,
$\mathrm{R}_{12}(u) = \mathrm{R}(u) \otimes I'$, etc.

In this Section we consider a special form of $\mathrm{L}$-operators which is related to simple Lie algebras
${\cal A}$ and their representations.
Let $X_a$ $(a=1,\dots,{\rm dim} {\cal A})$ be generators of ${\cal A}$
and $||{\sf g}_{ab}||$ -- matrix of the Killing form for ${\cal A}$ in the basis $\{ X_a \}$.
Introduce a polarized (or split) Casimir operator for ${\cal A}$
 \begin{equation}
 \label{polKaz}
{\bf r} = {\sf g}^{ab} \, X_a \otimes X_b  \; \in \; {\cal A} \otimes {\cal A} \; ,
 \end{equation}
 where ${\sf g}^{ab}$ is the inverse matrix of Killing form. Recall that
 quadratic Casimir operator $C_2 = {\sf g}^{ab} \, X_a \cdot X_b$ is
 the element of the enveloping algebra ${\cal U}({\cal A})$. The
operator ${\bf r}$ satisfies identity
 \begin{equation}
 \label{rrr}
 [{\bf r}_{12} + {\bf r}_{13}, \, {\bf r}_{23}] = 0 \; ,
 %\;\; \Rightarrow \;\;\; [{\bf r}_{13}, \, {\bf r}_{23}]
 %= \frac{1}{2} \, [{\bf r}_{12}, \; {\bf r}_{13} - {\bf r}_{23}] \; ,
 \end{equation}
where again we have used standard notations
$$
{\bf r}_{13} = {\sf g}^{ab} \, X_a \otimes {\bf 1} \otimes X_b \; , \;\;\;
{\bf r}_{12} =  {\sf g}^{ab} \, X_a \otimes  X_b \otimes {\bf 1}  \; , \;\;\;
{\bf r}_{23} =  {\sf g}^{ab} \, {\bf 1} \otimes  X_a \otimes X_b  \; ,
$$
and ${\bf 1}$ is the unit element in ${\cal U}({\cal A})$.

Let $T$ and $T'$ be
representations of ${\cal A}$ in vector spaces
 $V$ and $V'$, respectively. Further we investigate special
 solutions of equation (\ref{kaz-d10}) which can be represented
 in the form:
  \begin{equation}
 \label{LKaz}
 \mathrm{L}(u) = (T \otimes T') (u \,{\bf 1} \otimes {\bf 1} +  {\bf r}) =
 u \, (I \otimes I') + {\sf g}^{ab}  \, (T_a \otimes T'_b) \;\;
 \in  \;\; {\rm End}(V \otimes V') \; ,
 \end{equation}
 where $T_a := T(X_a)$ and $T'_b := T'(X_b)$.
The matrix (\ref{LKaz}) is constructed by means of polarized Casimir operator
(\ref{polKaz}) for the algebra ${\cal A}$ and as we show in next
Sections this matrix is a solution of equation (\ref{kaz-d10})
only for the special choice of the algebra ${\cal A}$ and representations $T$ and $T'$.

\vspace{0.3cm}

\subsection{The case of the algebra ${\cal A} = s\ell(N,\mathbb{C})$\label{slnA}}

\mbox{} Consider Lie algebra $g\ell(N,\mathbb{C})$ with generators
$E_{ij}$ ($i,j=1,\dots,N$) which obey commutation relations
 \begin{equation}
 \label{defGL}
 \begin{array}{c}
 [E_{ij}, \; E_{k\ell}] = \delta_{jk} E_{i\ell} - \delta_{i\ell} E_{kj} \; .
 \end{array}
  \end{equation}
One can embed Lie algebra $s\ell(N,\mathbb{C})$ into
the algebra $g\ell(N,\mathbb{C})$ by choosing generators $X_a$ of $s\ell(N,\mathbb{C})$ as
$E_{ij}$ ($i\neq j$ and $i,j=1,\dots,N$) and
$H_{k} = E_{kk} - \frac{1}{N} \sum_m E_{mm}$,
where only $(N-1)$ elements $H_{k}$ are independent in view of the equation
$\sum_k H_k =0$. These generators satisfy
commutation relations
 \begin{equation}
 \label{defSL}
 \begin{array}{c}
 [E_{ij}, \; E_{k\ell}] = \delta_{jk} E_{i\ell} - \delta_{i\ell} E_{kj} \; , \;\;\;\;\;
 i \neq \ell \; {\rm or} \; j \neq k \; , \;\;\; [E_{ij}, \; E_{ji}] = H_i - H_j \; , \\ [0.2cm]
 [H_k , \; E_{ij}] = (\delta_{ki}  - \delta_{kj}) E_{ij} \; , \;\;\; [H_k , \; H_{j}] = 0 \; .
 \end{array}
  \end{equation}
In the defining representation $T$ of $s\ell(N,\mathbb{C})$ the elements
$E_{ij}$ and $H_{k}$ are $(N \times N)$ traceless matrices
 \begin{equation}
 \label{repSL}
 T(E_{ij})= e_{ij} \; , \;\;\;  T(H_{k})= e_{kk} - \frac{1}{N} I_N \equiv h_k \; ,
 \end{equation}
 where $e_{ij}$ are matrix units and $I_N = \sum_j e_{jj}$. Matrices (\ref{repSL})
 act in the $N$-dimensional vector space $V_N = \mathbb{C}^N$. Introduce
 permutation matrix $P_{12}$ which acts in the space $V_N \otimes V_N$ as following
 \begin{equation}
 \label{IdPe}
P_{12} \, w_1 \otimes w_2 = w_2 \otimes w_1 \; , \;\;\;\; \forall \, w_1,w_2 \in  V_N \; .
 \end{equation}
 %Let $T'$ be any representation of $sl(n,\mathbb{C})$.

\noindent
{\bf Proposition 1. \cite{KRS}}
 {\it The operator (cf. (\ref{LKaz}))
\begin{equation}
 \label{kaz-d13f}
\mathrm{L}(u) = u \, I_N \otimes {\bf 1} + \sum_i \, h_i \otimes  H_i +
\sum_{i \neq j} \, e_{ij} \otimes E_{ji}  \; ,
 \end{equation}
 is the universal $\mathrm{L}$-operator for the Lie algebra $s\ell(N,\mathbb{C})$.
  In other words the operator (\ref{kaz-d13f})
 satisfies intertwining relations (\ref{kaz-d10})
with Yangian $\mathrm{R}$-matrix
 \begin{equation}
 \label{kaz-d10b}
 \begin{array}{c}
  %R_{12}(u) := u I_{n} \otimes I_n + P_{12} \; ,
  \mathrm{R}_{12}(u) := u \, P_{12}  + I_{N} \otimes I_N \; ,
  \end{array}
 \end{equation}
 and the universality means that the second factors in (\ref{kaz-d13f})
 can be taken
in an arbitrary representation $T'$ of $s\ell(N,\mathbb{C})$ (cf.
(\ref{LKaz})).}

 \vspace{0.1cm}

 \noindent
 {\bf Proof.} First we write operator (\ref{kaz-d13f}) in terms of $g\ell(N,\mathbb{C})$ generators
 \begin{equation}
 \label{Lopgl}
\mathrm{L}(u) = (u - 1/N) \, I_N \otimes {\bf 1} +  e_{ij} \otimes E_{ji}   \; ,
 \end{equation}
 where the sum over all indices $i$ and $j$ is implied. Note that
 the split Casimir operator ${\bf r} = E_{ij} \otimes E_{ji}$ satisfies equations (\ref{rrr}) and we have
 \begin{equation}
 \label{TTee}
 e_{ij} \otimes E_{ji} = (T \otimes 1) \,{\bf r} \; ,
  \;\;\; (T \otimes T)\,{\bf r} = e_{ij} \otimes e_{ji} = P_{12}    \; .
 \end{equation}
 Substitution of (\ref{Lopgl}) into (\ref{kaz-d10}) and using (\ref{TTee}) gives relation
 $$
 (u-v) \, P_{12} \, (T \otimes T \otimes 1)
 \left(  [{\bf r}_{13}, \, {\bf r}_{23}] +
   [{\bf r}_{12}, \, {\bf r}_{23}] \right) = 0  \; ,
 $$
which is identity in view of (\ref{rrr}). Thus, $\mathrm{L}(u)$ (\ref{kaz-d13f})
satisfies intertwining relation (\ref{kaz-d10}) and it means that $\mathrm{L}(u)$ is
the $\mathrm{L}$-operator.  \hfill $\bullet$

\vspace{0.2cm}

Now we take the second factors of
universal $\mathrm{L}$-operator (\ref{kaz-d13f}) in the differential representation $\rho$
of $s\ell(N,\mathbb{C})$ (see \cite{DM} for details)
and make the redefinition:
\begin{equation}
 \label{3.13'}
(1 \otimes \rho)\mathrm{L}(u+1/N) \;\;\; \to \;\;\;
\mathrm{L}(u) =
u  \, I_N \otimes \rho({\bf 1}) +  e_{ij} \otimes \rho(E_{ji}) \; .
 \end{equation}
The representation $\rho$ is characterized by
parameters $(\rho_1, \dots, \rho_N)$ subject condition
$\sum\limits_{k=1}^N \rho_k = N(N-1)/2$ and it is important that
spectral parameter $u$ and parameters $\rho_k$
 %of the representation are always
 %combined
 are always collected in combinations
 \begin{equation}
 \label{3.13''}
 u_k = u -\rho_k \; .
 \end{equation}
The $\mathrm{L}$-operator (\ref{3.13'})
 %\marginpar{\bf Zamenil (\ref{kaz-d13f}) na (\ref{3.13'})}
can be written in the factorized form \cite{DM}
\begin{equation}
 \label{Lsl4}
 \mathrm{L}(u_1, \dots, u_N) =
  \mathrm{Z}\cdot \mathrm{D}(u_1, \dots, u_N) \cdot \mathrm{Z}^{-1}  \; ,
\end{equation}
where low-triangular  and upper-triangular
 $(N \times N)$ matrices  $\mathrm{Z}$
and $\mathrm{D}$ are
\begin{equation}
\label{Lsln02}
\mathrm{Z} = I_N + \sum_{k>m} z_{km} \, e_{km}  \; , \;\;\;
\mathrm{D}(u_1, \dots, u_N) = \sum_{k=1}^N  u_k \, e_{kk} + \sum_{i<j} D_{ij} \, e_{ij}  \; .
\end{equation}
Here we use notation
\begin{equation}
\label{notZD}
\begin{array}{c}
D_{ij} = - \partial_{ji} - \sum\limits_{k=j+1}^N \, z_{kj} \, \partial_{ki}
\; , \;\;\; \partial_{ji} \equiv \frac{\partial}{\partial z_{ji}}   \; ,  \;\;\;\; (i < j) \; .
\end{array}
\end{equation}
We stress that all elements of matrices $\mathrm{Z}$
and $\mathrm{D}$ have to be interpreted as operators acting in the space of functions $f(\mathrm{Z})$.
The important fact is that there exist
operators $\mathcal{T}_k$ ($k=1,\dots,N-1$)
which permute parameters $u_k$ and $u_{k+1}$ in $\mathrm{L}$-operator:
\begin{equation}
\label{s-op}
\begin{array}{c}
\mathcal{T}_k \cdot \mathrm{L}(u_1, \dots, u_k,  u_{k+1}, \dots , u_n)
=
\mathrm{L}(u_1, \dots, u_{k+1},  u_k, \dots , u_n) \cdot \mathcal{T}_k \; .
\end{array}
\end{equation}
One can find that
\begin{equation}
\label{SkD}
\mathcal{T}_k = (D_{k,k+1})^{u_{k+1}-u_k} \; ,
\end{equation}
where $D_{k,k+1}$ are the elements of the matrix $D$.

The operators $\mathcal{T}_k$ have clear group theoretical meaning
as intertwining operators~\cite{GN, Knapp} for equivalent representations which differ by the permutation of parameters $\rho_k$ and $\rho_{k+1}$.
These intertwining operators corresponds to the
elementary transpositions $s_k$ in the Weyl group.
In the case  under consideration the Weyl group
of $s\ell(N,\mathbb{C})$ is the group of permutation of parameters $(\rho_1, \dots, \rho_n)$:
 \begin{equation}\label{3.18'}
s_k: \;\; (\rho_1, \dots, \rho_k,  \rho_{k+1}, \dots , \rho_n) \;\; \to \;\;
(\rho_1, \dots, \rho_{k+1},  \rho_k, \dots , \rho_n) \; .
 \end{equation}
As an illustration we present $\mathrm{L}$-operator (\ref{Lsl4}) for the simplest $s\ell(2,\mathbb{C})$ case $(N=2)$.
In this case $\rho_1 + \rho_2 =1$ and
it is convenient to use standard spin parameters $\ell$
instead of parameters $\rho_1$ and $\rho_2$ :
\begin{equation}
\label{Lsl21}
\rho_1 = \ell+1 \ \ ,\ \ \rho_2 = -\ell\ ,\ \ u_1 = u - \ell - 1 \; , \;\;\; u_2 = u + \ell \; .
\end{equation}
Then we write operator $\mathrm{L}(u_1,u_2)$ (\ref{Lsl4}) for
$N=2$  in the form
\begin{equation}
\label{Lsl22}
\begin{array}{c}
\mathrm{L}(u_1,u_2)  =
\begin{pmatrix}1&0 \\ z & 1 \end{pmatrix}
\begin{pmatrix}u_1& - \partial_{z} \\ 0 & u_2 \end{pmatrix}
\begin{pmatrix}1&0 \\ - z & 1 \end{pmatrix} = \\ [0.4cm]
= u \, I_2 +
\begin{pmatrix}
z\, \partial_{z}  - \ell \; , & \; - \partial_{z} \\
z^2 \, \partial_{z} - 2 \ell \, z  \; , &
\; - z \, \partial_{z}  + \ell
\end{pmatrix} =
%u \, I_2 +  \begin{pmatrix} S_3 \; , & \; S_- \\ S_+ \; , & - S_3 \end{pmatrix} =
u \, I_2  + \mathbf{S}^{(\ell)}\; ,
\end{array}
\end{equation}
where $z = z_{21}$ and elements of matrix $\mathbf{S}^{(\ell)}$
are generators of $s\ell(2,\mathbb{C})$ in the standard differential realization.
One can directly check the identity
$$
\partial_{z}^{2 \ell+1} \cdot \mathbf{S}^{(\ell)} =
\mathbf{S}^{(-\ell-1)}\cdot \partial_{z}^{2 \ell+1} \; ,
$$
which corresponds to the permutation $\rho_1 \leftrightarrow \rho_2$
and justifies (\ref{s-op}) and (\ref{SkD}). In Section {\bf \ref{sec-sl4}} we
investigate $\mathrm{L}$-operator (\ref{Lsl4}) for the $s\ell(4,\mathbb{C})$ case.

\subsection{The case of the Lie algebra ${\cal A} =so(p+1,q+1)$. Spinorial $\mathrm{R}$-matrix}

\mbox{} Let $\Gamma_a$ $(a=0,\dots,n+1)$ be $2^{{n\over 2}+1}$-dimensional
gamma-matrices in $\mathbb{R}^{p+1,q+1}$ (\ref{cnfA4}),  where $n=p+q$. Operators $\Gamma_{a}$ are
generators of the Clifford algebra (\ref{cnfA40}) which, as a vector
space, has dimension $2^{n+2}$. The standard basis in this
space is formed by antisymmetrized products
of the $\Gamma$-matrices
\begin{equation}
\label{Gamma01}
\Gamma_{a_1\ldots a_k} = \frac{1}{k!}
\sum_{s }(-1)^{\mathrm{p}(s)} \Gamma_{s(a_1)} \cdots\Gamma_{s(a_k)} \equiv \Gamma_{A_k}
\;\;\;\;\; (\forall  k \leq n+2) \; , \;\;\;\;
\Gamma_{A_k} = 0 \;\;\;\;\; (\forall  k > n+2) \; ,
\end{equation}
where the summation is taken over all permutations $s$ of $k$ indices
$\{a_1, \dots, a_k\} \to \{s(a_1), \dots, s(a_k)\}$ and
$\mathrm{p}(s)$ denote the parity of the permutation $s$.
We start from the general $SO(p+1,q+1)$-invariant expression for the
$\mathrm{R}$-matrix which acts in the tensor product
of two vector spaces $V$ of $2^{{n\over 2}+1}$-dimensional spinor representations of $SO(p+1,q+1)$
\begin{equation}
\label{r-mtr01}
\mathrm{R}(u) = \sum_{k=0}^{\infty} \frac{\mathrm{R}_k(u)}{k!}\cdot \Gamma_{a_1\ldots a_k} \otimes
 \Gamma^{a_1\ldots a_k} \;\;
  \in  \;\; {\rm End}(V \otimes V ) \; .
\end{equation}
Note that in the r.h.s. of (\ref{r-mtr01})
the summation over $k$ is not run up to infinity since it is
automatically truncated by the condition $k \leq n+2$ (see (\ref{Gamma01})).
 We show in the Appendix B (see also \cite{Witten}, \cite{KarT}, \cite{Resh}, \cite{ZamL}) that
the $\mathrm{R}$-matrix (\ref{r-mtr01}) satisfies Yang-Baxter equation (\ref{Y-B1})
 %rewritten in the braid form \begin{equation} \label{Y-B2}
 %\mathrm{R}_{12}(u - v) \, \mathrm{R}_{23}(u) \,  \mathrm{R}_{12}(v) =
 %\mathrm{R}_{23}(v) \, \mathrm{R}_{12}(u)  \,  \mathrm{R}_{23}(u - v) \;\;
 %\in  \;\; {\rm End}(V \otimes V \otimes V) \; , \end{equation}
if coefficient functions $\mathrm{R}_k(u)$ are fixed as
%\marginpar{\bf Otmetil, chto $R_{2k}=0$}
 $\mathrm{R}_{k}(u) = 0$ for odd $k$ and obey the
recurrent relation
\begin{equation} \label{reccur}
\mathrm{R}_{k+2}(u) = -\frac{u+k}{u+n-k} \,\mathrm{R}_{k}(u) \, ,
\end{equation}
for even $k$.
 Recall now that the Lie algebra $so(p+1,q+1)$ is
generated by elements
 $M_{ab}$ subject
 commutation relations (\ref{cnfA5}).
 The operator $\mathrm{L}(u)$ (\ref{LKaz}) of $so(p+1,q+1)$-type can be written in the form
 \begin{equation}
 \label{LKaz-so}
 \mathrm{L}(u) = u \, {\bf I} \otimes {\rm 1} \; + \frac{1}{2} \,
 T(M_{ab})  \otimes M^{ab}\,,
 \end{equation}
where $T$ denote the spinor representation (\ref{cnfA6}) of $so(p+1,q+1)$
which acts in the space $V$.
 The generators $M_{ab}$ which are in the second factors of $\mathrm{L}(u)$
  %are generators of $so(p+1,q+1)$
  can be thought as taken in arbitrary representation $T'$.
Now we investigate the cases when operator $\mathrm{L}(u)$ defined in (\ref{LKaz-so})
satisfies intertwining equation (\ref{kaz-d10}).
After substitution of $\mathrm{L}(u)$ (\ref{LKaz-so}) into (\ref{kaz-d10})
(with $\mathrm{R}$-matrix (\ref{r-mtr01})) and
 some calculations (see Appendix B) equation (\ref{kaz-d10}) acquires the form
$$
\sum_{k=0}^{\infty} \frac{1}{k!}\biggl[
\left(u+n-k\right) \mathrm{R}_{k+2}(u)+
\left(u+k\right) \mathrm{R}_{k}(u) \biggl]\,
T'(M_{ab}) \,
\biggl[\Gamma^{ab c_1\ldots c_k}\otimes\Gamma_{ c_1\ldots c_k} - \Gamma_{ c_1\ldots c_k}\otimes \Gamma^{ab c_1\ldots c_k}\biggl] +
$$
 \begin{equation}\label{RLLbig}
- \frac{i}{2}\,\sum_{k=0}^{\infty} \frac{1}{(k)!}\biggl[
\mathrm{R}_{k+3}(u)+\mathrm{R}_{k+1}(u)\biggl]\,
T'(\left\{M^{ab}\,,M^{c}{}_{d}\right\}) \,
\biggl[\Gamma_{ab c c_1\ldots c_k}\otimes\Gamma^{d c_1\ldots c_k} + \Gamma^{d c_1\ldots c_k}\otimes
\Gamma_{ab c c_1\ldots c_k}\biggl]
= 0\, ,
 \end{equation}
where $\left\{A\,,B\right\} = A\cdot B + B \cdot A$.
The first term in the previous equation
turns to zero due to recurrence relation (\ref{reccur}). The second term could be equal to zero for special
choice of the representation $T'$ of generators $M_{ab}$ and for special projections
in spinor spaces $V$, e.g., Weyl projections $V \to V_{\pm} = \frac{{\bf I} \pm \Gamma_{n+3}}{2}\, V$
or choice of the Majorana representation for gamma-matrices.
We consider more restrictive condition which is
\begin{equation}\label{asym}
T' \left( \left\{M_{[ab}\,,M_{c]d}\right\} \right) = 0 \; .
\end{equation}
Here square brackets denote antisymmetrization.
Below we itemize some cases when the condition (\ref{asym}) is fulfilled
\begin{itemize}
\item The differential representation $T'$:
\begin{equation}\label{T2}
M_{ab} \to T'(M_{ab})  = i (y_{a} \, \partial_{b}   - y_{b}\, \partial_{a} ) \, ,
 \end{equation}
where $ \partial_{a} = \frac{\partial}{\partial y^a}$ and $y_a$ are coordinates in the space $\mathbb{R}^{p+1,q+1}$.
\item
 Fundamental (defining) $(n+2)$-dimensional representation $T'$:
 \begin{equation}\label{T1}
M_{ab} \to T'(M_{ab})  = i g (e_{ab}  - e_{ba}) \, ,
 \end{equation}
where $e_{ab}$ are matrix units and $g = ||g_{ab}||$. This case was considered
in \cite{Witten} and \cite{KarT}.

\item The differential representation
$T'=\rho$ (\ref{cnfA1}) for the scalar case $S_{\mu\nu}=0$ and
arbitrary $\Delta$:
\begin{equation}\label{T3}
M_{ab} \to T'(M_{ab})  = \rho(M_{ab}) \; , \;\;\;\;
 S_{\mu\nu}=0 \, .
 \end{equation}
Using relations (\ref{cnfA2}) the conditions (\ref{asym}) are written as
$$
\rho\left( \left\{ L_{[\mu \nu}  \,, L_{\lambda] \sigma} \right\} \right) = 0 \ ;\
\rho\left( \left\{L_{[\mu \nu}\,,P_{\lambda]}\right\}\right) = 0 \ ;\
\rho\left( \left\{L_{[\mu \nu}\,,K_{\lambda]}\right\} \right)= 0\, ;
$$
$$
\rho\left( \{\, L_{\mu \nu} \, , \, D \, \} \right) + \frac{1}{2}
\rho\left( \{\, K_{\mu} \, , \, P_{\nu} \, \} \right) -\frac{1}{2}
\rho\left( \{\, K_{\nu} \, , \, P_{\mu} \, \} \right) = 0\,.
$$
One can directly check that these conditions are identities. One can also check that
 the representations (\ref{T1}) and (\ref{T3}) can be extracted from the
 differential representation (\ref{T2}).

\end{itemize}

%\newpage

In the following Section it will be important that in particular case
of conformal algebra $so(2,4)$ of $4$-dimensional Minkowski space
($n=4$) the $\mathrm{RLL}$-relation (\ref{kaz-d10}) with
$\mathrm{R}$-matrix (\ref{r-mtr01}) and $\mathrm{L}$-operator
(\ref{LKaz-so}) can be satisfied for any representation $T'$ of the
generators $\{M_{ab}\}$ so that the condition (\ref{asym}) is
dispensable.
 Further we are going to prove it.
The recurrent relations (\ref{reccur}) for odd and even coefficients are independent.
Let us choose $\mathrm{R}_{0}(u) = (u+4)/8$ and $\mathrm{R}_1(u) = 0$.
The choice $\mathrm{R}_1(u) = 0$ is reasonable since all
odd coefficients in (\ref{r-mtr01}) are vanished in Weyl projection.
Hence due to (\ref{reccur}) $\mathrm{R}$-matrix (\ref{r-mtr01}) takes the form
\begin{equation} \label{Rn=4}
\mathrm{R}(u) =
\mathrm{R}_0(u) \cdot \mathbf{I} \otimes \mathbf{I} +
\frac{\mathrm{R}_2(u)}{2!} \cdot \Gamma_{a_1 a_2} \otimes
 \Gamma^{a_1 a_2} +
\frac{\mathrm{R}_4(u)}{4!} \cdot \Gamma_{a_1 \ldots a_4} \otimes
 \Gamma^{a_1 \ldots a_4}
+ \frac{\mathrm{R}_6(u)}{6!} \cdot \Gamma_{a_1 \ldots a_6} \otimes
 \Gamma^{a_1 \ldots a_6}
\end{equation}
where
$$\mathrm{R}_0 (u) = (u+4)/8 \; , \; \mathrm{R}_2 (u) = - u/8 \; , \; \mathrm{R}_4 (u) =  u/8 \; , \; \mathrm{R}_6 (u) = - (u + 4)/8$$
and
the last term in (\ref{RLLbig}) which is responsible for the condition (\ref{asym}) reduces to
\begin{equation} \label{acmn=4}
\frac{2}{3!}\biggl[
\mathrm{R}_{6}(u)+\mathrm{R}_{4}(u)\biggl]\,
\left\{M^{ab}\,,M^{c}{}_{d}\right\}\,
\biggl[\Gamma_{ab c c_1 c_2 c_3}\otimes\Gamma^{d c_1 c_2 c_3} + \Gamma^{d c_1 c_2 c_3}\otimes
\Gamma_{ab c c_1 c_2 c_3}\biggl]\;.
\end{equation}
All the other terms vanish because of the special form of coefficients $\mathrm{R}_k (u)$ and
owing to finiteness of the Clifford algebra of gamma-matrices.

Next we note that owing to
$ \Gamma_{a b c c_1 c_2 c_3} = \alpha \, \epsilon_{a b c c_1 c_2 c_3} \,\Gamma_7$ (\ref{cnfA4b})
and $\Gamma_7 \cdot \Gamma_7 = \mathbf{I}$
the gamma-matrix structure in (\ref{acmn=4}) can be transformed as follows
$$
\Gamma_{a b c c_1 c_2 c_3}\otimes\Gamma^{d c_1 c_2 c_3} = \Gamma_7 \otimes \Gamma_7 \cdot \Gamma_{ab c c_1 c_2 c_3} \cdot \Gamma^{d c_1 c_2 c_3} =
120 \cdot \Gamma_7 \otimes \Gamma_7 \cdot \left[ \, \delta^{d}_{a} \, \Gamma_{b c} - \delta^{d}_{b} \, \Gamma_{a c} + \delta^{d}_{c} \, \Gamma_{a b} \, \right]\;.
$$
Consequently (\ref{acmn=4}) which is proportional to
$$
\left\{M^{ab}\,,M^{c}{}_{d}\right\} \left[ \, \delta^{d}_{a} \, \Gamma_{b c} - \delta^{d}_{b} \, \Gamma_{a c} + \delta^{d}_{c} \, \Gamma_{a b} \,\right] = 2 \left\{M^{a (b}\,,M^{c)}{}_{a}\right\} \, \Gamma_{b c} = 0
$$
turns to zero. In the last expression the parentheses $(...)$ denote symmetrization.
Therefore $\mathrm{RLL}$-equation (\ref{kaz-d10}) is valid for
any representation of generators $\{M_{ab}\}$ of the algebra $so(2,4)$.

Let us rewrite the expression for $\mathrm{R}$-matrix (\ref{Rn=4}) in a more transparent form.
All gamma-matrix structures in (\ref{Rn=4})
have block-diagonal representation that can be seen from (\ref{cnfA4}). Therefore it is reasonable to consider
projections of (\ref{Rn=4}) on corresponding irreducible subspaces.
 %We introduce spaces $V_{+}$ and $V_{-}$ so that
 %matrices $\frac{1+\Gamma_7}{2}$ and $\frac{1-\Gamma_7}{2}$ are
 %projectors on $V_{+}$ and $V_{-}$ correspondingly.
  As before we introduce subspaces $V_{+}$ and $V_{-}$ obtained by Weyl projections:
 $V_+ = \frac{1+\Gamma_7}{2} V$ and $V_- =\frac{1-\Gamma_7}{2}V$.
At first we note that
$$
\left.\mathrm{R}(u)\right|_{V_{+}\otimes V_{-}} =
\left.\mathrm{R}(u)\right|_{V_{-}\otimes V_{+}} = 0
$$
because
$$
\left[ \mathbf{I} \otimes \mathbf{I} -
\frac{1}{6!} \cdot \Gamma_{A_6} \otimes \Gamma^{A_6} \right]_{V_{+}\otimes V_{-}} =
\left[ \frac{1}{2!} \cdot \Gamma_{A_2} \otimes \Gamma^{A_2} -
\frac{1}{4!} \cdot \Gamma_{A_4} \otimes \Gamma^{A_4} \right]_{V_{+}\otimes V_{-}} = 0\;.
$$
In order to perform projection on the space $V_{-}\otimes V_{-}$ we take into account that
$$
\left[ \mathbf{I} \otimes \mathbf{I} +
\frac{1}{6!} \cdot \Gamma_{A_6} \otimes \Gamma^{A_6} \right]_{V_{-}\otimes V_{-}} =
\left[ \frac{1}{2!} \cdot \Gamma_{A_2} \otimes \Gamma^{A_2} +
\frac{1}{4!} \cdot \Gamma_{A_4} \otimes \Gamma^{A_4} \right]_{V_{-}\otimes V_{-}} = 0
$$
and
$$
\left. - \frac{1}{8} \cdot \Gamma_{ab} \otimes \Gamma^{ab} \right|_{ V_{-}\otimes V_{-}} =
\frac{1}{2} \left( \ell_{\mu \nu} \otimes \ell^{\mu \nu} + k_{\mu} \otimes p^{\mu} + p_{\mu} \otimes k^{\mu}
\right) - d \otimes d = \mathrm{P} - \frac{1}{4} \cdot {\rm I} \otimes {\rm I}
$$
where $\mathrm{P}$ is a permutation operator.
Consequently we obtain Yang $\mathrm{R}$-matrix
\begin{equation}\label{YangR}
\left.\mathrm{R}(u)\right|_{V_{-}\otimes V_{-}} = \left[ 2 \;\mathrm{R}_0 (u) \cdot \mathbf{I} \otimes \mathbf{I} +
\mathrm{R}_2 (u) \cdot \Gamma_{a b} \otimes \Gamma^{a b} \right]_{V_{-}\otimes V_{-}} =
{\rm I} \otimes {\rm I} + u \cdot \mathrm{P}\;.
\end{equation}

The $\mathrm{L}$-operator (\ref{LKaz-so}) is also reducible since
$T(M_{ab})$ consists of two irreducible blocks (\ref{cnfA7}). Therefore
its projection to the subspace $V_{-}$
 \begin{equation}\label{LKaz-so1}
 \mathrm{L}(u) = u \, {\rm I} \otimes {\rm 1} \; + \; \frac{1}{2}\,
 T_1(M_{ab})  \otimes M^{ab} \; ,
 \end{equation}
in the case of 4-dimensional Minkowski space ($n=4$),
fulfils $\mathrm{RLL}$-relation with Yang $\mathrm{R}$-matrix.
In full analogy we obtain that
$$
\left.\mathrm{R}(u)\right|_{V_{+}\otimes V_{+}} = {\rm I} \otimes {\rm I} + u \cdot \mathrm{P}\;
$$
and the second
projection of the $\mathrm{L}$-operator (\ref{LKaz-so}) (on the subspace $V_{+}$)
fulfils Yangian relation.

%%%%%%%%%%%%%%%%%%%%%%%%%%%%%%%%%%%%%%%%%%%%%
 %\subsection{Factorization of operator $\mathrm{L}(u)$
 %(\ref{LKaz}) for the algebra $so(p+1,q+1)$ }
 %%%%%%%%%%%%%%%%%%%%%%%%%%%%%%%%%%%%%%%%%%%%%%%%%%%%%%%%%%%%%%%%%%%%%%%

 \vspace{0.5cm}

At the end of this Section we consider operator $\mathrm{L}(u)$ (\ref{LKaz})
  for the algebra $so(p+1,q+1)$
 for the special choice of the representations
 $T=T_1$ (cf. eq. (\ref{LKaz-so1})) and $T'=\rho$, where $T_1$ is the spinor representation (\ref{cnfA10}) and
 $\rho$ is the differential representation (\ref{cnfA1}). This operator $\mathrm{L}(u)$
 is written in the form:
\begin{equation}
 \label{kaz-d13}
\mathrm{L}^{(\rho)}(u) = u \, {\rm I} +  \frac{1}{2} T_1(M^{ab}) \otimes \rho(M_{ab}) = \left(
  \begin{array}{cc}
    u_+ \cdot \boldsymbol{1} + \mathbf{S} - \mathbf{p} \cdot  \mathbf{x} \; , & \;
    \mathbf{p} \\ [0.2cm]
  \mathbf{x} \cdot \mathbf{S} - \overline{\mathbf{S}} \cdot \mathbf{x}
    - \mathbf{x} \cdot \mathbf{p} \cdot \mathbf{x} +
     (\Delta-\frac{n}{2}) \cdot \mathbf{x}  \; , & \;\;
      u_- \cdot \boldsymbol{1} + \overline{\mathbf{S}} + \mathbf{x} \cdot \mathbf{p}
  \end{array}
\right) \; ,
 \end{equation}
where
\begin{equation} \label{u+u-}
u_+ = u + \frac{\Delta -n}{2} \; , \;\;\;  u_- = u - \frac{\Delta}{2} \; , \;\;\; n=p+q \; ,
\end{equation}
and we have used the expression (\ref{cnfA27}) for the matrix
 $\frac{1}{2} T_1(M^{ab}) \otimes \rho(M_{ab})$ which was introduced in (\ref{cnfA16}).
The important fact is:

\vspace{0.5cm}

\noindent
{\bf Proposition 2.} {\it The operator (\ref{kaz-d13})
is expressed in the following
  %\marginpar{\bf \tiny Otredaktiroval footnote}
factorized form \footnote{ The factorized form (\ref{cnfA17}) of the
$so$-type $\mathrm{L}$-operator for the scalar representation
($\mathbf{S} =\overline{\mathbf{S}} =0$) was obtained independently by
G.Korchemsky and V.Pasquier (private communication).}
\begin{equation}
\label{cnfA17}
\mathrm{L}^{(\rho)}( u ) = \left(
\begin{array}{cc}
  \boldsymbol{1} & \; \mathbf{0} \\
  \mathbf{x} & \; \boldsymbol{1}
\end{array}
\right) \cdot \left(
\begin{array}{cc}
  u_+ \cdot \boldsymbol{1} + \mathbf{S}  & \ \ \mathbf{p} \\
  \mathbf{0}  & u_- \cdot \boldsymbol{1} + \overline{\mathbf{S}}
\end{array}
\right) \cdot \left(
\begin{array}{cc}
  \boldsymbol{1} & \; \mathbf{0} \\
  - \mathbf{x} & \; \boldsymbol{1}
\end{array}
\right)  \; .
\end{equation}
}

\vspace{0.2cm}

\noindent
{\bf Proof.}  One can show that (\ref{cnfA17})
 is equivalent to (\ref{kaz-d13}) by direct calculation.
\hfill $\bullet$

\vspace{0.2cm}

\noindent
{\bf Remark 1.} The formula (\ref{cnfA17}) for the $so(p+1,q+1)$-type  operator $\mathrm{L}^{(\rho)}( u )$
can be considered as the recurrent formula if we interpret
operators $(u_+ \cdot \boldsymbol{1} + \mathbf{S})$ and
$(u_- \cdot \boldsymbol{1} + \overline{\mathbf{S}})$ as two smaller
$so(p,q)$-type operators $\mathrm{L}^{(\rho)}( u )$.

\vspace{0.2cm}

\noindent
{\bf Remark 2.} Consider $so(p+1,q+1)$-type
  %\marginpar{\bf \tiny Absolyutno novyj Remark 2.}
$\mathrm{L}$-operator (\ref{LKaz-so})
which satisfies $\mathrm{RLL}$-relations (\ref{kaz-d10}) with spinorial $\mathrm{R}$-matrix (\ref{r-mtr01}),
(\ref{reccur}):
 \begin{equation}
 \label{yang01}
 \mathrm{L}(u) = u \; {\bf I}  \; - \; \frac{1}{8} \, (\Gamma_a \, \Gamma_b - \Gamma_b \, \Gamma_a)
 (y^a \partial^b - y^b \partial^a) \; ,
 \end{equation}
 where for the generators $M^{ab}$ we have used the representation $T'$ given in (\ref{T2}).
 The $\mathrm{L}$-operator (\ref{yang01}) satisfies crossing symmetry relation
 %\marginpar{\bf Proverit' formulu}
 \begin{equation}
 \label{yang02}
  \mathrm{L}^T(u)  \cdot \mathrm{C} \cdot  \mathrm{L}(u') =
  \left(u \, u'   - \frac{1}{4} \, T'(C_2) \right) \, \mathrm{C} \; ,
 \end{equation}
 where $u'= u + \frac{n}{2}$, $n=(p+q)$, $\mathrm{C}$
 is the $2^{{n \over 2}+1}$-dimensional
 analog of matrix $\mathrm{C}$ introduced in (\ref{A4d}) and $C_2$ is the Cazimir operator (\ref{kaz}).
 Since the representation (\ref{cnfA6}) is reducible (see (\ref{cnfA7}))
 the operator $\mathrm{L}(u)$ (\ref{yang01}) has the block diagonal form
 \begin{equation}
\label{yang04}
 \mathrm{L}(u) = \left(
 \begin{array}{cc}
   \mathrm{L}_+(u) & \mathbf{0} \\
  \mathbf{0} & \mathrm{L}_-(u)
 \end{array}
 \right) \; ,
 \end{equation}
 and in view of relations (\ref{A4g}) one can rewrite relation (\ref{yang02})
 for blocks $\mathrm{L}_\pm(u)$ as following
 \begin{equation}
\label{yang03}
\begin{array}{l}
 1.) \;\;\; \frac{(n+2)(n+1)}{2} \; - \; {\rm even} \;\;\; \Rightarrow \;\;\;
  \mathrm{L}^T_\pm(u)  \cdot \mathrm{c} \cdot  \mathrm{L}_\pm(u') =
   z(u) \cdot \mathrm{c}  \\ [0.2cm]
 2.) \;\;\;  \frac{(n+2)(n+1)}{2} \; - \; {\rm odd} \;\;\; \Rightarrow \;\;\;
  \mathrm{L}^T_+(u)  \cdot \mathrm{g} \cdot  \mathrm{L}_-(u') =
   z(u) \cdot \mathrm{g} \; .
  \end{array}
\end{equation}
where $z(u) = \left(u \, u'   - \frac{1}{16} \, T'(C_2) \right)$. It is clear that
the irreducible parts $\mathrm{L}_\pm(u)$ of the operator (\ref{yang01}) satisfy
$\mathrm{RLL}$-relations (\ref{kaz-d10}):
 \begin{equation}
\label{yang05}
 \mathrm{R}^{(\pm)}_{12} (u-v) \, \mathrm{L}_{\pm 1}(u) \, \mathrm{L}_{\pm 2}(v) =
 \mathrm{L}_{\pm 1}(v) \, \mathrm{L}_{\pm 2}(u) \, \mathrm{R}^{(\pm)}_{12} (u-v) \; ,
 \end{equation}
 where $\mathrm{R}^{(\pm)}(u) = \left. \mathrm{R}(u)\right|_{V_{\pm}\otimes V_{\pm}}$
 and the matrix $\mathrm{R}(u)$ is given in (\ref{r-mtr01}).

Consider instead of operators $\mathrm{L}_\pm(u)$
defined in (\ref{yang01}), (\ref{yang04}) more general operators
\begin{equation}\label{yang06}
 \mathrm{L}_\pm(u) = {\rm I} \; + \; \sum_{k=1}^{\infty}
 \frac{1}{u^k}  \, L^{(k)}_\pm  \; .
 \end{equation}
Then relations (\ref{yang05})  will define the infinite-dimensional
 quadratic algebra  generated by the set of elements
 $\{(L^{(0)}_\pm)_{\alpha \beta}, (L^{(1)}_\pm)_{\alpha \beta}, \dots \}$,
 $(\alpha, \beta = 1,2,\dots, 2^{n \over 2})$.
 We denote this algebra as ${\rm Y} ({\sf spin}(n+2,\mathbb{C}))$.
  For the generators of
 ${\rm Y} ({\sf spin}(n+2,\mathbb{C}))$
 it is also necessary to add additional constraints (\ref{yang03}),
  where the function $z(u)$ have to be considered as a central element of
 ${\rm Y} ({\sf spin}(n+2,\mathbb{C}))$.
 The results of this Subsection show
 that the algebra ${\rm Y} ({\sf spin}(n+2,\mathbb{C}))$
 possesses evaluation representations
 when $(L^{(k)}_\pm)_{\alpha \beta} \to 0$
 for $k >1$ and $(L^{(1)}_\pm) \to \frac{1}{2} T_1(M_{ab}) T'(M^{ab})$.
 Here $M_{ab}$ are generators of ${\sf spin}(n+2,\mathbb{C})$ and
 special representations $T'$ are
 itemized in (\ref{T1}) -- (\ref{T3}). For the special case $n=4$ the
 matrix $\left.\mathrm{R}(u)\right|_{V_{-}\otimes V_{-}}$
 is the Yang $\mathrm{R}$-matrix (\ref{YangR}),
 all $4\times 4$ matrices $\left.(\Gamma^{a_1 \dots a_{2k}})\right|_{V_-}$
 form the basis for $s\ell(4)$ and we see that the algebra ${\rm Y} ({\sf spin}(6))$
 is isomorphic to the Yangian ${\rm Y} (s\ell(4))$.

\section{$\mathrm{L}$-operator for the conformal algebra in four dimensions}
\setcounter{equation}{0}

\mbox{} Now we restrict our consideration to the case of 4-dimensional
Minkowski space $\mathbb{R}^{1,3}$, i.e., $p=1$, $q=3$ and $n=4$. In
this case generators  (\ref{cnfA10}) are
 \begin{equation}
\label{cnfA11f}
\ell_{\mu \nu} = \frac{i}{4}[ \gamma_{\mu} , \gamma_{\nu} ] \; , \qquad
p_{\mu} = \gamma_{\mu} \frac{1 + \gamma_5}{2} \; , \qquad k_{\mu} =
\gamma_{\mu} \frac{1 - \gamma_5}{2} \; , \qquad d = - \frac{i}{2} \gamma_5 \; ,
\end{equation}
where $\gamma_{5} = -i \gamma_0 \gamma_1 \gamma_2 \gamma_3$ and we
choose common representation (\ref{cnfA11s})
 \begin{equation}
 \label{cnfA11sf}
\gamma_{\mu} = \left(
  \begin{array}{cc}
  \mathbf{0} & \boldsymbol{\sigma}_{\mu} \\
  \overline{ \boldsymbol{\sigma} }_{\mu} & \mathbf{0}
  \end{array}
\right) \, , \;\; \gamma_{5} =  \left(
  \begin{array}{cc}
  \mathrm{I}_2 & \mathbf{0} \\
  \mathbf{0} & - \mathrm{I}_2
  \end{array}
\right) \, , \;\; \frac{1 + \gamma_5 }{2} = \left(
  \begin{array}{cc}
  \mathrm{I}_2 & \mathbf{0} \\
  \mathbf{0} & \mathbf{0}
  \end{array}
\right) \, , \;\; \frac{1 - \gamma_5 }{2} = \left(
  \begin{array}{cc}
  \mathbf{0} & \mathbf{0} \\
  \mathbf{0} & \mathrm{I}_2
  \end{array}
\right) \; ,
 \end{equation}
 constructed by means of $2\times 2$-matrices $\boldsymbol{\sigma}_{\mu}$
 and $\overline{ \boldsymbol{\sigma} }_{\mu}$ (\ref{sigmaM}).
Note that in the representation (\ref{cnfA11sf}) we have identities
\begin{equation}
\label{cnfA12} \gamma_\mu^\dagger = \gamma_0 \,  \gamma_\mu \, \gamma_0
\; , \;\;\; \gamma_5^\dagger = \gamma_5 = - \gamma_0 \,  \gamma_5 \,
 \gamma_0 \; ,
\end{equation}
which are analogs of (\ref{cnfA4d}) and (\ref{cnfA4f})
 and correspond to the case 2.) stated in
(\ref{cnfA4g}) and (\ref{cnfA4r}) for ${\bf C}=\gamma_0$ and ${\bf
g}=I_2$.

It is known that fifteen matrices (\ref{cnfA11f}) form the basis in the
space ${\sf Mat}(4)$ of all traceless $4 \times 4$ matrices. Then one
can check by using (\ref{cnfA12}) that any $4 \times 4$ matrix
(\ref{s6s4})
 which belongs to a linear span of (\ref{cnfA11f}) satisfies equation
$A^\dagger \, \gamma_0 + \gamma_0 \, A = 0$ which defines Lie algebra
$su(2,2)$. This equation means that $4 \times 4$ matrices $A$
 (\ref{s6s4}) not only represent all elements of the conformal algebra
 $so(2,4)=spin(2,4)$ but also generate the Lie algebra $su(2,2)$.
 In other words we have established the well known isomorphism
$so(2,4) = su(2,2)$. For complexifications of these algebras we have
$so(6,\mathbb{C}) = s\ell(4,\mathbb{C})$. Below we use this isomorphism to
relate operators $\mathrm{L}(u)$ (\ref{3.13'}), (\ref{kaz-d13}) for the special
choices of algebras $s\ell(4,\mathbb{C})$ and $so(6,\mathbb{C})$. Then one can
 investigate the $so(6,\mathbb{C})$-type $\mathrm{L}$-operator
  by applying known
 facts \cite{DM} about $s\ell$-type $\mathrm{L}$-operators.

To proceed further we present explicitly the $\mathrm{L}$-operator for the
conformal algebra $so(2,4)$. This $\mathrm{L}$-operator is given by formulas
(\ref{kaz-d13}) and (\ref{cnfA17}) (for $n=p+q=4$):
 \begin{equation}
 \label{cnfA17f}
 \mathrm{L}^{(\rho)}( u ) = \left( \begin{array}{cc}
 \mathrm{I}_2 & \; \mathbf{0} \\
 \mathbf{x} & \; \mathrm{I}_2
 \end{array}\right) \cdot \left( \begin{array}{cc}
 u_+ \cdot \mathrm{I}_2 + \mathbf{S}  \, , & \ \ \mathbf{p} \\
 \mathbf{0} \; , &
 u_- \cdot \mathrm{I}_2 + \overline{\mathbf{S}}
 \end{array} \right) \cdot \left(
 \begin{array}{cc}   \mathrm{I}_2 & \; \mathbf{0} \\  - \mathbf{x} & \; \mathrm{I}_2
 \end{array} \right) =
 \end{equation}
\begin{equation}
 \label{cnfA27f}
=
\begin{pmatrix}
u_+ \cdot \mathrm{I}_2 + \mathbf{S} - \mathbf{p}\cdot \mathbf{x}  \; ,
& \mathbf{p}\\
 \mathbf{x}\cdot \left(u_+ \cdot \mathrm{I}_2 + \mathbf{S}\right)
 - \left(u_- \cdot \mathrm{I}_2 + \overline{\mathbf{S}}\right) \cdot \mathbf{x}
 -\mathbf{x} \cdot \mathbf{p} \cdot \mathbf{x}
\; , & \ \  u_- \cdot \mathrm{I}_2 + \overline{\mathbf{S}} + \mathbf{x} \cdot \mathbf{p}
\end{pmatrix} \; ,
\end{equation}
where
$
 u_+ = u + \frac{\Delta -4}{2}\, ,\, u_- = u - \frac{\Delta}{2}
$
and $2 \times 2$ matrices $\mathbf{p}$, $\mathbf{x}$, $\mathbf{S}$,
   $\overline{\mathbf{S}}$
 were defined in (\ref{TSPK}), (\ref{spinorS}), (\ref{spinorSbar}).
We stress that $2 \times 2$ matrices $u_+ \cdot \mathrm{I}_2 + \mathbf{S}$ and
$u_- \cdot \mathrm{I}_2 + \overline{\mathbf{S}}$ are two $\mathrm{L}$-operators
(see (\ref{spinorS}) -- (\ref{barsss}), (\ref{Lsl22})) for the case of
$s\ell(2,\mathbb{C})=so(1,3)$.
 We note that the basis of the algebra $so(2,4)$
 %\marginpar{\bf Novoe predlozhenie}
 (which is the real form
 of $so(6,\mathbb{C})$) is
 the basis of the algebra $so(6,\mathbb{C})$ and therefore the operator
 (\ref{kaz-d13}), (\ref{cnfA17f}) can be considered
 (after a complexification when all coordinates $x_\mu$ are complex numbers)
 as the $\mathrm{L}$-operator of the  algebra $so(6,\mathbb{C})$ as well.

 %%%%%%%%%%%%%%%%%%%%%%%%%%%%%%%%%%%%%%%%%%%%%%%%%%%%%%%%%%%%%%%%%%%%%%%%%%%%%%%%%%%%%%%%%%%%%
\subsection{$\mathrm{L}$-operators for $s\ell(4,\mathbb{C})$ and
$so(6,\mathbb{C})$\label{sec-sl4}}
%%%%%%%%%%%%%%%%%%%%%%%%%%%%%%%%%%%%%%%%%%%%%%%%%%%%%%%%%%%%%%%%%%%%%%%%%%%%%%%%%%%%%%%%%%%%%

\mbox{} Now using the construction of $\mathrm{L}$-operator for $s\ell(N,\mathbb{C})$ (see Section {\bf \ref{Defs}}) and
the isomor\-phism $so(6,\mathbb{C}) = s\ell(4,\mathbb{C})$ we
investigate relations of $\mathrm{L}$-operator for $s\ell(4,\mathbb{C})$ (which
satisfies (\ref{kaz-d10})) and $\mathrm{L}$-operator (\ref{cnfA17f}) for  the
algebra $so(2,4)$. The complexification of the last $\mathrm{L}$-operator
 %\marginpar{\bf Redaktorskaya pravka}
 is also given by
(\ref{3.13'}), (\ref{Lsl4}) but with the special choice of the basis $\{\rho(E_{ij}) \}
 \rightarrow \{L_{\mu\nu}, P_\mu, K_\nu, D\}$ and $\{ e_{ij} \}
 \rightarrow \{\ell_{\mu\nu}, p_\mu, k_\nu, d \}$ in the representations $\rho$
(\ref{cnfA1}) and $T_1$ (\ref{cnfA11f}).

Consider $\mathrm{L}$-operator (\ref{Lsl4}), (\ref{Lsln02}) for $s\ell(4,\mathbb{C})$
case, where
 weights $\rho_1, \dots, \rho_4$ are
 related by condition $\rho_1 + \dots + \rho_4 = 6$. The
 factorized form of this $\mathrm{L}$-operator  in the right hand side of (\ref{Lsl4})
 contains $(4\times 4)$ matrices $\mathrm{Z}$ and $\mathrm{D}$:
\begin{equation}
 \label{Lsl01}
\mathrm{Z}=
\begin{pmatrix}1&0&0&0\\
z_{21}&1&0&0\\
z_{31}&z_{32}&1&0\\
z_{41}&z_{42}&z_{43}&1
\end{pmatrix} \; , \;\;\;
\mathrm{D}=
%\begin{pmatrix} u_1&-\partial_{z_{21}} - z_{32}\partial_{z_{31}}- z_{42}\partial_{z_{41}}
%&-\partial_{z_{31}} - z_{43}\partial_{z_{41}}& -\partial_{z_{41}}\\
%0&u_2&-\partial_{z_{32}} - z_{43}\partial_{z_{42}}&-\partial_{z_{42}}\\
%0&0&u_3&-\partial_{z_{43}}\\ 0&0&0&u_4 \end{pmatrix}
 \begin{pmatrix}
 u_1& D_{12} & D_{13} & D_{14}  \\
 0& u_2 & D_{23} & D_{24} \\
 0 & 0 & u_3 & D_{34} \\
 0&0&0&u_4
 \end{pmatrix} \; ,
 %\;\;\;\; D_{ij} = - \sum\limits_{k=j}^4 \, z_{kj} \, \partial_{ki} \; ,
 \end{equation}
where elements $D_{ij}$ are differential operators defined in (\ref{notZD}).
 Note that  we have $D_{k4}= -\partial_{4k}$ $(k=1,2,3)$. In view of the isomorphism
$s\ell(4,\mathbb{C}) = so(6,\mathbb{C})$ one can expect that factorized form
(\ref{Lsl4}) for $N=4$ is transformed into the factorized form of the $\mathrm{L}$-operator (\ref{cnfA17f})
for conformal algebra $so(2,4)$. To obtain explicitly this transformation
we write $4 \times 4$ matrices (\ref{Lsl01}) and then (\ref{Lsl4}) in a $2 \times 2$
 block-matrix form with blocks
 \begin{equation}
 \label{Lsl02}
\mathrm{z}_1=
\begin{pmatrix}1&0\\
z_{21}&1
\end{pmatrix} \; , \;\;\; \mathrm{z}_2=
\begin{pmatrix}1&0\\
z_{43}&1
\end{pmatrix}\; , \;\;\; \mathrm{z}=
\begin{pmatrix}
z_{31}&z_{32}\\
z_{41}&z_{42}
\end{pmatrix} \; ,
\end{equation}
\begin{equation}
 \label{Lsl03}
\mathrm{d}_1=
\begin{pmatrix}u_1 & D_{12} \\
 0 & u_2
\end{pmatrix}\; , \;\;\;
\mathrm{d}_2 = \begin{pmatrix}
 u_3 & - \partial_{43} \\
 0 & u_4
\end{pmatrix}\; , \;\;\;
\mathrm{d} = -
\begin{pmatrix}
 \partial_{31} & \partial_{41} \\
 \partial_{32} & \partial_{42}
\end{pmatrix} \; .
\end{equation}
Indeed, using these blocks we first deduce factorized
expressions for $\mathrm{Z}$ and $\mathrm{D}$:
\begin{align}
\mathrm{Z}=
\begin{pmatrix}
 \mathrm{z}_1 & \mathbf{0} \\
 \mathbf{0} & \mathrm{I}_2
\end{pmatrix}\begin{pmatrix}
 \mathrm{I}_2 & \mathbf{0} \\
\mathrm{z}  & \mathrm{I}_2
\end{pmatrix} \begin{pmatrix}
 \mathrm{I}_2 & \mathbf{0} \\
 \mathbf{0} & \mathrm{z}_2
\end{pmatrix} \, , \;\;
\mathrm{D}=
\begin{pmatrix}
\mathrm{d}_{1} \; , &  \; \mathrm{d}\cdot\mathrm{z}_2 \\
\mathbf{0} \;\; , & \; \mathrm{d}_{2}
\end{pmatrix} =
\begin{pmatrix}
\mathrm{d}_{1} \; , &  \mathrm{d} \\
\mathbf{0} \;\; , &  \mathrm{d}_{2} \cdot\mathrm{z}_2^{-1}
\end{pmatrix}  \begin{pmatrix}
 \mathrm{I}_2 & \mathbf{0} \\
 \mathbf{0} & \mathrm{z}_2
\end{pmatrix} \, ,
\end{align}
 and then $s\ell(4,\mathbb{C})$-type $\mathrm{L}$-operator (\ref{Lsl4}) is
 also written, after multiplication of the matrices in the middle,
 in the factorized form
\begin{equation}
 \label{Lsl04}
\mathrm{L}(u) = \mathrm{Z} \cdot \mathrm{D} \cdot \mathrm{Z}^{-1} = \begin{pmatrix}
 \mathrm{z}_1& \mathbf{0}\\
 \mathbf{0} & \mathrm{I}_2
\end{pmatrix}\begin{pmatrix}
  \mathrm{I}_2 & \mathbf{0} \\
\mathrm{z}  & \mathrm{I}_2
\end{pmatrix}
\begin{pmatrix}\mathrm{d}_{1} \; , & \; \mathrm{d}\\
\mathbf{0} \; , & \; \mathrm{z}_2\cdot \mathrm{d}_{2}\cdot \mathrm{z}_2^{-1}
\end{pmatrix}
\begin{pmatrix}
 \mathrm{I}_2 & \mathbf{0} \\
- \mathrm{z}  & \mathrm{I}_2
\end{pmatrix}\begin{pmatrix}
 \mathrm{z}_1^{-1}&\mathbf{0} \\
 \mathbf{0} & \mathrm{I}_2
\end{pmatrix} \; .
\end{equation}
We note that here matrix $\mathrm{z}_2\cdot \mathrm{d}_{2}\cdot \mathrm{z}_2^{-1}$ is just the
usual $\mathrm{L}$-operator (\ref{Lsl22}) for $s\ell(2,\mathbb{C})$ case
\begin{equation}
 \label{ell2}
\mathrm{z}_2\,\mathrm{d}_{2}\,\mathrm{z}_2^{-1} = \begin{pmatrix}1&0\\
z_{43}&1
\end{pmatrix}\begin{pmatrix}u_3&-\partial_{43}\\
0&u_4
\end{pmatrix}\begin{pmatrix}1&0\\
-z_{43}&1
\end{pmatrix}  \; ,
\end{equation}
and
the whole dependence on $z_{43}$ in $\mathrm{L}(u)$ (\ref{Lsl04}) is absorbed only in
this operator (\ref{ell2}).
 %$\mathrm{z}_2\cdot \mathrm{d}_{2}\cdot \mathrm{z}_2^{-1}$.

Multiplication of all matrices in (\ref{Lsl04}) gives
\begin{equation}
 \label{Lsl05}
\mathrm{L}(u) =
\begin{pmatrix}\mathrm{z}_1\cdot \left(\mathrm{d}_{1}-\mathrm{d}\cdot \mathrm{z}\right)
 \cdot \mathrm{z}_1^{-1} \; ,
& \;\; \mathrm{z}_1\cdot \mathrm{d}\\
\mathrm{z}\cdot \left(\mathrm{d}_{1}-\mathrm{d}\cdot \mathrm{z}\right) \cdot \mathrm{z}_1^{-1}-
(\mathrm{z}_2\cdot \mathrm{d}_{2}\cdot \mathrm{z}_2^{-1}) \cdot \mathrm{z}\cdot \mathrm{z}_1^{-1} \; , & \;\;\;\;
\mathrm{z}\cdot \mathrm{d} + (\mathrm{z}_2\cdot \mathrm{d}_{2}\cdot \mathrm{z}_2^{-1})
\end{pmatrix} \; ,
\end{equation}
and comparing of this expression with $so(2,4)$-type $\mathrm{L}$-operator (\ref{cnfA27f})
 suggests the natural change of variables
\begin{equation}
 \label{Lsl06}
\mathbf{x} = \mathrm{z} \cdot \mathrm{z}_1^{-1}  \; , \;\;\;
\mathbf{p} =  \mathrm{z}_1 \cdot \mathrm{d}  \; , \;\;\;
\boldsymbol{\chi}_1 = \mathrm{z}_1  \; , \;\;\; \boldsymbol{\chi}_2 = \mathrm{z}_2  \; ,
\end{equation}
where in view of the explicit form of matrices $\mathrm{z}_1$ and $\mathrm{z}_2$
(\ref{Lsl02}) we have to fix
$$
\boldsymbol{\chi}_1=
\begin{pmatrix}1&0\\
\chi_1 &1
\end{pmatrix} \; , \;\;\;
\boldsymbol{\chi}_2=
\begin{pmatrix}1&0\\
\chi_2 &1
\end{pmatrix}\;\; \Rightarrow \;\; \chi_1 = z_{21}  \; , \;\;\; \chi_2 = z_{43}  \; .
$$
The inverse transformations with respect to (\ref{Lsl06}) are:
\begin{equation}
 \label{Lsl06op}
\mathrm{z}  = \mathbf{x} \cdot \boldsymbol{\chi}_1  \; , \;\;\;
\mathrm{d} =  \boldsymbol{\chi}_1^{-1} \cdot \mathbf{p}   \; , \;\;\;
\mathrm{z}_1 = \boldsymbol{\chi}_1   \; , \;\;\; \mathrm{z}_2 = \boldsymbol{\chi}_2  \; .
\end{equation}
Now we fix
 \begin{equation}
 \label{u-ell}
 \begin{array}{c}
u_1 = u_{+} - \ell - 1 \; , \;\;\; u_2 = u_{+} + \ell \; ,   \\ [0.2cm]
u_3 = u_{-} - \dot{\ell} - 1 \; , \;\;\; u_4 = u_{-} + \dot{\ell} \; .
 \end{array}
 \end{equation}
In terms of new variables $\mathbf{x}$, $\boldsymbol{\chi}_1$, $\boldsymbol{\chi}_2$
and $\mathbf{p}$ (\ref{Lsl06}) the $\mathrm{L}$-operator (\ref{Lsl04}), (\ref{Lsl05}) acquires the form(cf. (\ref{cnfA27f})):
 \begin{equation}
 \label{Lsl07}
 \mathrm{L}( u ) =
\begin{pmatrix}
u_+ \cdot \mathrm{I}_2 + \mathbf{S}^{(\ell)} - \mathbf{p}\cdot \mathbf{x}  \; ,
& \mathbf{p}\\
 \mathbf{x}\cdot \left(u_+ \cdot \mathrm{I}_2 + \mathbf{S}^{(\ell)}\right)
 - \left(u_- \cdot \mathrm{I}_2 + \overline{\mathbf{S}}^{(\dot{\ell})}\right) \cdot \mathbf{x}
 -\mathbf{x} \cdot \mathbf{p} \cdot \mathbf{x}
\; , & \ \  u_- \cdot \mathrm{I}_2 + \overline{\mathbf{S}}^{(\dot{\ell})} + \mathbf{x} \cdot \mathbf{p}
\end{pmatrix} \; ,
\end{equation}
where we have introduced two $s\ell(2,\mathbb{C})$-type $\mathrm{L}$-operators
\begin{equation}
 \label{Lsl08}
 \begin{array}{c}
\mathrm{z}_1\cdot \mathrm{d}_{1} \cdot \mathrm{z}_1
= \begin{pmatrix}1&0\\
z_{21}&1
\end{pmatrix}\begin{pmatrix}u_1& D_{12} \\
0&u_2
\end{pmatrix}\begin{pmatrix}1&0\\
-z_{21}&1
\end{pmatrix}
= \begin{pmatrix}1&0\\
\chi_1 & 1
\end{pmatrix}\begin{pmatrix}u_1& -\partial_{\chi_1} \\
0&u_2
\end{pmatrix}\begin{pmatrix}1&0\\
-\chi_1 &1
\end{pmatrix} =  u_+ \, \mathrm{I}_2 + \mathbf{S}^{(\ell)}  \; ,
\end{array}
\end{equation}
and
\begin{equation}
 \label{ell2a}
  \begin{array}{c}
\mathrm{z}_2\,\mathrm{d}_{2}\,\mathrm{z}_2^{-1} = \begin{pmatrix}1&0\\
z_{43}&1
\end{pmatrix}\begin{pmatrix}u_3&-\partial_{43}\\
0&u_4
\end{pmatrix}\begin{pmatrix}1&0\\
-z_{43}&1
\end{pmatrix}
=  \begin{pmatrix} 1&0\\ \chi_{2}& 1
\end{pmatrix}\begin{pmatrix}u_3&-\partial_{\chi_2}\\ 0&u_4
\end{pmatrix}\begin{pmatrix}1&0\\ -\chi_{2}& 1
\end{pmatrix} = u_{-} \, \mathrm{I}_2 + \mathbf{S}^{(\dot{\ell})}  \; .
 \end{array}
\end{equation}
Here we have used notations $\mathbf{S}^{(\ell)}$  (\ref{sss}) and $\mathbf{S}^{(\dot{\ell})}$
 (\ref{barsss}) for the matrices of $s\ell(2,\mathbb{C})$
  generators (\ref{Lsl22}) and interpret $\mathbf{S}^{(\ell)}$ and $\mathbf{S}^{(\dot{\ell})}$ as matrices
$\mathbf{S}$ and $\overline{\mathbf{S}}$ of spin operators
$S_{\mu\nu}$ (see (\ref{TSPK})) appeared in the differential representation
(\ref{cnfA1}) of conformal algebra $so(2,4)$. The generators of two $s\ell(2,\mathbb{C})$ algebras
which were packed into the matrices (\ref{sss}) and (\ref{barsss})
are differential operators over variables $\chi_1$ and $\chi_2$ and act in spaces
 %In view of this, irreducible representations of the algebras $\mathbf{S}^{(\ell)}$
 %and $\overline{\mathbf{S}}^{(\dot{\ell})}$ will be constructed as
 of functions of $\chi_1$ and $\chi_2$.
It is natural to call $\chi_1$ and $\chi_2$ as harmonic variables.

Let us summarize connection between variables in
the first and second approaches. From (\ref{TSPK}) and (\ref{Lsl06}) we have:
\begin{flushleft}
  $\chi_1 = z_{21}$, \\
  $(\mathbf{x})_{11} = -i (x_0 + x_3) = z_{31}-z_{32}z_{21}
   \; , \qquad  (\mathbf{x})_{12} = -i (x_1 - i x_2) = z_{32}$  \\
  $(\mathbf{x})_{21} = -i (x_1 + i x_2) = z_{41}-z_{42}z_{21}  \; ,
  \qquad (\mathbf{x})_{22} = -i (x_0 - x_3)  = z_{42}\; , \qquad \chi_2 =  z_{43}$
\end{flushleft}
In the next Subsection we also use light-cone coordinates
 \begin{equation}\label{lc-not}
x_{\pm}= -i (x_0 \pm x_3) \ ,\ x=-i (x_1 - i x_2) \ ,\
\bar{x} = -i (x_1 + i x_2) \ ,
%\ z_1 = \chi_1 \ ,\ z_2 = \chi_2 \; ,
 \end{equation}
  so that $2 \times 2$ blocks (\ref{Lsl06}) inside $\mathrm{L}$-operator (\ref{cnfA17f}) have the form
 \begin{equation}
 \label{bfxp}
\mathbf{x} = \left(
  \begin{array}{cc}
    x_{+} & x \\
    \bar{x} & \; x_{-}
  \end{array}
\right) \qquad \mathbf{p} = \left(
  \begin{array}{cc}
    -\partial_{x_{+}} & -\partial_{\bar{x}} \\
    -\partial_x & -\partial_{x_{-}}
  \end{array}
\right)  \equiv \mathbf{p}_x \; .
 \end{equation}
A solution of
equations (\ref{u+u-}), (\ref{3.13''}) (for $n=4$) and (\ref{u-ell}) gives the connection between
parameters $\rho_k$ and $\Delta , \ell , \dot{\ell}$
\begin{equation}\label{param}
\rho_1 = -\frac{\Delta}{2} +\ell+3\ \ \ \rho_2 =
-\frac{\Delta}{2} -\ell+2\ \ \ \rho_3 = \frac{\Delta}{2}
+\dot{\ell}+1\ \ \ \rho_4 = \frac{\Delta}{2} -\dot{\ell}
\end{equation}
Thus,
we have the bridge between two formulations of
$\mathrm{L}$-operator in $s\ell(4,\mathbb{C})$ and $so(6,\mathbb{C})$
(or $su(2,2)$ and $so(2,4)$) cases. In the next Subsection we shall reproduce all
constructions \cite{DM} of intertwining operators for the $s\ell(4,\mathbb{C})$ case
and apply them for $so(6,\mathbb{C})$ case.

\subsection{Intertwining operators
and star-triangle relation. The $so(6,\mathbb{C})=s\ell(4,\mathbb{C})$ case}

\mbox{} In Section {\bf \ref{slnA}}
 %\marginpar{\bf Ubral tvoj novyj remark v konec -- Remark 3}
we have introduced operators $\mathcal{T}_k$ which intertwine two
$s\ell(N,\mathbb{C})$-type
$\mathrm{L}$-operators (\ref{Lsl4}) and permute their spectral parameters as it is shown in
(\ref{s-op}). In this Subsection we consider intertwining operators for a
product
of two $s\ell(4,\mathbb{C})$-type $\mathrm{L}$-operators (\ref{Lsl4}):
 \begin{equation}\label{prodLL}
\mathrm{L}_1(u_1,u_2,u_3,u_4) \,\mathrm{L}_2(v_1,v_2,v_3,v_4) \in
 {\rm End}(\mathbb{C}^4  \otimes  V_{\Delta_1,\ell_1,\dot{\ell}_1}  \otimes V_{\Delta_2,\ell_2,\dot{\ell}_2}) \; .
 \end{equation}
Here operators $\mathrm{L}_1$ and $\mathrm{L}_2$ act in different quantum
spaces $V_{\Delta_1,\ell_1,\dot{\ell}_1}$ and $V_{\Delta_2,\ell_2,\dot{\ell}_2}$
(the spaces of the differential representations $\rho$)
and indices $1$ and $2$ indicate these spaces, respectively.
Recall the definition of the spectral parameters in operators
$\mathrm{L}_1$ and $\mathrm{L}_2$ (see (\ref{param})):
 \begin{equation}\label{param01}
 \begin{array}{c}
\displaystyle{
(u_1 , u_2 , u_3 , u_4) = \left( u + \frac{\Delta_1}{2}-\ell_1 - 3,
u + \frac{\Delta_1}{2}+\ell_1 - 2 , u - \frac{\Delta_1}{2}-\dot{\ell}_1 -1 ,
u - \frac{\Delta_1}{2}+\dot{\ell}_1 \right) } \; , \\ [0.3cm]
\displaystyle{(v_1 , v_2 , v_3 , v_4) = \left( v + \frac{\Delta_2}{2}-\ell_2 - 3,
v + \frac{\Delta_2}{2}+\ell_2 - 2 , v - \frac{\Delta_2}{2}-\dot{\ell}_2 -1 ,
v - \frac{\Delta_2}{2}+\dot{\ell}_2 \right) } \; ,
\end{array}
 \end{equation}
where $\Delta_1\,,\Delta_2$ are the scaling dimensions  and  $(\ell_1, \, \dot{\ell}_1)$, $(\ell_2, \, \dot{\ell}_2)$ are
the spin values.
For the general case of $s\ell(N,\mathbb{C})$-type $\mathrm{L}$-operators the intertwiners $\mathrm{S}$
such that
$$
\mathrm{S} \cdot \mathrm{L}_1(u_1,\dots,u_N) \,\mathrm{L}_2(v_1,\dots,v_N) =
\mathrm{L}_1(u'_1,\dots,u'_N) \,\mathrm{L}_2(v'_1,\dots,v'_N) \cdot \mathrm{S} \; ,
$$
\begin{equation}\label{pers}
 (v'_1,\dots,v'_N, u'_1,\dots,u'_N) = s(v_1,\dots,v_N, u_1,\dots,u_N) \; ,
\end{equation}
were constructed in \cite{DM}.
In equation (\ref{pers}) we
denote by $s$ any permutation of $2N$ spectral parameters $(v_1,\dots,v_N, u_1,\dots,u_N)$.
In this Subsection we briefly discuss the intertwining operators $\mathrm{S}$ for the product
$\mathrm{L}_1( u_1 , \cdots , u_4 ) \,
\mathrm{L}_2 ( v_1 , \cdots , v_4 )$ of two $s\ell(4,\mathbb{C})$-type
$\mathrm{L}$-operators which permute parameters
inside the set $\mathbf{u} = ( v_1 ,\cdots , v_4
, u_1 ,\cdots , u_4 )$. First, we choose
the following variables for the operator $\mathrm{L}_1$: light-cone coordinates
 $\vec{x}_1=(y_{+},y_{-}, y ,\bar{y})$  for
 space-time vector (see (\ref{lc-not})) and $\chi_1$ and $\chi_2$ for harmonic variables. For
the operator $\mathrm{L}_2$ we choose $\vec{x}_2=(z_{+},z_{-}, z ,\bar{z})$ for space-time vector
 and $\eta_1$ and $\eta_2$ for harmonic variables.
In terms of these variables
the differential operators $D_{k,k+1}$ (\ref{notZD}) for $\mathrm{L}_1$ and $\mathrm{L}_2$ have
 the following representations
 \begin{equation}
 \label{DyDz}
 \begin{array}{l}
\mathrm{L}_1: \;\;\;  \mathrm{D}_{12} \to \partial_{\chi_1} \; , \;\;\;
\mathrm{D}_{23} \to \mathrm{D}_{y} =\partial_{y} + \chi_2 \partial_{y_{-}} - \chi_1
\partial_{y_{+}} - \chi_1 \chi_2 \partial_{\bar{y}} \; , \;\;\;
\mathrm{D}_{34} \to \partial_{\chi_2} \; , \\ [0.2cm]
\mathrm{L}_2: \;\;\;  \mathrm{D}_{12} \to \partial_{\eta_1} \; , \;\;\;
\mathrm{D}_{23}  \to \mathrm{D}_{z} =\partial_{z} + \eta_2 \partial_{z_{-}} - \eta_1
\partial_{z_{+}} - \eta_1 \eta_2 \partial_{\bar{z}}\ , \;\;  \mathrm{D}_{34} \to \partial_{\eta_2} \; .
\end{array}
 \end{equation}
Then, according to the results of \cite{DM} (see also Section {\bf \ref{slnA}}), the intertwining operators
$\mathcal{T}_k$ (\ref{s-op}), (\ref{SkD}) which separately permute the spectral parameters $(v_1 ,\cdots , v_4)$
in $\mathrm{L}_2$  and $(u_1 ,\cdots , u_4)$ in $\mathrm{L}_1$ are
\begin{equation}\label{s1s2s3}
 \mathrm{L}_2: \;\;\; \mathcal{T}_1( \mathbf{u} ) = \partial^{v_2 - v_1}_{\eta_1} \qquad
 \mathcal{T}_2(\mathbf{u} ) = \mathrm{D}^{v_3 - v_2}_z \qquad
 \mathcal{T}_3( \mathbf{u} ) =
\partial^{v_4 - v_3}_{\eta_2} \; ,
 \end{equation}
\begin{equation}\label{s5s6s7}
\mathrm{L}_1: \;\;\; \mathcal{T}_5( \mathbf{u} ) = \partial^{u_2 - u_1}_{\chi_1} \qquad
 \mathcal{T}_6(\mathbf{u} ) = \mathrm{D}^{u_3 - u_2}_y \qquad
  \mathcal{T}_7( \mathbf{u} ) =
\partial^{u_4 - u_3}_{\chi_2} \; .
 \end{equation}
The middle intertwining operator which correspond to the permutation
$u_1 \leftrightarrow v_4$ in the
product of two $\mathrm{L}$-operators (\ref{prodLL}) is \cite{DM}
$$
\mathcal{T}_4( \mathbf{u} ) = \mathrm{S}(\vec{x}_1-\vec{x}_2)^{u_1 - v_4} \; ,
$$
where
 \begin{equation}
 \label{Sx1x2}
\mathrm{S}(\vec{x}_1-\vec{x}_2) = (\bar{y} - \bar{z}) + \chi_1 ( y_{-} -
z_{-} ) + \eta_{2} ( z_{+} - y_{+} ) + \chi_1 \eta_2 ( z - y ) \; .
 \end{equation}

Next we construct the composite intertwining operators $\mathrm{S}_1$ and
 $\mathrm{S}_2$. The first operator $\mathrm{S}_1$ interchanges pairs $(v_1,v_2)$ and
 $(v_3,v_4)$: $(\underline{v_1 , v_2}, v_3 , v_4 ) \rightarrow (v_3 , v_4 , \underline{v_1, v_2} )$.
 In terms of physical parameters this permutation is written as
 $(\Delta,\ell_2,\dot{\ell}_2)\rightarrow (4-\Delta,\dot{\ell}_2,\ell_2)$.
 We explain the choice of this intertwining operator at the end of this Subsection (see Remark 2).
 According to (\ref{s1s2s3})
 the
explicit form of $\mathrm{S}_1$ is
\begin{equation}\label{U1}
\mathrm{S}_1 = \mathcal{T}_2(s_1 s_3 s_2 \mathbf{u}) \, \mathcal{T}_1(s_3 s_2 \mathbf{u}) \,
 \mathcal{T}_3( s_2 \mathbf{u}) \, \mathcal{T}_2(\mathbf{u})
= \mathrm{D}^{v_4 - v_1}_z \,
 \partial^{v_4 - v_2}_{\eta_2} \, \partial^{v_3 - v_1}_{\eta_1} \, \mathrm{D}^{v_3 - v_2}_z \; .
\end{equation}
We stress that in (\ref{U1}) for each $\mathcal{T}_k$
the previous permutations $s_m$ (\ref{3.18'}) of the spectral parameters should be taken into account.

\vspace{0.2cm}

The second intertwining operator $\mathrm{S}_2$ interchanges pairs $(v_3,v_4)$ and $(u_1,u_2)$:
$$
(v_1 , v_2, \underline{v_3 , v_4} , \underline{\underline{u_1 , u_2}} , u_3 , u_4) \rightarrow (v_1 ,
v_2 , \underline{\underline{u_1 , u_2}} , \underline{v_3 , v_4} , u_1 , u_2) \; ,
$$
and explicit form of $\mathrm{S}_2$ is
 \begin{equation}
 \label{U2}
\mathrm{S}_2 = \mathcal{T}_4(s_5 s_3 s_4 \mathbf{u})
 \mathcal{T}_5( s_3 s_4 \mathbf{u})
 \mathcal{T}_3(s_4 \mathbf{u}) \mathcal{T}_4(\mathbf{u}) =
 \mathrm{S}(\vec{x}_1-\vec{x}_2)^{u_2 - v_3}\,  \partial^{u_2 - v_4}_{\chi_1} \,
  \partial^{u_1 - v_3}_{\eta_2} \, \mathrm{S}(\vec{x}_1-\vec{x}_2)^{u_1 - v_4} \; .
 \end{equation}
The remarkable fact \cite{DM} is that the operators $\mathrm{S}_1$ and $\mathrm{S}_2$ satisfy the
braid relation
\begin{equation}\label{U3}
\mathrm{S}_1 \, \mathrm{S}_2 \, \mathrm{S}_1 = \mathrm{S}_2 \, \mathrm{S}_1 \, \mathrm{S}_2 \; .
\end{equation}
In next Section {\bf \ref{Lybe}} we interpret the identity (\ref{U3})
 as the star-triangle relation for propagators of spin
particles in certain conformal field theory.
 %The identity (\ref{U3}) is the main result of this paper and
 %we prove it in the next Section {\bf \ref{Lybe}}.

 %\add{In the next section we go to the Euclidean space and in this
 %subsection we work with complexification -- $SL(6,C)$ so that what
 %we should to do with Lorentz covariance and $SO(1,3)$ below.}

 \vspace{0.2cm}

\noindent
{\bf Remark 1.} One can try to write operators
 %\marginpar{\bf Novaya vstavka}
$\mathrm{D}_y$, $\mathrm{D}_z$ in (\ref{DyDz}) and
$S(\vec{x}_1 - \vec{x}_2)$ in (\ref{Sx1x2}) in covariant form
(under the transformations of the subgroup $SO(4,\mathbb{C}) \subset SO(6,\mathbb{C})$
with generators $\rho(L_{\mu\nu})$ (\ref{cnfA1}))
by means of introducing new homogeneous variables
 $\lambda_\alpha,\tilde{\lambda}^{\dot{\alpha}},
 \mu_\alpha,\tilde{\mu}^{\dot{\alpha}}$ (see Subsection {\bf \ref{spinSS}}):
 $$
 %\begin{array}{c}
 \chi_1 = \frac{\lambda_2}{\lambda_1} \; , \;\;\;
 \chi_2 = \frac{ \tilde{\lambda} ^{\dot{2}} }{\tilde{\lambda}^{\dot{1}}} \; , \;\;\;
 \eta_1 = \frac{\mu_2}{\mu_1} \; , \;\;\;
 \eta_2 = \frac{ \tilde{\mu} ^{\dot{2}} }{\tilde{\mu}^{\dot{1}}} \; ,
 %\end{array}
 $$
 $$
 \partial_{\chi_1} = \lambda_1 \, \partial_{\lambda_2} \; , \;\;\;
 \partial_{\eta_1} = \mu_1 \, \partial_{\mu_2} \; , \;\;\;
 \partial_{\eta_2} = \tilde{\mu}^{\dot{1}} \, \partial_{\tilde{\mu}^{\dot{2}}} \; ,
 $$
 In terms of these new variables we have
 \begin{equation}
 \label{DzDy}
 D_y = \frac{1}{(\lambda_1 \tilde{\lambda}^{\dot{1}})} \lambda^\alpha \; ({\bf p}_y)_{\alpha \dot{\alpha}}
 \tilde{\lambda}^{\dot{\alpha}}  \; , \;\;\;
  D_z = \frac{1}{(\mu_1 \tilde{\mu}^{\dot{1}})} \mu^\alpha \; ({\bf p}_z)_{\alpha \dot{\alpha}}
 \tilde{\mu}^{\dot{\alpha}}  \; ,
  %\end{array}
 \end{equation}
 $$
 S(\vec{x}_1 - \vec{x}_2)  = \frac{1}{(\lambda_1 \tilde{\mu}^{\dot{1}})} \tilde{\mu}_{\dot{\alpha}}
 (\mathbf{y} - \mathbf{z})^{\dot{\alpha} \alpha} \; \lambda_\alpha  \; ,
 $$
 where $\lambda^\alpha = \lambda_\beta \varepsilon^{\beta \alpha}$,
 $\mu^\alpha = \mu_\beta \varepsilon^{\beta \alpha}$,
 $\tilde{\mu}_{\dot{\alpha}}=\tilde{\mu}^{\dot{\beta}} \varepsilon_{\dot{\beta}\dot{\alpha}}$
 ($\varepsilon^{\beta \alpha}$ and $\varepsilon_{\dot{\beta}\dot{\alpha}}$ --- antisymmetric
 tensors)
  and ${\bf p}_y,{\bf p}_z, \mathbf{y},\mathbf{z}$ were defined in (\ref{bfxp}).
 Then the operators (\ref{U1}), (\ref{U2}) is represented in the form
 \begin{equation}
 \label{U11}
 \mathrm{S}_1 = (\mu \; {\bf p}_z \; \tilde{\mu})^{v_4-v_1} \;
  \left( \frac{1}{\mu_1} \partial_{\tilde{\mu}^{\dot{2}} } \right)^{v_4-v_2} \;
  \left( \frac{1}{\tilde{\mu}^{\dot{1}} }\partial_{\mu_2} \right)^{v_3-v_1} \;
  (\mu \; {\bf p}_z \; \tilde{\mu})^{v_3-v_2}  \; ,
  \end{equation}
 \begin{equation}
 \label{U22}
 \mathrm{S}_2 =
 (\tilde{\mu} (\mathbf{y} - \mathbf{z})\; \lambda)^{u_2 - v_3} \;
 \left( \frac{1}{\tilde{\mu}^{\dot{1}} }\partial_{\lambda_2} \right)^{u_2-v_4} \;
  \left( \frac{1}{\lambda_1} \partial_{\tilde{\mu}^{\dot{2}} } \right)^{u_1-v_3} \;
  (\tilde{\mu} (\mathbf{y} - \mathbf{z})\; \lambda)^{u_1 - v_4} \; .
 \end{equation}
The covariance
of the operators $\mathrm{S}_1$ (\ref{U11}) and $\mathrm{S}_2$ (\ref{U22})
under $SO(4,\mathbb{C})$ transformations
 (or the Lorentz covariance for real forms of $\mathrm{S}_1$ and $\mathrm{S}_2$)
 %and take $SO(1,3)$ instead of $SO(4,\mathbb{C})$)
is broken in view of the presence of noncovariant
operators $\frac{1}{\mu_1} \partial_{\tilde{\mu}^{\dot{2}} }$,
$\frac{1}{\tilde{\mu}^{\dot{1}} }\partial_{\lambda_2}$ etc. in
(\ref{U11}) and (\ref{U22}).
In next Section {\bf \ref{Lybe}}, using slightly
different approach, we derive another operators $\mathrm{S}_1$, $\mathrm{S}_2$ and
$\mathrm{S}_3$  which are represented in the Lorentz covariant form
and therefore their physical interpretation as propagators of
spining particles will be clarified.

\vspace{0.2cm}

 %\add{The same remark about Euclidean space. The expression for the
 %$\rho(C_3)$ I have changed.}

\noindent
{\bf Remark 2.}  The irreducible representations
 %\marginpar{\bf Redaktorskaya pravka}
 of the algebra $so(6,\mathbb{C})$
(complexification of the conformal algebra
$so(2,4)$) in the differential realization (\ref{cnfA1}) is characterized by the
conformal dimension $\Delta$ and spin parameters ($\ell$, $\dot{\ell}$)
which are labels of the representations of the subalgebra $so(4,\mathbb{C})=
s\ell(2,\mathbb{C})+ s\ell(2,\mathbb{C})$
\footnote{There is also parameter which is eigenvalue of the operator
$\hat{\ell}_{\mu\nu}\hat{\ell}^{\mu\nu}$ (see (\ref{cnfA2z})) but this
additional parameter is
not important for our consideration.}.
If all Casimir operators for two such representations of $so(6,\mathbb{C})$ coincide then
these representations are equivalent
 %\marginpar{\bf V etom Remark ya ne stal zamenyat' slovo equivalent}
(or partially equivalent) and the intertwining operator
between these representations should exist. For the algebra
$so(6,\mathbb{C})$ (\ref{cnfA1}) there are three Casimir operators: the first one is $\rho(C_2)$ (\ref{cnfA2z}) and
two others are
$$
\rho(C_3) =  \epsilon^{abcdef}\,\rho(M_{ab} \, M_{cd} \, M_{ef}) \; , \;\;\;
\rho(C_4) =  \rho(M_{ab} \, M^{bc} \, M_{cd} \, M^{da}) \; .
$$
In view of the isomorphism $so(6,\mathbb{C}) = s\ell(4,\mathbb{C})$,
the eigenvalues of these Casimir operators are elementary symmetric polynomials in four variables
$(\rho_1,\rho_2,\rho_3,\rho_4)$ (\ref{param})
and therefore any permutations of these variables lead to the equivalent representations.
Consider the spectral parameters (\ref{u-ell}), (\ref{param01}):
$$
 (u_1 , u_2 , u_3 , u_4) = (u-\rho_1 , \; u-\rho_2 , \; u-\rho_3 , \; u-\rho_4) =
  \left( u_+ -\ell - 1, \; u_{+} + \ell, \; u_{-} - \dot{\ell} -1 , \; u_{-} +\dot{\ell} \right) \; ,
$$
instead of parameters (\ref{param}). Note that the permutation $u_1 \leftrightarrow u_2$
is equivalent to the transformation $\ell \to -1-\ell$ while permutation $u_3 \leftrightarrow u_4$
is equivalent to $\dot{\ell} \to -1-\dot{\ell}$. Both permutation are not appropriate for
us since we would like to work with the finite dimensional representations of spin
algebras (\ref{sss}) and (\ref{barsss}) when
parameters $2\ell$ and $2\dot{\ell}$ are nonnegative integers. The other permutations
of $(u_1 , u_2 , u_3 , u_4)$ include
the interchanging $u_+ \leftrightarrow u_-$. In this case we have two possibilities
$\ell \to -1-\dot\ell$ or $\ell \to \dot{\ell}$. Again the first possibility is
 not appropriate for us since it is not compatible with the finite dimensional representations of spin
algebras. As the final result we have only one variant of intertwining operator
which permutes $u_+ \leftrightarrow u_-$, $\ell \to \dot{\ell}$
 and therefore corresponds to the permutation of pairs of the spectral parameters $(u_1,u_2)$ and  $(u_3,u_4)$.
Precisely this intertwining operator was constructed in (\ref{U1}) and will be
investigated in the next Section.

\vspace{0.3cm}

\noindent
{\bf Remark 3.}
In the paper \cite{DM} the complex group $SL(N,C)$ were considered
 and there we have
$\frac{N(N-1)}{2}$ complex variables $z_{ik}$ and $\frac{N(N-1)}{2}$ complex conjugate variables $\bar{z}_{ik}$.
In the case of $SL(4,C)$ we have 6 complex variables and 6
 complex conjugate variables. In Subsection {\bf \ref{sec-sl4}} all operators
  are well defined because we work with the differential operators and one
 can restrict everything to complex variables and forget about
 complex conjugated variables -- the holomorphic and antiholomorphic
 sectors can be separated. In this Subsection the situation is different
 because operators like $\partial_z^{\alpha}$ (i.e., the operators
 $D_z^\alpha\sim (\mu  {\bf p}_z  \tilde{\mu})^\alpha$ in (\ref{U1}), (\ref{U11})) for noninteger
$\alpha$ needs antiholomorphic part $\partial_{\bar{z}}^{\bar{\alpha}}$ so that only the product
$\partial_z^{\alpha}\cdot\partial_{\bar{z}}^{\bar{\alpha}}$ can be defined as usual integral operator acting on the functions $f(z,\bar{z})$ defined on $\mathbb{R}^2$.
We omit the antiholomorphic part everywhere in this Subsection so that intertwining
operators are not properly defined and  can be treated only as formal operators.
 %have only formal and illustrative meaning.\\
 %Really I like the purpose of this section -- the explanation
 %why we have to joint elementary intertwining operators in
 %$\mathrm{S}_1$ and $\mathrm{S}_2$ and how all this algebra
 %looks from the point of view of the general $SL(6,C)$-picture.
 %But it is not very clear what are minimal modifications to improve this Subsection.
 This is another reason why in next Section we develop slightly different approach.

%%%%%%%%%%%%%%%%%%%%%%%%%%%%%%%%%%%%%%%%%%%%%%%%%%%%%%%%%%%%%%%%%%%%%%%%%%%%%%%%%%%%%%%%%%
%%%%%%%%%%%%%%%%%%%%%%%%%%%%%%%%%%%%%%%%%%%%%%%%%%%%%%%%%%%%%%%%%%%%%%%%%%%%%%%%%%%%%%%%%%

\section{General $\mathrm{R}$-operator \label{Lybe}}
\setcounter{equation}{0}

\mbox{} In this Section we are going to construct
$\mathrm{R}$-operator as solution of the defining $\mathrm{RLL}$-equation~\cite{KRS,Fad}
$$
\CCR_{12}(u-v)\,\mathrm{L}_1(u)\,\mathrm{L}_2(v) =
\mathrm{L}_1(v)\,\mathrm{L}_2(u)\,\CCR_{12}(u-v)
$$
with conformal $\mathrm{L}$-operator (\ref{cnfA17}).
 Here indices $1,2$ correspond to two infinite-dimensional spaces of differential
representation $\rho$ of the conformal algebra
$\textsf{conf}(\mathbb{R}^{n})$ (\ref{cnfA1}) and we consider two cases:
\begin{itemize}
  \item Dimension $n$ of the Euclidean
  %\marginpar{\bf redaktorskaya pravka}
  space $\mathbb{R}^{n}$ is arbitrary and representations of the conformal algebra are restricted to the case of scalars: $\mathbf{S}=0$ and $\bar{\mathbf{S}}=0$.
  \item Dimension $n$ of the space $\mathbb{R}^{n}$ is fixed by $n=4$ and representations of the
   conformal algebra are generic: $\mathbf{S}\neq 0$ and $\bar{\mathbf{S}}\neq 0$.
\end{itemize}

\subsection{$n$-dimensional scalar case}

\mbox{} In this case the defining $\mathrm{RLL}$-equation
has the form
\begin{equation} \label{PermRLL}
\CCR_{12}(u-v)\,\mathrm{L}_1(u_+,u_-)\,\mathrm{L}_2(v_+,v_-) =
\mathrm{L}_1(v_+,v_-)\,\mathrm{L}_2(u_+,u_-)\,\CCR_{12}(u-v)\,,
\end{equation}
where
$$
\mathrm{L}_1( u_+,u_- ) = \left(
\begin{array}{cc}
  \boldsymbol{1} & \; \mathbf{0} \\
  \mathbf{x}_1 & \; \boldsymbol{1}
\end{array}
\right) \cdot \left(
\begin{array}{cc}
  u_+ \cdot \boldsymbol{1}  & \ \ \mathbf{p}_1 \\
  \mathbf{0}  & u_- \cdot \boldsymbol{1}
\end{array}
\right) \cdot \left(
\begin{array}{cc}
  \boldsymbol{1} & \; \mathbf{0} \\
  - \mathbf{x}_1 & \; \boldsymbol{1}
\end{array}
\right)  \,,
$$
$$
\mathrm{L}_2( v_+,v_- ) = \left(
\begin{array}{cc}
  \boldsymbol{1} & \; \mathbf{0} \\
  \mathbf{x}_2 & \; \boldsymbol{1}
\end{array}
\right) \cdot \left(
\begin{array}{cc}
  v_+ \cdot \boldsymbol{1}  & \ \ \mathbf{p}_2 \\
  \mathbf{0}  & v_- \cdot \boldsymbol{1}
\end{array}
\right) \cdot \left(
\begin{array}{cc}
  \boldsymbol{1} & \; \mathbf{0} \\
  - \mathbf{x}_2 & \; \boldsymbol{1}
\end{array}
\right)  \; ,
$$
and $u_+ = u + \frac{\Delta_1 -n}{2}$ , $u_- = u - \frac{\Delta_1}{2}$ ,
$v_+ = v + \frac{\Delta_2 -n}{2}$ , $v_- = v - \frac{\Delta_2}{2}$ (\ref{u+u-}).

The $\CCR$-operator in (\ref{PermRLL}) interchanges a pair of parameters $(u_+,u_-)$ in the first
$\CCL$-operator
with a pair $(v_+,v_-)$ from the second $\CCL$-operator. It seems to
be reasonable to consider also operators which perform the
other interchanges of four parameters.
In  order to carry out it systematically we
joint them in the set $\mathbf{u} = (v_+,v_-,u_+,u_-)$.
Then $\CCR$-operator represents the permutation $s$ such that
\begin{equation} \label{RPerm}
s \mapsto \CCR(u-v) \ ;\ s\,\mathbf{u} =
(\underline{u_+,u_-},\underline{v_+,v_-}).
\end{equation}
An arbitrary permutation can be builded from elementary transpositions
$\mathrm{s}_{1}$, $\mathrm{s}_{2}$ and
$\mathrm{s}_{3}$
$$
s_{1}\mathbf{u} = (\underline{v_-},\underline{v_+},u_+,u_-)\ ;\ s_{2}\mathbf{u}
 = (v_+,\underline{u_+},\underline{v_-},u_-) \ ;\
s_{3}\mathbf{u} = (v_+,v_-,\underline{u_-},\underline{u_+})\,.
$$
In particular: $s = s_2 s_1 s_3 s_2$.
Thus we reduce the problem to construction of
operators $\mathrm{S}_i(\mathbf{u})$ ($i=1,2,3$)
which represent elementary transpositions
\begin{equation} \label{S1}
(\underline{v_+\ ,\ v_-},u_+,u_-)\ :\
\mathrm{S}_1(\mathbf{u})\,\mathrm{L}_2(v_+,v_-) =
\mathrm{L}_2(v_-,v_+) \, \mathrm{S}_1(\mathbf{u})
\end{equation}
\begin{equation} \label{S2}
(v_+,\underline{v_-\ ,\ u_+}, u_-)\ :\
\mathrm{S}_2(\mathbf{u})\,\mathrm{L}_1(u_+,u_-)\,\mathrm{L}_2(v_+,v_-)=
\mathrm{L}_1(v_-,u_-)\,\mathrm{L}_2(v_+,u_+) \, \mathrm{S}_2(\mathbf{u})\,
\end{equation}
\begin{equation} \label{S3}
(v_+,v_-,\underline{u_+\ ,\ u_-})\ :\
\mathrm{S}_3(\mathbf{u})\,\mathrm{L}_1(u_+,u_-) =
\mathrm{L}_1(u_-,u_+) \, \mathrm{S}_3(\mathbf{u})
\end{equation}
We have
the correspondence
\begin{equation}
s_i \mapsto \mathrm{S}_i(\mathbf{u})\ \ ;\ \ s_i s_j \mapsto
\mathrm{S}_i(s_j\mathbf{u})\,\mathrm{S}_j(\mathbf{u})\ \ ;\ \ s_i s_j s_k \mapsto
\mathrm{S}_i(s_js_k\mathbf{u})\,\mathrm{S}_j(s_k\mathbf{u})\,
\mathrm{S}_k(\mathbf{u})\ \ ;\ \ \cdots
\end{equation}
and for the proof that it is indeed the representation of the permutation
group of four parameters we have to check the corresponding
defining (Coxeter) relations
\begin{equation}\label{def1}
s_i s_i = \II \longrightarrow
\mathrm{S}_i(s_i\mathbf{u})\,\mathrm{S}_i(\mathbf{u})= \II\ ;\
s_1s_3 = s_3s_1 \longrightarrow
\mathrm{S}_1(s_3\mathbf{u})\,\mathrm{S}_3(\mathbf{u})=
\mathrm{S}_3(s_1\mathbf{u})\,\mathrm{S}_1(\mathbf{u})
\end{equation}
\begin{equation}\label{def2}
s_1 s_2 s_1 = s_2 s_1 s_2 \longrightarrow
\mathrm{S}_1(s_2s_1\mathbf{u})\,\mathrm{S}_2(s_1\mathbf{u})\,
\mathrm{S}_1(\mathbf{u})=
\mathrm{S}_2(s_1s_2\mathbf{u})\,\mathrm{S}_1(s_2\mathbf{u})\,
\mathrm{S}_2(\mathbf{u})
\end{equation}
\begin{equation}\label{def3}
s_2 s_3 s_2 = s_3 s_2 s_3 \longrightarrow
\mathrm{S}_2(s_3s_2\mathbf{u})\,\mathrm{S}_3(s_2\mathbf{u})\,
\mathrm{S}_2(\mathbf{u})=
\mathrm{S}_3(s_2s_3\mathbf{u})\,\mathrm{S}_2(s_3\mathbf{u})\,
\mathrm{S}_3(\mathbf{u})
\end{equation}
Then $\CCR$-operator can be constructed form these building blocks:
\begin{equation} \label{Rs}
\CCR(\mathbf{u}) = \mathrm{S}_2(s_1 s_3 s_2 \mathbf{u}) \, \mathrm{S}_1(s_3 s_2 \mathbf{u}) \,
\mathrm{S}_3(s_2 \mathbf{u}) \,\mathrm{S}_2(\mathbf{u})\,
\end{equation}
We will see that operators $\mathrm{S}_i$ depend on their parameters
in a special way
\begin{equation}\label{depend}
\mathrm{S}_1(\mathbf{u}) = \mathrm{S}_1(v_--v_+)\ ;\
\mathrm{S}_2(\mathbf{u}) = \mathrm{S}_2(u_+-v_-)\ ;\
\mathrm{S}_3(\mathbf{u}) = \mathrm{S}_3(u_--u_+),
\end{equation}
so that the operator $\mathrm{R}(\mathbf{u})$ depends on
the difference of spectral parameters $u-v$ as it should
\begin{equation}
\mathrm{R}(\mathbf{u}) =
\mathrm{S}_2(u_--v_+)\,
\mathrm{S}_1(u_+-v_+)\,\mathrm{S}_3(u_--v_-)\,
\mathrm{S}_2(u_+-v_-)\,.
\end{equation}
The Yang-Baxter relation for this $\CCR$-operator is the direct consequence of the Coxeter relations for the building blocks $\mathrm{S}_i(\mathbf{u})$.
In explicit notations relations~(\ref{def2}) and (\ref{def3}) have the form
$$
\mathrm{S}_1(u_+-v_-)\,\mathrm{S}_2(u_+-v_+)\,
\mathrm{S}_1(v_--v_+)=
\mathrm{S}_2(v_--v_+)\,\mathrm{S}_1(u_+-v_+)\,
\mathrm{S}_2(u_+-v_-)\,,
$$
$$
\mathrm{S}_2(u_--u_+)\,\mathrm{S}_3(u_--v_-)\,
\mathrm{S}_2(u_+-v_-)=
\mathrm{S}_3(u_+-v_-)\,\mathrm{S}_2(u_--v_-)\,
\mathrm{S}_3(u_--u_+)\,,
$$
and are particular examples of the general relations
\begin{equation}\label{Cox}
\mathrm{S}_1(a)\,\mathrm{S}_2(a+b)\,
\mathrm{S}_1(b)=
\mathrm{S}_2(b)\,\mathrm{S}_1(a+b)\,
\mathrm{S}_2(a)\ \ ;\ \
\mathrm{S}_2(a)\,\mathrm{S}_3(a+b)\,
\mathrm{S}_2(b)=
\mathrm{S}_3(b)\,\mathrm{S}_2(a+b)\,
\mathrm{S}_3(a)\,.
\end{equation}

We are going to construct operators $\mathrm{S}_i(\mathbf{u})$ and
at the first stage we consider operators $\mathrm{S}_1$ and $\mathrm{S}_3$ which
are examples of the operator $\mathrm{S}$
being defined by the equation
\begin{equation} \label{SL=LS}
\hat{\mathrm{S}} \cdot \mathrm{L}(u_+,u_-) = \mathrm{L}(u_-,u_+)\cdot \hat{\mathrm{S}} \,
\end{equation}
As soon as here we deal with a scalar case differential
representation of the conformal algebra
is parameterized by one parameter --
conformal dimension $\Delta$. We denote it by
$\rho^{\Delta}$.
Taking in mind the definition of the parameters $u_+$ and $u_-$ we see
that their transposition corresponds to
$\Delta \to n-\Delta$. Since $\CCL$-operator is linear on
spectral parameter and in view of equation (\ref{SL=LS}) we conclude that $\mathrm{S}$
intertwines
 %\marginpar{\bf "equivalent" zamenil na "two"}
 two representations of the conformal algebra:
$\rho^{\Delta}$ and $\rho^{n-\Delta}$.
 %Indeed
 Note that such a change of conformal dimension do
preserve the Casimir operator (\ref{cnfA2z}).

Let us represent the intertwining operator as an integral operator acting on fields
$\Phi(x)$ where $x\in \mathbb{R}^{p,q}$
$$
\left[\mathrm{S}\,\Phi\right](x) =
\int \mathrm{d}^n y\, \mathrm{S}\left(x, y\right)\,\Phi(y)\,,
$$
then defining equation for $\mathrm{S}$ (\ref{SL=LS}) is equivalent to
the set of equations
$$
\int \mathrm{d}^n y \, \mathrm{S}(x,y) \, G^{\Delta}_y \, \Phi(y) = \int \mathrm{d}^n y \, G^{n-\Delta}_x \, \mathrm{S}(x,y) \, \Phi(y)\,,
$$
which can be rewritten as the set of differential equations for the kernel
$\mathrm{S}\left(x, y\right)$
\begin{equation}
\label{sec3b}
\left( G_y^{\Delta} \right)^{T} \mathrm{S}\left(x, y\right) =
G_x^{n-\Delta}\,\mathrm{S}\left(x, y\right)\,.
\end{equation}
Here $G^{\Delta}_x$ denotes generators of conformal group in scalar ($S_{\mu \nu} = 0$)
differential representation (\ref{cnfA1})
and $T$ stands for transposition arising after integration by parts
$$
\int \mathrm{d}^n y \, \mathrm{S}(x,y) \, G_y \, \Phi(y) = \int \mathrm{d}^n y \, \left[ G_y^{T}\, \mathrm{S}(x,y)\right] \, \Phi(y)\,.
$$
We obtain the following equations:
\begin{itemize}
\item translation
 \begin{equation}\label{equa1}
\left(\frac{\partial}{\partial x_{\mu}}
+ \frac{\partial}{\partial y_{\mu}}\right)
\mathrm{S}\left(x, y\right) = 0\,,
 \end{equation}
\item Lorentz rotation
\begin{equation}\label{equa2}
\left( y_{\nu} \frac{\partial}{\partial y_{\mu}} - y_{\mu} \frac{\partial}{\partial y_{\nu}} \right) \mathrm{S}(x,y) =
\left( x_{\mu} \frac{\partial}{\partial x_{\nu}} - x_{\nu} \frac{\partial}{\partial x_{\mu}} \right) \mathrm{S}(x,y)\,,
\end{equation}
\item dilatation
\begin{equation}\label{equa3}
\left(x_{\mu}\frac{\partial}{\partial x_{\mu}}
+ y_{\mu}\frac{\partial}{\partial y_{\mu}}\right)
\mathrm{S}\left(x,y\right) =
-2\left(n-\Delta\right)\,\mathrm{S}\left(x,y\right)\,,
\end{equation}
\item conformal boost
\begin{equation}\label{equa4}
\left( x^2 \frac{\partial}{\partial x_{\mu}} - 2 x_{\mu} x_{\nu} \frac{\partial}{\partial x_{\nu}} +
y^2 \frac{\partial}{\partial y_{\mu}} - 2 y_{\mu} y_{\nu} \frac{\partial}{\partial y_{\nu}} \right) \mathrm{S}(x,y) =
2(n- \Delta) \, (x_{\mu} + y_{\mu}) \, \mathrm{S}(x,y)\, .
\end{equation}
 Note that in the scalar case $S_{\mu \nu} = 0$ the conformal boost equation is dispensable since it can
  be derived from (\ref{equa1}) -- (\ref{equa3}).
\end{itemize}
The set of equations for the kernel of $\hat{\mathrm{S}}$ coincides
with the set of equations for the Green function of
the two scalar fields with equal scaling dimensions in conformal field theory~\cite{Zaikov}. The solution is well known
$$
\mathrm{S}(x,y) = \frac{c}{(x-y)^{2(n-\Delta)}}\,,
$$
and is fixed up to an arbitrary multiplicative constant.
The action of the integral operator with the kernel
$\mathrm{S}(x,y)$ on the function $\Phi(x)$ can be represented
in different forms
\begin{equation}\label{Sscal}
[ \hat{\mathrm{S}} \, \Phi ] ( x )  =
c\,\int \frac{\mathrm{d}^n y }{(x-y)^{2(n-\Delta)}}
\cdot \Phi( y ) = c\,\int \frac{\mathrm{d}^n y }{y^{2(n-\Delta)}}
\cdot \Phi( x-y ) =
c\,\int \frac{\mathrm{d}^n y \,\, e^{iy \hat{p}}}{y^{2(n-\Delta)}}
\cdot \Phi\bigl( x)\,,
\end{equation}
where $\hat{p}_\nu = -i\partial_{x^\nu}$.
 %\marginpar{\bf pomenyal znak u $\hat{p}_\nu$}
There exists useful expression for this operator
\begin{equation} \label{Sdiff}
 \hat{\mathrm{S}}(u_- - u_+) =
{\hat{p}}^{\,2(u_- - u_+)} =
{\hat{p}}^{\,2\left(\frac{n}{2}-\Delta\right)}\,.
\end{equation}
Indeed, using the well-known formula for the Fourier transformation
$$
\int \mathrm{d}^n y\,
\frac{\mathrm{e}^{ -i\, y p}}
{y^{2\left(\frac{n}{2}-\alpha\right)}} = \frac{a(\alpha)}{p^{2\alpha}}\ \ ;\ \ a(\alpha)\equiv
\pi^{\frac{n}{2}}\, 4^{\alpha}\,
\frac{\Gamma\left(\alpha\right)}
{\Gamma\left(\frac{n}{2}-\alpha\right)}\,,
$$
it is possible to present the integral
operator of considered type as
 %in a following form
$$
\int \, \frac{\mathrm{d}^n y }
{(x-y)^{2\left(\frac{n}{2}-\alpha\right)}}\,\Phi(y) = a(\alpha)\,\hat{p}^{\,-2\alpha}\,\Phi(x)\,.
$$
In our case $\alpha = \Delta-\frac{n}{2}$, so that it remains to
choose the normalization constant $c$ in (\ref{Sscal}) in a special way
$$
c = \frac{1}{a\left(\Delta-\frac{n}{2}\right)} =
4^{\frac{n}{2}-\Delta} \pi^{\frac{n}{2}} \frac{\Gamma(n-\Delta)}{\Gamma(\Delta-\frac{n}{2})} \; ,
 %\,\,\longrightarrow\,\,  \hat{\mathrm{S}} =
 %{\hat{p}}^{\,2\left(\frac{n}{2}-\Delta\right)}\,.
$$
 to fix operator $\hat{\mathrm{S}}$ in the form (\ref{Sdiff}).
 %\marginpar{\bf \tiny ubral pod \% tvoyu novuyu frazu
 %pro equivalentnost', t.k. $p^\alpha = p^{\alpha -2} p^2$
 %i esli $\alpha$ ne celoe, to u $p^\alpha$ i $p^{\alpha -2}$
 %nulevykh mod kak by net, a u $p^2$ est'???? U Kollera ya ne nashel
 %utverzhdeniya ob equivalentnosti etikh predstavlenij.
 %V literature est' ponyatie slaboj equivalentnosti dlya
 %predstavlenij $C^*$ algebr von-Neumanna, s etim nado razbirat'sya}
 %\add{Note that in generic situation when the power
 %$(\frac{n}{2} - \Delta)$ is not positive integer the
 %operator $\mathrm{S}$ has not nontrivial kernel and indeed gives the isomorphism of
 %equivalent representations \cite{Koller}.}
Thus we have constructed operators $\mathrm{S}_1$ and $\mathrm{S}_3$
using solely their representation theory meaning. Explicit expressions
are the following
$$
\mathrm{S}_1(v_- - v_+) = \hat{p}_2^{\,2(v_- - v_+)} \ \ ;\ \
\mathrm{S}_3(u_- - u_+) = \hat{p}_1^{\,2(u_- - u_+)} \; .
$$

\vspace{0.2cm}

\noindent
{\bf Remark.} Solution (\ref{Sdiff}) can be obtained directly
if write equations (\ref{equa1}) -- (\ref{equa4})
 %\marginpar{\bf  Etot novyj Remark mozhno
 %vybrosit' -- na vashe usmotrenie}
 in the operator form (cf. (\ref{SL=LS})):
 \begin{equation}\label{equa1a}
\bullet \;\; {\rm translation:} \;\;\;\; \;\;\;\;\;\;\;\;
 [ \hat{p}_\mu \, , \; \hat{\mathrm{S}}] = 0 \,, %\\ [0.2cm]
 \end{equation}
 %\item
 \begin{equation}\label{equa2a}
\bullet \;\; {\rm Lorentz \;\; rotation:} \;\;\;\;
  [ x_\nu \, \hat{p}_\mu - x_\mu \, \hat{p}_\nu \, , \; \hat{\mathrm{S}}] = 0 \,, %\\ [0.2cm]
  \end{equation}
  %\item dilatation
\begin{equation}\label{equa3a}
  \bullet \;\; {\rm dilatation:} \;\;\;\;
   \left( x^\mu \, \hat{p}_\mu  - i (n-\Delta) \right) \; \hat{\mathrm{S}}
   = \hat{\mathrm{S}} \; \left( x^\mu \, \hat{p}_\mu  - i \Delta \right) \,,
\end{equation}
\begin{equation}\label{equa4a}
 \bullet \;\; {\rm conformal \;\; boost:} \;\;\;\;
 %\begin{array}{l}
 \left(  x_\mu  (x^\nu \, \hat{p}_\nu - 2 i (n-\Delta) )  - x^2 \hat{p}_\mu  \right) \; \hat{\mathrm{S}}
 %= \\ [0.2cm]
   = \hat{\mathrm{S}} \;
   \left(  x_\mu  (x^\nu \, \hat{p}_\nu - 2 i \Delta)  - x^2 \hat{p}_\mu  \right)  \, .
 %\end{array}
\end{equation}
Then equation (\ref{equa1a}) gives that $\hat{\mathrm{S}}$ depends only on $\hat{p}_{\mu}$,
from (\ref{equa2a}) we obtain that $\hat{\mathrm{S}}$ depends on the Lorentz invariant combination
$\hat{p}^2$ and (\ref{equa3a}) leads to the solution (\ref{Sdiff}) up to a normalization
constant. Equation (\ref{equa4a}) is
optional since operator (\ref{Sdiff}) satisfies (\ref{equa4a}) automatically.

\vspace{0.2cm}

It remains to construct the last building block
for $\CCR$-operator -- operator $\mathrm{S}_2$. It happens that
it can be produced directly from the operator $\mathrm{S}$
obtained above using some kind of duality transformation
$$
p \to x_{2}-x_{1} \equiv x_{21}\ ;\ u_+ \to v_-\ ;\ u_- \to u_+\,,
$$
so that $\mathrm{S}_2$ is the operator
of multiplication by the function
$$
\mathrm{S}_2(u_+ - v_-) = x_{12}^{2(u_+ - v_-)}\,.
$$
To explain the origin of these duality we start
from the defining equation (\ref{SL=LS}) for $\mathrm{S}$
\begin{equation} \label{sec3a}
\mathrm{S}\, \left(
\begin{array}{cc}
  \boldsymbol{1} & \; \mathbf{0} \\
  \mathbf{x} & \; \boldsymbol{1}
\end{array}
\right) \left(
\begin{array}{cc}
  u_+ \cdot \boldsymbol{1}  & \ \ \mathbf{p} \\
  \mathbf{0}  & u_- \cdot \boldsymbol{1}
\end{array}
\right) \left(
\begin{array}{cc}
  \boldsymbol{1} & \; \mathbf{0} \\
  - \mathbf{x} & \; \boldsymbol{1}
\end{array}
\right)
= \left(
\begin{array}{cc}
  \boldsymbol{1} & \; \mathbf{0} \\
  \mathbf{x} & \; \boldsymbol{1}
\end{array}
\right) \left(
\begin{array}{cc}
  u_- \cdot \boldsymbol{1} & \ \ \mathbf{p} \\
  \mathbf{0}  & u_+ \cdot \boldsymbol{1}
\end{array}
\right) \left(
\begin{array}{cc}
  \boldsymbol{1} & \; \mathbf{0} \\
  - \mathbf{x} & \; \boldsymbol{1}
\end{array}
\right)
\mathrm{S}\,,
\end{equation}
and show that the defining equation for $\mathrm{S}_2$
can be obtained from considered ones by the duality transformation.
For this purpose we rewrite the defining equation (\ref{S2}) for the
operator $\mathrm{S}_2$ in an explicit form
using factorization of the $\CCL$-operator
($\mathbf{x}_{21}\equiv \mathbf{x}_2-\mathbf{x}_1$)
$$
\mathrm{S}_2\underline{\left(%
\begin{array}{cc}
  \mathbf{1} & \mathbf{0} \\
  \mathbf{x}_1 & \mathbf{1} \\
\end{array}%
\right)\left(\begin{array}{cc}
  \mathbf{1} & \mathbf{0} \\
  \mathbf{0} & u_- \cdot \mathbf{1} \\
\end{array}%
\right)}\left(%
\begin{array}{cc}
  u_+ \cdot \mathbf{1} & \mathbf{p}_{1} \\
  \mathbf{0} & \mathbf{1} \\
\end{array}%
\right)\left(%
\begin{array}{cc}
  \mathbf{1} & \mathbf{0} \\
  \mathbf{x}_{21} & \mathbf{1} \\
\end{array}%
\right)\left(%
\begin{array}{cc}
  \mathbf{1} & \mathbf{p}_{2} \\
  \mathbf{0} & v_- \cdot \mathbf{1} \\
\end{array}%
\right)\underline{\left(\begin{array}{cc}
  v_+ \cdot \mathbf{1} & \mathbf{0} \\
  \mathbf{0} & \mathbf{1} \\
\end{array}%
\right)\left(%
\begin{array}{cc}
  \mathbf{1} & \mathbf{0} \\
  -\mathbf{x}_2 & \mathbf{1} \\
\end{array}%
\right)} =
$$
$$
= \underline{\left(%
\begin{array}{cc}
  \mathbf{1} & \mathbf{0} \\
  \mathbf{x}_1 & \mathbf{1} \\
\end{array}%
\right)\left(\begin{array}{cc}
  \mathbf{1} & \mathbf{0} \\
  \mathbf{0} & u_- \cdot \mathbf{1} \\
\end{array}%
\right)}\left(%
\begin{array}{cc}
  v_- \cdot \mathbf{1} & \mathbf{p}_{1} \\
  \mathbf{0} & \mathbf{1} \\
\end{array}%
\right)\left(%
\begin{array}{cc}
  \mathbf{1} & \mathbf{0} \\
  \mathbf{x}_{21} & \mathbf{1} \\
\end{array}%
\right)\left(%
\begin{array}{cc}
  \mathbf{1} & \mathbf{p}_{2} \\
  \mathbf{0} & u_+ \cdot \mathbf{1} \\
\end{array}%
\right)\underline{\left(\begin{array}{cc}
  v_+ \cdot \mathbf{1} & \mathbf{0} \\
  \mathbf{0} & \mathbf{1} \\
\end{array}%
\right)\left(%
\begin{array}{cc}
  \mathbf{1} & \mathbf{0} \\
  -\mathbf{x}_2 & \mathbf{1} \\
\end{array}%
\right)}\mathrm{S}_2
$$
By the condition
$
[\mathrm{S}_2,\mathbf{x}_1] =
[\mathrm{S}_2,\mathbf{x}_2] = 0\,
$
it is possible to cancel the underlined factors so that
equation is transformed to the much more simple form
$$
\mathrm{S}_2\left(%
\begin{array}{cc}
  \mathbf{1} & \mathbf{p}_{1} \\
  \mathbf{0} & \mathbf{1} \\
\end{array}%
\right)\left(%
\begin{array}{cc}
  u_+ \cdot \mathbf{1} & \mathbf{0} \\
  \mathbf{x}_{21} & v_- \cdot \mathbf{1} \\
\end{array}%
\right)\left(%
\begin{array}{cc}
  \mathbf{1} & \mathbf{p}_{2} \\
  \mathbf{0} & \mathbf{1} \\
\end{array}%
\right)
= \left(%
\begin{array}{cc}
  \mathbf{1} & \mathbf{p}_{1} \\
  \mathbf{0} & \mathbf{1} \\
\end{array}%
\right)\left(%
\begin{array}{cc}
  v_- \cdot \mathbf{1} & \mathbf{0} \\
  \mathbf{x}_{21} & u_+ \cdot \mathbf{1} \\
\end{array}%
\right)\left(%
\begin{array}{cc}
  \mathbf{1} & \mathbf{p}_{2} \\
  \mathbf{0} & \mathbf{1} \\
\end{array}%
\right)\mathrm{S}_2
$$
One more requirement we impose is the
translation invariance: $[\mathrm{S}_2, \mathbf{p}_{1}+\mathbf{p}_{2}] = 0$,
so that we obtain
\begin{equation}
 \label{sec3c}
\mathrm{S}_2\left(%
\begin{array}{cc}
  \mathbf{1} & \mathbf{p}_{1} \\
  \mathbf{0} & \mathbf{1} \\
\end{array}%
\right)\left(%
\begin{array}{cc}
  u_+ \cdot \mathbf{1} & \mathbf{0} \\
  \mathbf{x}_{21} & v_- \cdot \mathbf{1} \\
\end{array}%
\right)\left(%
\begin{array}{cc}
  \mathbf{1} & -\mathbf{p}_{1} \\
  \mathbf{0} & \mathbf{1} \\
\end{array}%
\right)
= \left(%
\begin{array}{cc}
  \mathbf{1} & \mathbf{p}_{1} \\
  \mathbf{0} & \mathbf{1} \\
\end{array}%
\right)\left(%
\begin{array}{cc}
  v_- \cdot \mathbf{1} & \mathbf{0} \\
  \mathbf{x}_{21} & u_+ \cdot \mathbf{1} \\
\end{array}%
\right)\left(%
\begin{array}{cc}
  \mathbf{1} & -\mathbf{p}_{1} \\
  \mathbf{0} & \mathbf{1} \\
\end{array}%
\right)\mathrm{S}_2\,.
\end{equation}
Next we perform similarity transformation
of the previous equation
by means of the matrix
%$\sigma_1 \otimes \sigma_0$
$\begin{pmatrix} 0 & \II \\ \II & 0 \end{pmatrix}$:
\begin{equation}
 \label{sec3d}
\mathrm{S}_2\left(%
\begin{array}{cc}
  \mathbf{1} & \mathbf{0} \\
  \mathbf{p}_1 & \mathbf{1} \\
\end{array}%
\right)\left(%
\begin{array}{cc}
  v_- \cdot \mathbf{1} & \mathbf{x}_{21} \\
  \mathbf{0} & u_+ \cdot \mathbf{1} \\
\end{array}%
\right)\left(%
\begin{array}{cc}
  \mathbf{1} & \mathbf{0} \\
  -\mathbf{p}_1 & \mathbf{1} \\
\end{array}%
\right)
= \left(%
\begin{array}{cc}
  \mathbf{1} & \mathbf{0} \\
  \mathbf{p}_1 & \mathbf{1} \\
\end{array}%
\right)\left(%
\begin{array}{cc}
  u_+ \cdot \mathbf{1} & \mathbf{x}_{21} \\
  \mathbf{0} & v_- \cdot \mathbf{1} \\
\end{array}%
\right)\left(%
\begin{array}{cc}
  \mathbf{1} & \mathbf{0} \\
  -\mathbf{p}_1 & \mathbf{1} \\
\end{array}%
\right)\mathrm{S}_2\,.
\end{equation}
It remains to compare this equation with the defining
equation (\ref{sec3a}) for the intertwining operator $\mathrm{S}$
which suggests that the change
$$%\begin{equation}
% \label{sec3f}
\mathbf{x} \to \mathbf{p}_1\ ;\
\mathbf{p} \to \mathbf{x}_{21}\ ;\ u_+ \to v_-\ ;\ u_- \to u_+
%\end{equation}
$$
transforms (\ref{sec3a}) into (\ref{sec3d}).
Thus we have that $\mathrm{S}_2$ is the operator of multiplication by the function
$$
\mathrm{S}_2(u_+ - v_-) = x_{12}^{2(u_+ - v_-)}\,.
$$
Coxeter relations~(\ref{def1}) are evident and Coxeter relations~(\ref{Cox}) have
the following explicit forms
\begin{equation} \label{S1S2S1}
\hat{p}_2^{\,2a} \, x_{12}^{2(a+b)} \, \hat{p}_2^{\,2b} =
x_{12}^{2b} \,\, \hat{p}_2^{\,2(a+b)} \, x_{12}^{2b}\ \ ;\ \
\hat{p}_1^{\,2a} \, x_{12}^{2(a+b)} \, \hat{p}_1^{\,2b} =
x_{12}^{2b} \,\, \hat{p}_1^{\,2(a+b)} \, x_{12}^{2a}\,,
\end{equation}
and are both equivalent to the operator identity~\cite{Isaev2,Isaev2a}: 
\begin{equation} \label{scStTrOp}
\hat{p}^{\,2 a} \, x^{2(a+b)} \, \hat{p}^{\,2 b} =
x^{2 b} \, \hat{p}^{\,2(a+b)} \, x^{2 a}\,,
\end{equation}
which can be rewritten in the standard integral form
\begin{equation} \label{scStTrInt}
\int \mathrm{d}^n w \frac{1}{(x-w)^{2\alpha}(y-w)^{2\beta}(z-w)^{2\gamma}} =
\mathrm{V}(\alpha,\beta,\gamma)\cdot
\frac{1}{(y-z)^{2 \alpha'}(x-z)^{2 \beta'}(x-y)^{2 \gamma'}}\,,
\end{equation}
where

$$
\mathrm{V}(\alpha,\beta,\gamma) = \pi^{\frac{n}{2}} \, \frac{\Gamma(\alpha')\,\Gamma(\beta')\,\Gamma(\gamma')}
{\Gamma(\alpha)\,\Gamma(\beta)\,\Gamma(\gamma)}
\ \ ; \ \ \alpha' = \frac{n}{2} - \alpha \,,\; \beta' = \frac{n}{2} - \beta \,,\; \gamma' = \frac{n}{2} - \gamma
$$
and parameters respect the uniqueness condition
$$\alpha+\beta+\gamma=n\;.$$
This integral identity is
a well-known star-triangle
relation~\cite{FrP,Vas,Vas1}. It is useful to
represent the identity in the picture where marked vertex represents the integration over the variable $w$.

\medskip

\begin{center}
\includegraphics{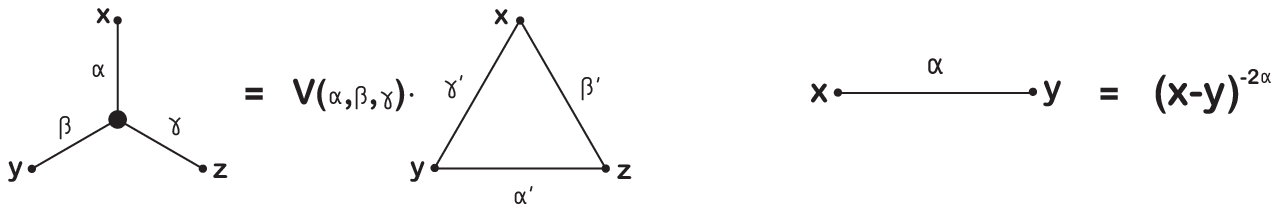}
\end{center}

\medskip

Finally we find explicit expression for $\CCR$-operator using (\ref{Rs})
\begin{equation}\label{Rscalar}
\CCR_{12}(u-v)= x_{12}^{2(u_- - v_+)} \, \hat{p}_2^{\,2(u_+ - v_+)} \, \hat{p}_1^{\,2(u_- - v_-)} \, x_{12}^{2(u_+ - v_-)}\,.
\end{equation}
This operator can be rewritten as an integral one
$$
[\mathrm{R}_{12}(u-v)\;\Phi](x_1,x_2) = c \cdot \int
\frac{\mathrm{d}^n y_1 \mathrm{d}^n y_2 \; \Phi(y_1,y_2)}{x_{12}^{2(v_+ - u_-)} (x_2 - y_2)^{2(u_+ - v_+ + \frac{n}{2})}
(x_1 - y_1)^{2(u_- - v_- + \frac{n}{2})} y_{12}^{2(v_- - u_+)} }
$$
where
$$
c = 4^{u_+ + u_- - v_+ - v_-} \pi^{-n} \frac{\Gamma(u_+ - v_+ + \frac{n}{2})\,\Gamma(u_- - v_- + \frac{n}{2})}
{\Gamma(v_+ - u_+)\,\Gamma(v_- - u_-)}
$$
We depict its kernel (up to a constant function of spectral parameters)
 as follows using the graphical rules outlined above
\begin{center}
\includegraphics{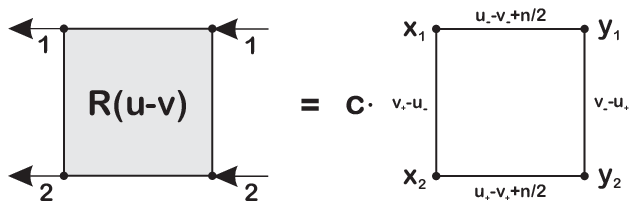}
\end{center}

Coxeter relations (\ref{S1S2S1}) are basic relations which
enable to establish Yang-Baxter equation for $\CCR$-operator (\ref{Rscalar}).
Corresponding prove is rather straightforward and in our notations it
repeats literally the one presented in~\cite{DKM} for the case of $\mathrm{SL}(2,\mathbb{C})$.
Here we illustrate the prove
\begin{center}
\includegraphics[scale=0.7]{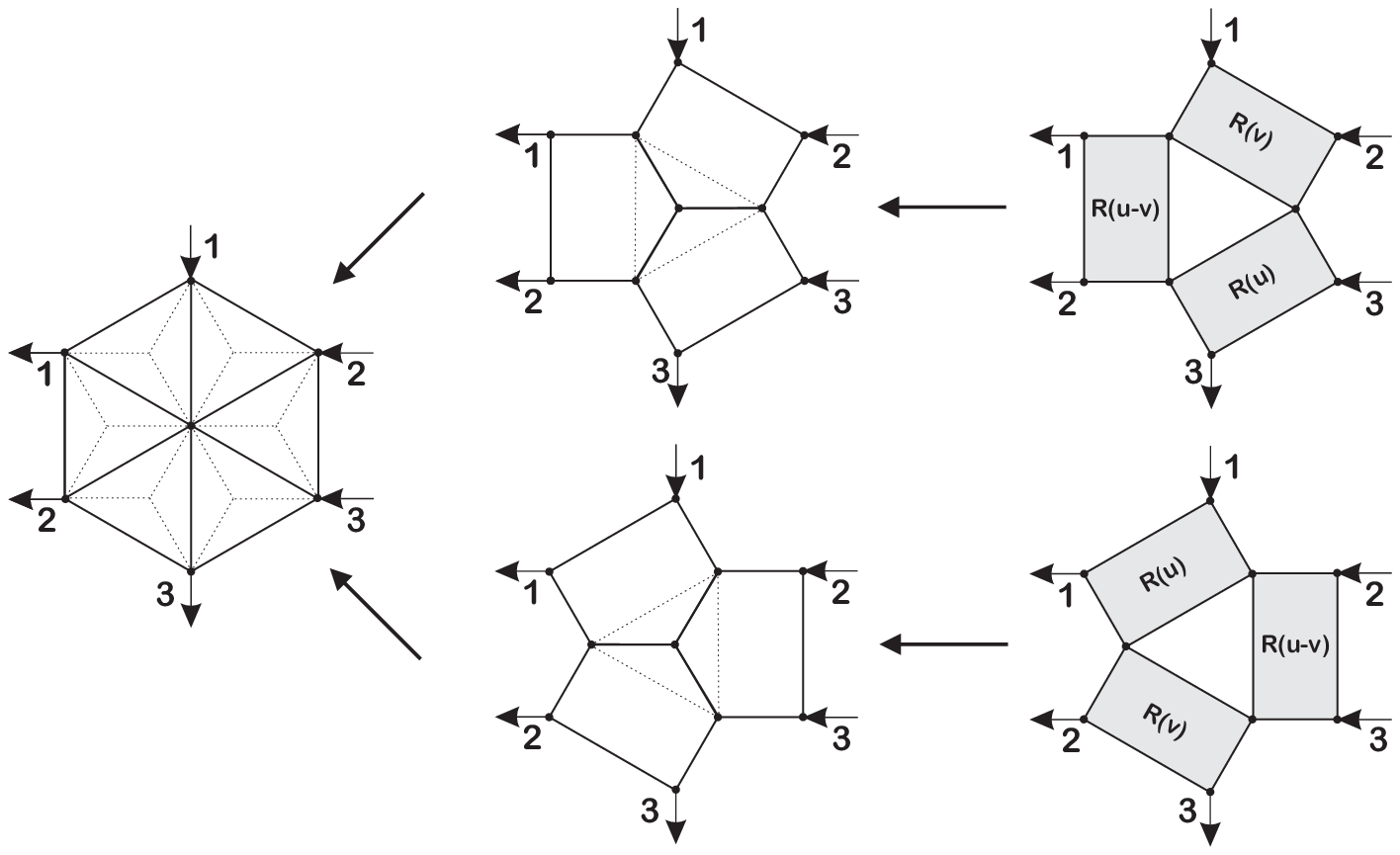}
\end{center}
The sequence of transformations in the picture is performed by means of
the star-triangle relation.

Using $\mathrm{R}$-matrix (\ref{Rscalar})
  %\marginpar{\bf Novaya vstavka}
which satisfies the
Yang-Baxter equation one can construct, by using the standard method,
the set of commuting
operators (Hamiltonians) and formulate the corresponding quantum
integrable system on the chain.
Here  we obtain one of these Hamiltonians and describe the integrable
chain model.
Consider the chain model with $N$ sites. The states of this chain are the
vectors in the space $V_{\Delta_1} \otimes \cdots \otimes V_{\Delta_N}$,
where each $V_{\Delta_a}$ is the vector space of the
differential representation $\rho$ of the conformal algebra
${\sf conf}(\mathbb{R}^n)$ (\ref{cnfA1}).
 For $\Delta_a=\Delta$ $(\forall a)$ the $\mathrm{R}$-matrix
 (\ref{Rscalar}) is
written in the form
 \begin{equation}
 \label{hamR}
\begin{array}{c}
\mathrm{R}_{ab}(\alpha;\xi):= x_{ab}^{2(\alpha+\xi)} \;
\hat{p}_{a}^{\; 2\alpha} \; \hat{p}_{b}^{\; 2\alpha}
\; x_{ab}^{2(\alpha-\xi)} = 1 + \alpha \,
\mathrm{h}_{a,b}(\xi) + \alpha^2 \dots \; ,  \\ [0.2cm]
a,b = 1,2,\dots,N \; ,
\end{array}
 \end{equation}
where $\xi = \frac{n}{2}-\Delta$, parameter $\alpha =u-v$ is taken as a small one and operators
 \begin{equation}
 \label{hamL}
\begin{array}{c}
 \mathrm{h}_{a,b}(\xi) =  2 \, \ln x_{ab}^{2}  + x_{ab}^{2\xi} \; \ln (\hat{p}_{a}^{\; 2} \;
 \hat{p}_{b}^{\; 2} ) \, x_{ab}^{-2\xi}  = \\ [0.3cm] = \hat{p}_{a}^{-2\xi} \; \ln(x_{ab}^{2})  \;
 \hat{p}_{a}^{\; 2\xi} + \hat{p}_{b}^{-2\xi} \; \ln(x_{ab}^{2})  \;
 \hat{p}_{b}^{\; 2\xi} + \ln (\hat{p}_{a}^{\; 2} \, \hat{p}_{b}^{\; 2} ) \; ,
\end{array}
 \end{equation}
are interpreted for $b=a+1$ as local Hamiltonian densities.
 The second expression for $\mathrm{h}_{a,b}(\xi)$ in (\ref{hamL}) is deduced from
 the $\mathrm{R}$-matrix (\ref{hamR}) which is written
 by means of (\ref{S1S2S1}) in another form
 $$
 \mathrm{R}_{ab}(\alpha;\xi):= \hat{p}_{a}^{\; - 2\xi} \; x_{ab}^{2\alpha} \;
\hat{p}_{a}^{\; 2(\alpha+\xi)} \; \hat{p}_{b}^{\; 2(\alpha-\xi)}
\; x_{ab}^{2\alpha} \; \hat{p}_{b}^{\; 2\xi}  \; .
 $$
Then the whole
Hamiltonian for the integrable chain model is given by the operator
 $$
 H(\xi) = \sum\limits_{a=1}^{N-1} \, \mathrm{h}_{a,a+1}(\xi) \; ,
 $$
 where $N$ is the length of the chain.
This operator is the high-dimensional
analog of the Hamiltonian for the integrable model which was considered in
\cite{Lipat}.
For $n=1$
  %\marginpar{\bf Popravil! A.Isaev-S.Derkachov}
and special choice of $\xi=1/2$ this operator formally reproduces
(up to an additional constant) the
holomorphic part of the Hamiltonian \cite{Lipat}.
The whole Hamiltonian \cite{Lipat} is the sum of holomorphic and anti-holomorphic parts and is obtained from (\ref{hamL}) for $n=2$ and $\xi=1$.
The more general two-dimensional model was considered in \cite{DKM}.

Another
example of the integrable lattice model based on the particular
star-triangle relation (\ref{scStTrInt}) was formulated in~\cite{Zam1}.

 %\add{ I have removed the note about Korchemsky-Pasquier on the page 17 and due to this I would like to propose to %remove all this absatz about "folklore".\\
 %In the conclusion of this Section we should mention
 %that some of these results in one or another form are known to specialists and probably are
 %the part of the folklore \footnote{For example the integrable model based on
 %this particular star-triangle relation is formulated in~\cite{Zam1}}.
 %In fact we just collect everything in one place and we hope that the fact that
 %$\mathrm{L}$-operators obey the local $\mathrm{RLL}$-relation with spinorial $\mathrm{R}$-matrix
 %is a new one. This local relation allows to construct the family of commuting operators --
 %transfer-matrices builded from $\mathrm{L}$-operators in a standard way
 %\begin{equation} \label{transfer} \mathrm{t}(u) = \tr \mathrm{L}_{1}(u)\,
 %\mathrm{L}_{2}(u)\,\cdots\, \mathrm{L}_{n}(u)  \longrightarrow \mathrm{t}(u)\,\mathrm{t}(v) =
 %\mathrm{t}(v)\,\mathrm{t}(u)\,,\end{equation}
 %where the trace is taken in the spinor representation.}

\subsection{General $\mathrm{R}$-operator in the case $so(5,1)$}

\mbox{} In the previous Section we have described the general strategy for the simplest nontrivial example. Now we repeat everything step by step in more complicated situation explaining the needed modifications at each stage.

All modifications are due to the use of the
more complicated representations of conformal group.
The scalar representation is characterized by one
parameter -- scaling dimension $\Delta$ so that
operator $\mathrm{L}(u)$ contains two parameters $u$ and $\Delta$, which are combined in a natural linear combinations
$u_+ = u + \frac{\Delta -n}{2}$ and $u_- = u - \frac{\Delta}{2}$.

The tensor representation is characterized by three
parameters -- scaling dimension $\Delta$ and two spins
$\ell$, $\dot{\ell}$ and now the
operator $\mathrm{L}(u)$ contains four parameters $u$ and $\Delta$, $\ell$, $\dot{\ell}$.
These parameters are combined in a
pairs $\mathbf{u}_{+}$ and $\mathbf{u}_{-}$ which are analogs of $u_{+}$ and $u_{-}$
$$
\mathbf{u}_{+}\equiv\bigl(u_{+}\,,\,\ell\bigl) =
\left(u + \frac{\Delta -n}{2}\,,\,\ell\right)\ \ \ ;\ \ \ \mathbf{u}_{-}\equiv\bigl(u_{-}\,,\,\dot{\ell}\bigl) =
\left(u - \frac{\Delta}{2}\,,\,\dot{\ell}\right)\,.
$$
We have the following expression for the operator $\mathrm{L}(u)$
$$
\mathrm{L}( \mathbf{u}_+\,,\,\mathbf{u}_- ) = \left(
\begin{array}{cc}
  \boldsymbol{1} & \; \mathbf{0} \\
  \mathbf{x} & \; \boldsymbol{1}
\end{array}
\right) \cdot \left(
\begin{array}{cc}
  u_+\cdot\mathbf{1} + \mathbf{S}^{( \ell )} & \ \ \mathbf{p} \\
  \mathbf{0}  & u_-\cdot\mathbf{1} + \overline{\mathbf{S}}^{( \dot{\ell} )}
\end{array}
\right) \cdot \left(
\begin{array}{cc}
  \boldsymbol{1} & \; \mathbf{0} \\
  - \mathbf{x} & \; \boldsymbol{1}
\end{array}
\right)  \,,
$$
and the defining $\mathrm{RLL}$-relation has the form
\begin{equation} \label{PRLL}
\CCR_{12}(u-v)\,\mathrm{L}_1(\mathbf{u}_+,\mathbf{u}_-)\,
\mathrm{L}_2(\mathbf{v}_+,\mathbf{v}_-) =
\mathrm{L}_1(\mathbf{v}_+,\mathbf{v}_-)\,
\mathrm{L}_2(\mathbf{u}_+,\mathbf{u}_-)\,\CCR_{12}(u-v)
\end{equation}
where (\ref{cnfA17})
$$
\mathrm{L}_1( u_+,u_- ) = \left(
\begin{array}{cc}
  \boldsymbol{1} & \; \mathbf{0} \\
  \mathbf{x}_1 & \; \boldsymbol{1}
\end{array}
\right) \cdot \left(
\begin{array}{cc}
  u_+\cdot\mathbf{1} + \mathbf{S}_1^{( \ell_1 )}  & \ \ \mathbf{p}_1 \\
  \mathbf{0}  & u_-\cdot\mathbf{1} + \overline{\mathbf{S}}_1^{( \dot{\ell}_1 )}
\end{array}
\right) \cdot \left(
\begin{array}{cc}
  \boldsymbol{1} & \; \mathbf{0} \\
  - \mathbf{x}_1 & \; \boldsymbol{1}
\end{array}
\right)  \,,
$$
$$
\mathrm{L}_2( v_+,v_- ) = \left(
\begin{array}{cc}
  \boldsymbol{1} & \; \mathbf{0} \\
  \mathbf{x}_2 & \; \boldsymbol{1}
\end{array}
\right) \cdot \left(
\begin{array}{cc}
  v_+\cdot\mathbf{1} + \mathbf{S}_2^{( \ell_2 )}  & \ \ \mathbf{p}_2 \\
  \mathbf{0}  & v_-\cdot\mathbf{1} + \overline{\mathbf{S}}_2^{( \dot{\ell}_2 )}
\end{array}
\right) \cdot \left(
\begin{array}{cc}
  \boldsymbol{1} & \; \mathbf{0} \\
  - \mathbf{x}_2 & \; \boldsymbol{1}
\end{array}
\right)
$$
To avoid misunderstanding we collect all parameters
$$
u_+ = u + \frac{\Delta_1 -n}{2}\,,\, u_- = u - \frac{\Delta_1}{2}
\,,\, v_+ = v + \frac{\Delta_2 -n}{2}\,,\, v_- = v - \frac{\Delta_2}{2}
$$
$$
\mathbf{u}_{+}\equiv\bigl(u_{+}\,,\,\ell_1\bigl)\,,\, \mathbf{u}_{-}\equiv\bigl(u_{-}\,,\,\dot{\ell}_1\bigl)\,,\,
\mathbf{v}_{+}\equiv\bigl(v_{+}\,,\,\ell_2\bigl)\,,\, \mathbf{v}_{-}\equiv\bigl(v_{-}\,,\,\dot{\ell}_2\bigl)\,.
$$
We construct $\CCR$-operator from basic building blocks
$\mathrm{S}_1(\mathbf{u}) , \mathrm{S}_2(\mathbf{u})$ and $\mathrm{S}_3(\mathbf{u})$ which satisfy more simple relations like
(\ref{S1}), (\ref{S2}), (\ref{S3}) with substitution
$(v_+ , v_- , u_+ , u_- )\to(\mathbf{v}_+ , \mathbf{v}_- , \mathbf{u}_+ , \mathbf{u}_- )$ and
represent elementary transpositions in the set of four pairs of parameters
$\mathbf{u} = (\mathbf{v}_+ , \mathbf{v}_- , \mathbf{u}_+ , \mathbf{u}_- )$.

Let us start with operators $\mathrm{S}_1$ and $\mathrm{S}_3$ which are two copies of the operator
$\mathrm{S}$ defined by the equation
\begin{equation} \label{boldSL=LS}
\mathrm{S}\,\mathrm{L}(\mathbf{u}_+ , \mathbf{u}_-) = \mathrm{L}(\mathbf{u}_- , \mathbf{u}_+)\,\mathrm{S}\,
\end{equation}
The exchange $\mathbf{u}_+ \leftrightarrow \mathbf{u}_-$ is equivalent to $u_+ \leftrightarrow u_-$ and $\ell \leftrightarrow \dot{\ell}$, i.e. $\Delta \leftrightarrow 4-\Delta$ and
$\ell \leftrightarrow \dot{\ell}$ .
Differential representation of the conformal algebra $\textsf{conf}(\mathbb{R}^{4})$
is parameterized by three numbers
$\Delta ,\ \ell , \ \dot{\ell}$ and we denote it by $\rho^{\Delta, \ell , \dot{\ell}}$ .
Thus operator $\mathrm{S}$ intertwines  two
 %equivalent
 %\marginpar{\bf zamenil "equivalent" na "two"}
 representations\footnote{It is easy to see
that values of the Casimir operator (\ref{cnfA2z}) coincides for these two representations.}
$\rho^{\Delta, \ell , \dot{\ell}} \sim \rho^{4-\Delta, \dot{\ell} , \ell}$ .
As in the previous Section operator $\mathrm{S}$ has transparent
representation theory meaning.

We consider representation of the conformal algebra on tensor fields
$\Phi_{\alpha_1 \cdots \alpha_{2\ell}}^
{\dot{\alpha}_1 \cdots \dot{\alpha}_{2\dot{\ell}}}(x)$
of the type $(\ell , \dot{\ell})$
where dotted and undotted indexes are symmetric separately
and where $x\in \mathbb{R}^{4}$.
In this situation it is convenient to work with the generating functions
$$
\Phi(x , \lambda , \tilde{\lambda}) =
\Phi_{\alpha_1 \cdots \alpha_{2\ell}}^
{\dot{\alpha}_1 \cdots \dot{\alpha}_{2\dot{\ell}}}(x)
\, \lambda^{\alpha_1} \cdots \lambda^{\alpha_{2\ell}}
\, \tilde{\lambda}_{\dot{\alpha}_1} \cdots
\tilde{\lambda}_{\dot{\alpha}_{2\dot{\ell}}} \; ,
$$
where $\lambda$ and $\tilde{\lambda}$ are auxiliary spinors.
Let us introduce the convolution
$$
\mathrm{F}(\lambda,\tilde{\lambda}) \ast \mathrm{G}(\lambda,\tilde{\lambda}) =
\mathrm{F}(\partial_{\lambda},\partial_{\tilde{\lambda}}) \left.\mathrm{G}(\lambda,\tilde{\lambda})\right|_{\lambda=0,\tilde{\lambda}=0}
$$
and use it to represent the intertwining operator as an
integral operator acting on generating functions
$$
\left[\mathrm{S} \, \Phi\right](X) =
\int \mathrm{d}^4 y \, \mathrm{S}(X,Y) * \Phi(Y)
$$
where we combine space-time coordinates and two
spinors in one compact notation $X = ( x , \lambda,\tilde{\lambda} )$ ,
$Y=( y , \eta , \tilde{\eta})$ and denote generating function
by $\Phi(X)$.

The defining equation for $\mathrm{S}$ is equivalent to
the set of differential equations for its kernel
$\mathrm{S}\left(X, Y\right)$
\begin{equation}
\label{sec4b}
\left( G_Y^{\Delta,\ell,\dot{\ell}} \right)^{T} \mathrm{S}\left(X , Y\right) =
G_X^{4-\Delta,\dot{\ell},\ell}\,\mathrm{S}\left(X , Y \right)\,.
\end{equation}
Here $G_X^{4-\Delta,\dot{\ell},\ell}$ denotes generators of conformal group in representation $\rho^{4-\Delta, \dot{\ell} , \ell}$,  $G_Y^{\Delta,\ell,\dot{\ell}}$ -- generators in representation $\rho^{\Delta, \ell , \dot{\ell}}$. The generators $S_{\mu \nu}$ are taken in the
form (\ref{SigmaMN})
$$
S_{\mu \nu} =  \lambda \, \boldsymbol{\sigma}_{\mu \nu} \, \partial_{\lambda} +
\tilde{\lambda} \, \overline{\boldsymbol{\sigma}}_{\mu \nu} \, \partial_{\tilde{\lambda}}\,.
$$
$T$ stands for transposition
$$
\int \mathrm{d}^4 y \, \mathrm{S}(X,Y)
\ast  G_Y^{\Delta,\ell,\dot{\ell}} \, \Phi(Y) =
\int \mathrm{d}^4 y \, \left[
\left( G_Y^{\Delta,\ell,\dot{\ell}} \right)^{T}\,
\mathrm{S}(X,Y)\right] \ast \, \Phi(Y)\,.
$$
arising after
integration by parts and using some evident properties of convolution like
$$
\begin{array}{c}
\mathrm{F}(\lambda) \ast \partial_{\lambda} \mathrm{G}(\lambda) =
\lambda\, \mathrm{F}(\lambda) \ast \mathrm{G}(\lambda) \\ [0.2cm]
\mathrm{F}(\lambda) \ast \lambda\, \mathrm{G}(\lambda) = \partial_{\lambda} \mathrm{F}(\lambda) \ast \mathrm{G}(\lambda)
\end{array}
$$
After substitution of explicit expression for generators one
obtains the following set of equations
\begin{itemize}
\item Translation
$$
\left(\frac{\partial}{\partial x_{\mu}}
+ \frac{\partial}{\partial y_{\mu}}\right)
\mathrm{S}\left(X, Y\right) = 0
$$
\item Lorentz rotations
$$
\left[i \left( y_{\nu} \frac{\partial}{\partial y_{\mu}} - y_{\mu} \frac{\partial}{\partial y_{\nu}} \right)
+ \eta \, \boldsymbol{\sigma}^{T}_{\mu \nu} \, \partial_{\eta} +
\tilde{\eta} \, \overline{\boldsymbol{\sigma}}^{T}_{\mu \nu} \, \partial_{\tilde{\eta}} \right]\mathrm{S}(X,Y) =
$$
$$=
\left[i\left( x_{\mu} \frac{\partial}{\partial x_{\nu}} - x_{\nu} \frac{\partial}{\partial x_{\mu}} \right)
 +\lambda \, \boldsymbol{\sigma}_{\mu \nu} \, \partial_{\lambda} +
\tilde{\lambda} \, \overline{\boldsymbol{\sigma}}_{\mu \nu} \, \partial_{\tilde{\lambda}}\right]\mathrm{S}(X,Y)
$$
\item Dilatation
$$
\left(x_{\mu}\frac{\partial}{\partial x_{\mu}}
+ y_{\mu}\frac{\partial}{\partial y_{\mu}}\right)
\mathrm{S}\left(X,Y\right) =
-2\left(4-\Delta\right)\,\mathrm{S}\left(X,Y\right)\,
$$
\item{Conformal boosts}
$$
\left(
- i y^2 \frac{\partial}{\partial y_{\mu}} + 2 i y_{\mu} y_{\nu} \frac{\partial}{\partial y_{\nu}}
+2 y^{\nu} ( \eta \, \boldsymbol{\sigma}^{T}_{\nu \mu} \, \partial_{\eta} +
\tilde{\eta} \, \overline{\boldsymbol{\sigma}}^{T}_{\nu \mu} \, \partial_{\tilde{\eta}} )
+ 2 i (4- \Delta) \, y_{\mu} \right) \mathrm{S}(X,Y) =
$$
$$ =
\left( i x^2 \frac{\partial}{\partial x_{\mu}} - 2 i x_{\mu} x_{\nu} \frac{\partial}{\partial x_{\nu}}
 +2 x^{\nu} ( \lambda \, \boldsymbol{\sigma}_{\nu \mu} \, \partial_{\lambda} +
\tilde{\lambda} \, \overline{\boldsymbol{\sigma}}_{\nu \mu} \, \partial_{\tilde{\lambda}} )
- 2 i (4- \Delta) \, x_{\mu} \right) \, \mathrm{S}(X,Y)
$$
\end{itemize}
This set of equations for the kernel of $\mathrm{S}$ coincides
with the set of equations for a Green function for two fields of the types $(\ell , \dot{\ell})$ and $(\dot{\ell} , \ell)$ in conformal field theory and the solution is well known~\cite{Zaikov}
$$
\mathrm{S}(X,Y) = \frac{1}{(2\ell)!}\frac{1}{(2\dot{\ell})!}\,\frac{\Bigl(\tilde{\lambda} \,
(\overline{\mathbf{x}-\mathbf{y}}) \, \eta \Bigr)^{2\ell}
\Bigl(\lambda \, (\mathbf{x}-\mathbf{y}) \, \tilde{\eta} \Bigr)^{2\dot{\ell}}}{(x-y)^{2(4-\Delta)}}\,.
$$
In this Section for simplicity we shall use compact notation
\begin{equation} \label{notation}
\mathbf{x} = \sigma_{\mu}\frac{\,\,x^{\mu}}{|x|}\ ;\ \overline{\mathbf{x}} = \overline{\sigma}_{\mu} \frac{\,\,x^{\mu}}{|x|}\,
\end{equation}
where $\frac{\,\,x^{\mu}}{|x|}$ is the unit vector in the direction $x^{\mu}$.
Formula for the kernel $\mathrm{S}(X,Y)$ leads to the following explicit expression for the action of operator $\mathrm{S}$ on the generating function
(shadow transformation \cite{Shadow})
$$
[ \mathrm{S} \, \Phi ] ( X ) =
\int \frac{ \mathrm{d}^4 y \,\, \Phi\left(y\,,\, \tilde{\lambda} \, \overline{(\mathbf{x}-\mathbf{y})} \, ,\,
\lambda \, (\mathbf{x}-\mathbf{y})\, \right)}{(x-y)^{2(4-\Delta)}}
$$
which can be represented in more transparent form (remember that $u_--u_+ = 2-\Delta$)
\begin{equation}\label{S}
[ \mathrm{S}(u_--u_+) \, \Phi ] ( X )  =
\int \frac{\mathrm{d}^4 y }{y^{2(u_--u_++2)}}
\cdot \Phi( x-y\,,\, \tilde{\lambda} \, \overline{\mathbf{y}} \,,\, \lambda \, \mathbf{y} \,) =
\int \frac{\mathrm{d}^4 y \,\, e^{iy \hat{p}}}{y^{2(u_--u_++2)}}
\cdot \Phi\bigl( x\,,\, \tilde{\lambda} \, \overline{\mathbf{y}} \,,\, \lambda \, \mathbf{y} \,\bigl)\,
\end{equation}
where $\hat{p} = i\partial_{x}$.
These formulae clearly show the analogy and difference in comparison to the  considered scalar case.
The last formula with operator $\hat{p}$ is very similar to~(\ref{Sscal})
but there exists additional action of operator $\mathrm{S}$ on the spinor
variables $\lambda$ and $\tilde{\lambda}$.
These spinors are transformed by the matrices $\mathbf{y}$ and $\overline{\mathbf{y}}$ and after transformations are interchanged :
$\lambda \to \tilde{\lambda} \, \overline{\mathbf{y}} \,;\,\tilde{\lambda} \to \lambda \, \mathbf{y}$.

The operators $\mathrm{S}_1$ and $\mathrm{S}_3$
act on the function $\Phi(X_1 ; X_2)$ in a similar manner
$$
[\mathrm{S}_1(v_- - v_+) \, \Phi](X_1 ; X_2) =
\int \,\frac{\mathrm{d}^4 y\,\, e^{iy \hat{p}_2}}{y^{2(v_- - v_++ 2)}}
\,\, \Phi(X_1 ; x_2\,,\, \tilde{\lambda}_2 \, \overline{\mathbf{y}} \,,\,
\lambda_2 \, \mathbf{y} )
$$
\begin{equation} \label{SS3}
[\mathrm{S}_3(u_- - u_+) \, \Phi](X_1 ; X_2) =
\int \,\frac{\mathrm{d}^4 y\,\, e^{iy \hat{p}_1}}{y^{2(u_- - u_+ + 2)}}
\,\, \Phi(x_1\,,\, \tilde{\lambda}_1 \, \overline{\mathbf{y}} \,,\,
\lambda_1 \, \mathbf{y} \, ; X_2)
\end{equation}
In order to construct operator $\mathrm{S}_2$ we take into account the same observation as in a scalar case:
it can be produced directly from the operator $\mathrm{S}$ using duality transformation
$$
y \to p\ ;\
p \to x_{2}-x_{1} \equiv x_{21} \ ;\ \mathbf{u}_+ \to \mathbf{v}_-\ ;\ \mathbf{u}_- \to \mathbf{u}_+ \,,
$$
The change $\mathbf{u}_+ \to \mathbf{v}_-\ ;\ \mathbf{u}_- \to \mathbf{u}_+$ implies the corresponding change of spinors so that the expression for the action of operator $\mathrm{S}_2$ on the generating function $\Phi\bigl(X_1\,;\,X_2\bigl)$ is
\begin{equation}\label{S2spin}
[\mathrm{S}_2(u_+ - v_-) \, \Phi]\bigl(X_1\,;\,X_2\bigl) =
\int \frac{\mathrm{d}^4 p\,\,e^{i p \, x_{21}}}
{p^{2(u_+ - v_- + 2)}}
\,
\Phi\bigl(x_1\,,\,\tilde{\lambda}_2 \, \overline{\mathbf{p}} \,,\,\tilde{\lambda}_1\, ;
\, x_2\,,\,\lambda_2\,,\, \lambda_1 \, \mathbf{p} \bigl)\,.
\end{equation}
In the case of scalars there is no dependence on $\lambda_{1},\lambda_{2}$
and $\tilde{\lambda}_{1},\tilde{\lambda}_{2}$ so that integral over $p$ can be calculated explicitly and operator $\mathrm{S}_2$ reduces to the
operator of the multiplication to the function $x_{12}^{v_--u_+}$.

The proof of the duality rules is almost the same as in a scalar case.
We rewrite
the defining equation (\ref{boldSL=LS}) for operator $\mathrm{S}$ in the factorized form (\ref{cnfA17})
\begin{equation} \label{sec4a}
\mathrm{S}\, \left(
\begin{array}{cc}
  \boldsymbol{1} & \; \mathbf{0} \\
  \mathbf{x} & \; \boldsymbol{1}
\end{array}
\right) \left(
\begin{array}{cc}
  u_{+}\cdot \boldsymbol{1} + \mathbf{S}^{(\ell)}   & \ \ \mathbf{p} \\
  \mathbf{0}  & u_{-}\cdot \boldsymbol{1} + \overline{\mathbf{S}}^{(\dot{\ell})}
\end{array}
\right) \left(
\begin{array}{cc}
  \boldsymbol{1} & \; \mathbf{0} \\
  - \mathbf{x} & \; \boldsymbol{1}
\end{array}
\right)
=
\end{equation}
$$
=
\left(
\begin{array}{cc}
  \boldsymbol{1} & \; \mathbf{0} \\
  \mathbf{x} & \; \boldsymbol{1}
\end{array}
\right) \left(
\begin{array}{cc}
  u_{-}\cdot \boldsymbol{1} + \mathbf{S}^{(\dot{\ell})} & \ \ \mathbf{p} \\
  \mathbf{0}  & u_{+}\cdot \boldsymbol{1} + \overline{\mathbf{S}}^{(\ell)}
\end{array}
\right) \left(
\begin{array}{cc}
  \boldsymbol{1} & \; \mathbf{0} \\
  - \mathbf{x} & \; \boldsymbol{1}
\end{array}
\right)
\mathrm{S}\,.
$$
Using the same argumentation as in the previous Section one can easily see that
the defining equation for operator $\mathrm{S}_2$
$$
\mathrm{S}_2\,\mathrm{L}_1(\underline{\mathbf{u}_+},\mathbf{u}_-)\,
\mathrm{L}_2(\mathbf{v}_+,\underline{\mathbf{v}_-})=
\mathrm{L}_1(\underline{\mathbf{v}_-},\mathbf{u}_-)\,
\mathrm{L}_2(\mathbf{v}_+,\underline{\mathbf{u}_+}) \, \mathrm{S}_2\,
$$
can be transformed to the form
\begin{equation}
 \label{sec4d}
\mathrm{S}_2\left(%
\begin{array}{cc}
  \mathbf{1} & \mathbf{0} \\
  \mathbf{p}_1 & \mathbf{1} \\
\end{array}%
\right)\left(%
\begin{array}{cc}
  v_-\cdot \boldsymbol{1} + \mathbf{S}_2^{(\dot{\ell}_2)}  & \mathbf{x}_{21} \\
  \mathbf{0} & u_{+}\cdot \boldsymbol{1} + \overline{\mathbf{S}}_1^{(\ell_1)} \\
\end{array}%
\right)\left(%
\begin{array}{cc}
  \mathbf{1} & \mathbf{0} \\
  -\mathbf{p}_1 & \mathbf{1} \\
\end{array}%
\right)
=
\end{equation}
$$
=
\left(%
\begin{array}{cc}
  \mathbf{1} & \mathbf{0} \\
  \mathbf{p}_1 & \mathbf{1} \\
\end{array}%
\right)\left(%
\begin{array}{cc}
  u_{+}\cdot \boldsymbol{1} + \mathbf{S}_2^{(\ell_1)}  & \mathbf{x}_{21} \\
  \mathbf{0} & v_{-}\cdot \boldsymbol{1} +
  \bar{\mathbf{S}}_1^{(\dot{\ell}_2)}  \\
\end{array}%
\right)\left(%
\begin{array}{cc}
  \mathbf{1} & \mathbf{0} \\
  -\mathbf{p}_1 & \mathbf{1} \\
\end{array}%
\right)\mathrm{S}_2\,,
$$
if we require that
$$
[\mathrm{S}_2,\mathbf{x}_1] =
[\mathrm{S}_2,\mathbf{x}_2] =
[\mathrm{S}_2,\mathbf{u}_-] =
[\mathrm{S}_2,\mathbf{v}_+] = 0\,.
$$
Comparing equations (\ref{sec4a}) and (\ref{sec4d})
we conclude that the change
\begin{equation}
 \label{sec3f}
\mathbf{x} \to \mathbf{p}_1\ ;\
\mathbf{p} \to \mathbf{x}_{21}\ ;\ \mathbf{u}_+ \to \mathbf{v}_-\ ;\ \mathbf{u}_- \to \mathbf{u}_+
\end{equation}
transforms (\ref{sec4a}) into (\ref{sec4d}).
Using the formula~(\ref{S}) with momentum $\hat{p}$ and
duality rules~(\ref{sec3f}) we obtain the expression~(\ref{S2spin})
for the action of operator $\mathrm{S}_2(\mathbf{u})$ on the generating function $\Phi\bigl(X_1\,;\,X_2\bigl)$.

Thereby we have constructed operator representation
of elementary transpositions $s_1,s_2,s_3$ :
$$
s_{1}\mathbf{u} = (\underline{\mathbf{v}_-},\underline{\mathbf{v}_+},\mathbf{u}_+,\mathbf{u}_-)\ ;\
s_{2}\mathbf{u} = (\mathbf{v}_+,\underline{\mathbf{u}_+},\underline{\mathbf{v}_-},\mathbf{u}_-) \ ;\
s_{3}\mathbf{u} = (\mathbf{v}_+,\mathbf{v}_-,\underline{\mathbf{u}_-},\underline{\mathbf{u}_+})\,
$$
following the line of the previous Section. The corresponding Coxeter
relations have the more complicated form in comparison to the scalar case.
The first triple relation
$$
\mathrm{S}_1 (a) \, \mathrm{S}_2 (a+b) \, \mathrm{S}_1 (b) =
\mathrm{S}_2 (b) \, \mathrm{S}_1 (a+b) \, \mathrm{S}_2 (a)
$$
in explicit form looks as follows
$$
\int \frac{\mathrm{d}^4 z\, \mathrm{d}^4 k\, \mathrm{d}^4 y\,\,
e^{i \, z \, \hat{p}_2} e^{i \, k \, x_{21}} e^{i \, y \, \hat{p}_2}}
{
z^{2 \left( a + 2\right)}
k^{2 \left( a+b + 2\right)}
y^{2 \left( b + 2\right)}
}
\cdot\Phi(x_1\,,\,\lambda_2 \, \mathbf{z} \, \overline{\mathbf{k}} \,,\,
\tilde{\lambda}_1
\,;\, x_2\,,\,\lambda_1 \, \mathbf{k} \, \overline{\mathbf{y}} \,,\,
\tilde{\lambda}_2 \, \overline{\mathbf{z}} \, \mathbf{y} \, ) =
$$
\begin{equation}
\label{spin1}
= \int \frac{\mathrm{d}^4 q\, \mathrm{d}^4 y\, \mathrm{d}^4 k\,\,
e^{i \, q \, x_{21}} e^{i \, y \, \hat{p}_2} e^{i \, k \, x_{21}}}
{
q^{2 \left( b + 2\right)}
y^{2 \left( a+b + 2\right)}
k^{2 \left( a + 2\right)}
}
\cdot\Phi(x_1\,,\,\lambda_2 \, \mathbf{y}\,\overline{\mathbf{k}} \,,\, \tilde{\lambda}_1\,;
\,x_2\,,\, \lambda_1 \, \mathbf{q} \, \overline{\mathbf{y}}
\,,\, \tilde{\lambda}_2 \overline{\mathbf{q}} \, \mathbf{k}  \, )\,,
\end{equation}
and the second triple relation
$$
\mathrm{S}_3 (a) \, \mathrm{S}_2 (a+b) \, \mathrm{S}_3 (b) =
\mathrm{S}_2 (b) \, \mathrm{S}_3 (a+b) \, \mathrm{S}_2 (a)\,
$$
is equivalent to the similar integral relation
$$
\int \frac{\mathrm{d}^4 z\, \mathrm{d}^4 k\, \mathrm{d}^4 y\,e^{i \, z \, \hat{p}_1} e^{i \, k \, x_{21}} e^{i \, y \, \hat{p}_1}}
           {
             z^{2 \left( a + 2\right)}
             k^{2 \left( a+b + 2\right)}
             y^{2 \left( b + 2\right)}
           }
\, \Phi(x_1\,,\, \lambda_1\, \mathbf{z} \, \overline{\mathbf{y}} \, ,\,
           \tilde{\lambda}_2\, \overline{\mathbf{k}} \, \mathbf{y} \, ;
           \, x_2\,,\, \lambda_2\,,\, \tilde{\lambda}_1 \, \overline{\mathbf{z}} \, \mathbf{k} \, ) =
$$
\begin{equation}
\label{spin2}
= \int \frac{\mathrm{d}^4 q\, \mathrm{d}^4 y\, \mathrm{d}^4 k\,
e^{i \, q \, x_{21}} e^{i \, y \, \hat{p}_1} e^{i \, k \, x_{21}}}
           {
             q^{2 \left( b + 2\right)}
             y^{2 \left( a+b + 2\right)}
             k^{2 \left( a + 2\right)}
           }
\, \Phi(x_1\,,\, \lambda_1\, \mathbf{q} \, \overline{\mathbf{k}} \, ,\,
           \tilde{\lambda}_2\, \overline{\mathbf{q}} \, \mathbf{y} \, ;\,
           x_2\,,\, \lambda_2\,,\, \tilde{\lambda}_1 \, \overline{\mathbf{y}} \, \mathbf{k} \, ).
\end{equation}
These relations are equivalent to the following generalization
of the scalar star-triangle relation (\ref{scStTrOp})
\medskip
\begin{equation}\label{star-tr}
\frac{\hat{p}^{\mu_1}\cdots \hat{p}^{\mu_m}}{\hat{p}^{\,2 (a+m)}} \, \frac{\mathrm{A}_{\mu_1\nu_1}\cdots\mathrm{A}_{\mu_m\nu_m}}
{x^{2(a+b+m)}} \, \frac{\hat{p}^{\nu_1}\cdots
\hat{p}^{\nu_m}}{\hat{p}^{\,2(b+m)}} =
\frac{x^{\mu_1}\cdots x^{\mu_m}}{x^{\,2 (b+m)}} \, \frac{\mathrm{A}_{\mu_1\nu_1}\cdots\mathrm{A}_{\mu_m\nu_m}}
{\hat{p}^{2(a+b+m)}} \, \frac{x^{\nu_1}\cdots x^{\nu_m}}{x^{\,2(a+m)}}\,
\end{equation}
\medskip
where $m=0,1,2,\cdots$ and
the matrix $\mathrm{A}$ respects
properties $\mathrm{A}_{\mu\nu}\,\mathrm{A}^{\mu}_{\,\,\lambda} = 0 \ ;\ \mathrm{A}_{\nu\mu}\,\mathrm{A}_{\lambda}^{\,\,\mu} = 0$.
The equivalence of relations (\ref{spin1}), (\ref{spin2}) and~(\ref{star-tr}) and the validity of relation~(\ref{star-tr}) are proven in Appendix A.

Coxeter relations (\ref{spin1}) and (\ref{spin2}) are basic relations which
enable to prove Yang-Baxter equation for $\CCR$-operator
constructed from the basic building blocks
\begin{equation} \label{Rspin}
\begin{array}{c}
\displaystyle{ [\CCR_{12}\, \Phi](X_1 ; X_2)
= \int
\frac{ \mathrm{d}^4 q \, \mathrm{d}^4 k \, \mathrm{d}^4 y \, \mathrm{d}^4 z
\,\,\, e^{i \, (q + k)\, x_{21}}\, e^{i \, k \, (y -z)}
}
{
q^{2 \left( u_- - v_++2\right)}
z^{2 \left( u_+ - v_++2\right)}
y^{2 \left( u_- - v_-+2\right)}
k^{2 \left( u_+ - v_-+2\right)}
}\cdot } 
  \\ [0.4cm]
\cdot \Phi(x_1-y\,,\,\lambda_2\, \mathbf{z}\,\overline{\mathbf{k}}\,,\,
\tilde{\lambda}_2 \, \overline{\mathbf{q}}\,\mathbf{y}\,
;\, x_2-z\,,\, \lambda_1\,\mathbf{q}\,\overline{\mathbf{z}}\,,\,
\tilde{\lambda}_1 \, \overline{\mathbf{y}} \, \mathbf{k} \, )\,.
\end{array}
\end{equation}

To conclude this Subsection we 
would like to stress that in the case of the conformal algebra $so(5,1)$ of $4$-dimensional Euclidean
space we proved the new star-triangle relations (\ref{spin1}) and (\ref{spin2})
for generic representations of
the type $\rho_{\Delta,\ell,\dot{\ell}}$ included spin degrees of
 freedom, i.e. we generalize the scalar star-triangle
relation to the star-triangle relation for the propagators of particles with any spin
\footnote{Other generalizations of the
scalar star-triangle relation and special star-triangle identities which include $\gamma$-matrices and
propagators of spin particles were also considered in \cite{FrP}, \cite{Isaev2} (see eqs.(27)) and
\cite{Mitr}.}.
It seems that the integrable models of the type \cite{Lipat}, \cite{DKM} or \cite{Zam1} 
related to the spinorial $\mathrm{R}$-matrix (\ref{Rspin}) and 
spinorial star-triangle relations (\ref{spin1}) and (\ref{spin2}) are not known.

%\newpage

%%%%%%%%%%%%%%%%%%%%%%%%%%%%%%%%%%%%%%%%%%%%%%%%%%%%%%%%%%%%%%%%%%%%%%%%%%%%%%%%5
\section*{Acknowledgment}

\mbox{} We are grateful to A.Manashov, G.Korchemsky and V.Tarasov for discussions and
critical remarks.
The work of S.D. is supported by RFBR grants 11-01-00570, 11-01-12037, 12-02-91052
and Deutsche Forschungsgemeinschaft (KI 623/8-1).
The work of A.P.I. was supported in parts by the RFBR grant 11-01-00980-a and
by the NRU HSE Academic Fund Program (grant no.12-09-0064).
The work of D.C. is supported by the Chebyshev Laboratory
(Department of Mathematics and Mechanics, St.-Petersburg State University)
under RF government grant 11.G34.31.0026,
and by Dmitry Zimin's "Dynasty" Foundation.

%%%%%%%%%%%%%%%%%%%%%%%%%%%%%%%%%%%%%%%%%%%%%%%%%%%%%%%%%%%%%%%%%%%%%%%%%%%%%%%%%%%%%%%%%%%%

%%%%%%%%%%%%%%%%%%%%%%%%%%%%%%%%%%%%%%%%%%%%%%

%%%%%%%%%%%%%%%%%%%%%%%%%%%%%%%%%%%%%%%%%%%%%%%%%%%%%%%%%%%%%%%%%%%%%%%%%%%%%%%%%%%%%%%%%%%%

%%%%%%%%%%%%%%%%%%%%%%%%%%%%%%%%%%%%%%%%%%%%%%

\appendix
\renewcommand{\theequation}{\Alph{section}.\arabic{equation}}
\setcounter{table}{0}
\renewcommand{\thetable}{\Alph{table}}

\section*{Appendices}

%%%%%%%%%%%%%%%%%%%%%%%%%%%%%%%%%%%%%%%%%%%%%%%%%%%%%%%%%%%%%%%%%%%%%%%%%%%%%%%%%%%%%%%%%%%%

%%%%%%%%%%%%%%%%%%%%%%%%%%%%%%%%%%%%%%%%%%%%%%

%%%%%%%%%%%%%%%%%%%%%%%%%%%%%%%%%%%%%%%%%%%%%%%%%%%%%%%%%%%%%%%%%%%%%%%%%%%%%%%%%%%%%%%%%%%%
\section{Appendix: Direct proof of the star-triangle relation}
\setcounter{equation}{0}

\mbox{} In this Appendix we prove identities (\ref{spin1}) and (\ref{spin2}) which
are the corner stone of our construction.
The first relation~(\ref{spin1}) can be rewritten as an operator identity
$$
\int \frac{\mathrm{d}^4 z\, \mathrm{d}^4 k\, \mathrm{d}^4 y\,\,
           (\lambda_2  \, \mathbf{z} \, \overline{\mathbf{k}} \, \eta_1)^{2\ell_1}
           (\lambda_1 \, \mathbf{k} \, \overline{\mathbf{y}} \, \eta_2)^{2\ell_2}
           (\tilde{\lambda}_2 \, \overline{\mathbf{z}} \, \mathbf{y} \, \tilde{\eta}_2)^{2\dot{\ell}_2}}
           {
             z^{2 \left( a + 2\right)}
             k^{2 \left( a+b + 2\right)}
             y^{2 \left( b + 2\right)}
           }
\, e^{i \, z \, \hat{p}_2} e^{i \, k \, x_{21}} e^{i \, y \, \hat{p}_2} =
$$
\begin{equation}\label{1}
= \int \frac{\mathrm{d}^4 q\, \mathrm{d}^4 y\, \mathrm{d}^4 k\,\,
           (\lambda_2  \, \mathbf{y} \, \overline{\mathbf{k}} \, \eta_1)^{2\ell_1}
           (\lambda_1 \, \mathbf{q} \, \overline{\mathbf{y}} \, \eta_2)^{2\ell_2}
           (\tilde{\lambda}_2 \, \overline{\mathbf{q}} \, \mathbf{k} \, \tilde{\eta}_2)^{2\dot{\ell}_2}}
           {
             q^{2 \left( b + 2\right)}
             y^{2 \left( a+b + 2\right)}
             k^{2 \left( a + 2\right)}
           }
\, e^{i \, q \, x_{21}} e^{i \, y \, \hat{p}_2} e^{i \, k \, x_{21}}
\end{equation}
and the second one~(\ref{spin2}) as

$$
\int \frac{\mathrm{d}^4 z\, \mathrm{d}^4 k\, \mathrm{d}^4 y\,
           (\lambda_1 \, \mathbf{z} \, \overline{\mathbf{y}} \, \eta_1)^{2\ell_1}
           (\tilde{\lambda}_2 \, \overline{\mathbf{k}} \, \mathbf{y} \,  \tilde{\eta}_1)^{2\dot{\ell}_1}
           (\tilde{\lambda}_1 \, \overline{\mathbf{z}} \, \mathbf{k} \,  \tilde{\eta}_2)^{2\dot{\ell}_2}}
           {
             z^{2 \left( a + 2\right)}
             k^{2 \left( a+b + 2\right)}
             y^{2 \left( b + 2\right)}
           }
\, e^{i \, z \, \hat{p}_1} e^{i \, k \, x_{21}} e^{i \, y \, \hat{p}_1} =
$$
\begin{equation}
\label{2}
= \int \frac{\mathrm{d}^4 q\, \mathrm{d}^4 y\, \mathrm{d}^4 k\,
           (\lambda_1 \, \mathbf{q} \, \overline{\mathbf{k}} \, \eta_1)^{2\ell_1}
           (\tilde{\lambda}_2 \, \overline{\mathbf{q}} \, \mathbf{y} \, \tilde{\eta}_1)^{2\dot{\ell}_1}
           (\tilde{\lambda}_1 \, \overline{\mathbf{y}} \, \mathbf{k} \, \tilde{\eta}_2)^{2\dot{\ell}_2}}
           {
             q^{2 \left( b + 2\right)}
             y^{2 \left( a+b + 2\right)}
             k^{2 \left( a + 2\right)}
           }
\, e^{i \, q \, x_{21}} e^{i \, y \, \hat{p}_1} e^{i \, k \, x_{21}} \,
\end{equation}
where we use compact notations (\ref{notation}).
In a particular case $\ell_1 = \ell_2 = \dot{\ell}_1 = \dot{\ell}_2 = 0$ (i.e. scalar one)
spinor variables disappear from (\ref{1}) and (\ref{2})
and corresponding integrals can be easily evaluated. Therefore these
identities reduce to (\ref{S1S2S1}).

Both previous relations are equivalent to the following generating
integral identity
$$
\int \mathrm{d}^4 z\, \mathrm{d}^4 k\,
\frac{
\langle(x-z)\,\mathrm{A}\,(y-z)\rangle^{m}
\,\mathrm{e}^{ -i\,\langle k \bigl[z -\alpha\,\mathrm{B}(y-z) -\beta\,\mathrm{C}(x-z)\bigl]\rangle}}
{(x-z)^{2 a} \,
k^{2 b} \,
(y-z)^{2 c}}
=
$$
\begin{equation} \label{starTri}
=\frac{1}{(x-y)^{2 b}}\,\int \mathrm{d}^4 k\, \mathrm{d}^4 p\,
\frac{
\langle p\, \mathrm{A} k\rangle^{m}\,
\mathrm{e}^{ i\,\langle p \bigl[x +\alpha\,\mathrm{B}(x-y)\bigl]\rangle}
\,\mathrm{e}^{ -i\,\langle k \bigl[y +\beta\,\mathrm{C}(y-x)\bigl]\rangle}
}
{
k^{2 a}\,
p^{2 c}
}
\end{equation}
provided that parameters respect uniqueness condition
\begin{equation} \label{uniqueness}
a + c - b = 2 + m\; , \; m = 0,1,2,\cdots
\end{equation}
Here $\alpha$, $\beta$ are numerical parameters, matrices
$\mathrm{A} \, , \, \mathrm{B}\, , \, \mathrm{C}$ fulfil the following properties
\begin{equation} \label{prop}
\begin{array}{c}
A_{\mu \nu} A^{\mu}{}_{ \lambda} = A_{\nu \mu} A_{ \lambda}{}^{\mu} =0 \ ; \
B_{\mu \nu} B^{\mu}{}_{ \lambda} = B_{\nu \mu} B_{ \lambda}{}^{\mu} =0 \ ; \
C_{\mu \nu} C^{\mu}{}_{ \lambda} = C_{\nu \mu} C_{ \lambda}{}^{\mu} =0 \ ; \\ [0.3 cm]
A_{\mu \nu} + A_{\nu \mu} = 2 \,g_{\mu \nu} \tr A \ ; \
B_{\mu \nu} + B_{\nu \mu} = 2 \,g_{\mu \nu} \tr B \ ; \
C_{\mu \nu} + C_{\nu \mu} = 2 \,g_{\mu \nu} \tr C \ ; \\ [0.3 cm]
\mathrm{B}_{\mu \nu} \mathrm{C}_{\lambda}{}^{\nu} + \mathrm{B}_{\lambda \nu} \mathrm{C}_{\mu}{}^{\nu}
= 2 \,g_{\mu \lambda} \tr B C \ ; \
\mathrm{A}_{\mu \nu} \mathrm{B}_{\lambda} {}^{\nu} = \mathrm{B}_{\nu \mu} \mathrm{A}^{\nu} {}_{\lambda} \ ; \
\mathrm{A}_{\mu \nu} \mathrm{C}_{\lambda}{}^{\nu} = \mathrm{C}_{\mu \nu} \mathrm{A}_{\lambda}{}^{\nu}
\end{array}
\end{equation}
where we normalize the trace such that $\tr( \II )= 1$.
We also use shortcut notations
$
\langle x\,\mathrm{M}\,y \rangle = x_{\mu}\,y_{\nu}\, \mathrm{M}_{\mu\nu}$ .

In order to obtain (\ref{1}) we have to take in (\ref{starTri})
$$
\left(\tilde{\lambda}_2 \, \overline{\boldsymbol{\sigma}}_{\mu} \, \boldsymbol{\sigma}_{\nu} \,  \tilde{\eta}_2\right) = \mathrm{A}_{\mu\nu} \; ; \;
\left(\lambda_1 \, \boldsymbol{\sigma}_{\mu} \, \overline{\boldsymbol{\sigma}}_{\nu} \, \eta_2 \right) = \mathrm{B}_{\mu\nu} \; ; \;
\left(\lambda_2 \, \boldsymbol{\sigma}_{\nu} \, \overline{\boldsymbol{\sigma}}_{\mu} \, \eta_1 \right) = \mathrm{C}_{\mu\nu} \; ; \;
m = 2 \dot{\ell}_2 \;
$$
and apply $\partial^{2 \ell_2}_{\alpha} \partial^{2 \ell_1}_{\beta} |_{\alpha = \beta =0}$ .
To obtain (\ref{2}) we take in (\ref{starTri})
$$
\left(\lambda_1  \, \boldsymbol{\sigma}_{\mu} \, \overline{\boldsymbol{\sigma}}_{\nu} \, \eta_1\right) = \mathrm{A}_{\mu\nu} \; ; \;
\left(\tilde{\lambda}_2 \, \overline{\boldsymbol{\sigma}}_{\mu} \, \boldsymbol{\sigma}_{\nu} \, \tilde{\eta}_1 \right) = \mathrm{B}_{\mu\nu} \; ; \;
\left(\tilde{\lambda}_1 \, \overline{\boldsymbol{\sigma}}_{\nu} \, \boldsymbol{\sigma}_{\mu} \, \tilde{\eta}_2 \right) = \mathrm{C}_{\mu\nu} \; ; \;
m = 2 \ell_1 \;
$$
and apply
$\partial^{2 \dot{\ell}_1}_{\alpha} \partial^{2 \dot{\ell}_2}_{\beta} |_{\alpha = \beta =0}$ .
Using (\ref{idsi}) and Fierz identity
$\mathbf{\sigma}_{\mu} \otimes \overline{\mathbf{\sigma}}^{\mu} = 2 \,\mathrm{P}$
it is easy to check that previous expressions for matrices $\mathrm{A}\, , \,\mathrm{B}\, , \,\mathrm{C}$
fulfil relations (\ref{prop}).

Thus our aim is to prove (\ref{starTri}) that we will perform in two steps.
On the first step we
implement ceratin change of variables which enables to remove matrices
$\mathrm{B}$ and $\mathrm{C}$ from (\ref{starTri}) obtaining an integral relation
equivalent to the operator identity (\ref{star-tr}). On the second step we prove (\ref{star-tr}).

Let us consider the integral in the right hand side of (\ref{starTri})
$$
\int \mathrm{d}^4 k\, \mathrm{d}^4 p\,
\frac{ \langle p\, \mathrm{A} k\rangle^{m}\,
\mathrm{e}^{ i\,\langle p \bigl[x +\alpha\,\mathrm{B}(x-y)\bigl]\rangle}
\,\mathrm{e}^{ -i\,\langle k \bigl[y +\beta\,\mathrm{C}(y-x)\bigl]\rangle}
}
{
k^{2 a}\,
p^{2 c}
} =
$$
$$
= \langle \partial_{w_1} \mathrm{A} \, \partial_{w_2}\rangle^{m}\,
\left. \int \mathrm{d}^4 k\, \mathrm{d}^4 p\,
\frac{
\mathrm{e}^{ i\,\langle p \bigl[x + w_1 +\alpha\,\mathrm{B}(x-y)\bigl]\rangle}
\,\mathrm{e}^{ -i\,\langle k \bigl[y + w_2 +\beta\,\mathrm{C}(y-x)\bigl]\rangle}
}
{
k^{2 a}\,
p^{2 c}
} \right|_{w_1=w_2 =0} =
$$
then we implement Fourier transform (here $a' \equiv 2-a \, , \, b' \equiv 2 - b \, , \, c' \equiv 2-c$)
$$
= \left.4^{a'+c'} \pi^4 \frac{\Gamma(a') \, \Gamma(c')}{\Gamma(a)\,\Gamma(c)}
\langle \partial_{w_1} \mathrm{A} \, \partial_{w_2}\rangle^{m}\,
\frac{1}{\bigl[x + w_1 +\alpha\,\mathrm{B}(x-y)\bigl]^{2 c'}\bigl[y + w_2 +\beta\,\mathrm{C}(y-x)\bigl]^{2 a'}}
\right|_{w_1=w_2=0} =
$$
and perform differentiation
$$
= 4^{a'+c'+m} \pi^4 \frac{\Gamma(a'+m) \, \Gamma(c'+m)}{\Gamma(a)\,\Gamma(c)}
\frac{\langle \left[x +\alpha\,\mathrm{B}(x-y)\right] \mathrm{A} \left[y +\beta\,\mathrm{C}(y-x) \right]\rangle^{m}}
{\bigl[x +\alpha\,\mathrm{B}(x-y)\bigl]^{2 (c'+m)}\bigl[y +\beta\,\mathrm{C}(y-x)\bigl]^{2 (a'+m)}} \;.
$$
Further we introduce new variables which absorb matrices $\mathrm{B}$ and $\mathrm{C}$
\begin{equation} \label{XY}
X \equiv x +\alpha\,\mathrm{B}(x-y) \ ; \ Y \equiv y +\beta\,\mathrm{C}(y-x) \;.
\end{equation}
Then $X-Y = \mathrm{S} \cdot (x-y)$ where
\begin{equation} \label{Lin}
\mathrm{S} \equiv \II+\alpha\,\mathrm{B}+\beta\,\mathrm{C}\;.
\end{equation}
Using properties (\ref{prop}) one can easily obtain that
\begin{equation} \label{SS}
%\mathrm{S}_{\rho \mu} \, \mathrm{S}^{\rho}{}_{\nu} = \lambda \cdot \eta_{\mu \nu}
\mathrm{S} \cdot \mathrm{S}^{T} = \lambda \cdot \II
\; ; \;  \lambda \equiv
1 + 2 \alpha \tr B + 2 \beta \tr C + 2 \alpha \beta \tr B C
\end{equation}
therefore
$$
(X-Y)^2 = \lambda \cdot (x-y)^2
$$
and the right hand side of (\ref{starTri})
takes the form
\begin{equation} \label{RHS}
4^{b'} \pi^4 \lambda^b \,\frac{\Gamma(a'+m) \, \Gamma(c'+m)}{\Gamma(a)\,\Gamma(c)}
\frac{\langle X A Y \rangle^m}{X^{2(c'+m)}(X-Y)^{2b}Y^{2(a'+m)}} \;.
\end{equation}

Then we consider the left hand side of (\ref{starTri}) where we shift the integration variable
$$
\int \mathrm{d}^4 z\, \mathrm{d}^4 k\,
\frac{
\langle(z-x+y)\,\mathrm{A}\,z\rangle^{m}
\,\mathrm{e}^{ -i\,\langle k \bigl[z+y +\alpha\,\mathrm{B} z + \beta\,\mathrm{C}(z-x+y)\bigl]\rangle}}
{(x-y-z)^{2 a} \,
k^{2 b} \,
z^{2 c}} =
$$
and perform Fourier transform
$$
= 4^{b'} \pi^{2} \frac{\Gamma(b')}{\Gamma(b)}
\int \mathrm{d}^4 z\,
\frac{
\langle(z-x+y)\,\mathrm{A}\,z\rangle^{m}}
{(x-y-z)^{2 a} \,
\bigl[z+y +\alpha\,\mathrm{B} z + \beta\,\mathrm{C}(z-x+y)\bigl]^{2 b'}
z^{2 c}}\;.
$$

Further we change the integration variables $Z = \mathrm{S} \cdot z$ (\ref{Lin})
in the previous integral
and introduce variables (\ref{XY}) instead of $x$ and $y$.
Let us note that $X-Y+Z = \mathrm{S} \cdot (x-y+z)$ , consequently due to (\ref{SS})
$$
Z^2 = \lambda \cdot z^2 \; ; \; (X-Y-Z)^2 = \lambda \cdot (x-y-z)^2 \;.
$$
From (\ref{SS}) it follows that the Jacobian of the linear change is equal to
$\left|\det \mathrm{S}\right| = \lambda^2$.
Using (\ref{prop}) it is possible to deduce that
$\mathrm{S} \cdot \mathrm{A} \cdot \mathrm{S}^{T} = \lambda \cdot \mathrm{A}$ ,
thus
$$
(z-x+y)\,\mathrm{A}\,z = (Z-X+Y) \, \mathrm{S}^{-1 T} \cdot \mathrm{A} \cdot \mathrm{S}^{-1} \, Z =
\frac{1}{\lambda} \cdot (Z-X+Y) \,\mathrm{A}\, Z\;.
$$
Therefore the left-hand side of (\ref{starTri}) takes the form
\begin{equation} \label{LHS}
4^{b'} \pi^2 \frac{\Gamma(b')}{\Gamma(b)} \, \lambda^{a+c-2-m} \int \mathrm{d}^4 Z
\frac{\langle (Z-X+Y) \,\mathrm{A}\, Z \rangle^m}{(X-Y-Z)^{2a}(Z+Y)^{2b'} Z^{2c}}
\end{equation}
Finally equating (\ref{RHS}) with (\ref{LHS}),
performing a shift of integration variable in the later and taking
into account uniqueness condition (\ref{uniqueness})
we obtain that (\ref{starTri}) is equivalent to
\begin{equation} \label{star-trInt}
\int \mathrm{d}^4 Z
\frac{\langle (Z-X) \,\mathrm{A}\, (Z-Y) \rangle^m}{(X-Z)^{2a}Z^{2b'} (Z-Y)^{2c}} =
\pi^2 \,\frac{\Gamma(a'+m) \, \Gamma(b) \,\Gamma(c'+m)}{\Gamma(a)\,\Gamma(b')\,\Gamma(c)}
\frac{\langle X A Y \rangle^m}{X^{2(c'+m)}(X-Y)^{2b}Y^{2(a'+m)}}\;
\end{equation}
where
$$
a + c - b = 2 + m\,.
$$

(\ref{star-trInt}) is an integral form of the operator identity (\ref{star-tr})
in the same way as (\ref{scStTrInt}) is an integral form
of the scalar star-triangle identity (\ref{scStTrOp}).

Now we are going to prove (\ref{star-trInt}). At first let us evaluate the integral
$$
I(x,y) \equiv \int \mathrm{d}^4 z \,\frac{\langle (z-x) \,\mathrm{A} \,z\rangle^m}{(x-z)^{2a} (z-y)^{2b'} z^{2c}}\;.
$$
by means of inversion transform
$$
\begin{array}{c}
x \to \frac{x}{x^2}\ ;\
y \to \frac{y}{y^2}\ ;\
z \to \frac{z}{z^2}\ ;\
\mathrm{d}^4 z \to \frac{\mathrm{d}^4 z}{z^{8}}
\ ;\ (x-z)^2\to \frac{(x-z)^2}{x^2\,z^2}
\ ;\ (z-y)^2\to \frac{(z-y)^2}{z^2\,y^2} \\ [0.3 cm]
\langle (z-x) \, \mathrm{A} \, z \rangle \to
\langle \left( \frac{z}{z^2} - \frac{x}{x^2} \right)\, \mathrm{A} \, \frac{z}{z^2}\rangle =
\frac{\langle x \, \mathrm{A} \, (x-z)\rangle}{x^2 z^2}
\end{array}
$$
In the previous transformation we take into account (\ref{prop}).
Then due to uniqueness condition (\ref{uniqueness})
$$
I\left( \frac{x}{x^2} , \frac{y}{y^2} \right) = x^{2a} y^{2 b'}
\int \mathrm{d}^4 z \,\frac{\langle \frac{x}{x^2} \,\mathrm{A} \, (x-z)\rangle^m}{(x-z)^{2a} (z-y)^{2b'}}\;.
$$
In order to evaluate the previous integral we take into account a well-known formula for
convolution of two "`propagators"'
$$
\int \mathrm{d}^4 z \,\frac{1}{(x-w-z)^{2(a-m)} (z-y)^{2b'}} =
\pi^2 \frac{\Gamma(a'+m)\,\Gamma(b)\,\Gamma(c')}{\Gamma(a-m) \, \Gamma(b') \, \Gamma(c)}
\frac{1}{(x-y-w)^{2c'}}
$$
and apply to it $\left.\langle \frac{x}{x^2} \,\mathrm{A}\, \partial_{w}\rangle^m\right|_{w=0}$.
Thus we have
\begin{equation} \label{Iinv}
I\left( \frac{x}{x^2} , \frac{y}{y^2} \right) =
\pi^2 \frac{\Gamma(a'+m)\,\Gamma(b)\,\Gamma(c'+m)}{\Gamma(a) \, \Gamma(b') \, \Gamma(c)}\,
x^{2a} y^{2 b'} \frac{\langle \frac{x}{x^2} \,\mathrm{A} \, (x-y)\rangle^m}{(x-y)^{2(c'+m)}}
\end{equation}
To obtain $I(x,y)$
we perform inverse transform
$$
x \to \frac{x}{x^2}\ ;\
y \to \frac{y}{y^2}\ ;\
(x-y)^2\to \frac{(x-y)^2}{x^2\,y^2} \ ; \
\langle \frac{x}{x^2} \, \mathrm{A} \, (x-y) \rangle \to
\langle (y-x) \, \mathrm{A} \, \frac{y}{y^2} \rangle
$$
in (\ref{Iinv})
\begin{equation} \label{I}
I(x,y) = \pi^2 \frac{\Gamma(a'+m)\,\Gamma(b)\,\Gamma(c'+m)}{\Gamma(a) \, \Gamma(b') \, \Gamma(c)}\,
\frac{\langle (y-x) \,\mathrm{A} \, y \rangle^m}{(y-x)^{2(c'+m)} \,x^{2b}\, y^{2(a'+m)}}\;.
\end{equation}
Finally we note that (\ref{star-trInt}) coincides with (\ref{I})
at $x \to X-Y \; , \; y \to -Y$.

%%%%%%%%%%%%%%%%%%%%%%%%%%%%%%%%%%%%%%%%%%%%%%
\section{Appendix: Clifford algebra}
\setcounter{equation}{0}

\mbox{} Let $\Gamma_{a}$, $a=0,1,\ldots,n+1$, be a set of
$n+2$ operators satisfying the standard relations
\begin{equation}\label{def}
\Gamma_{a}\Gamma_{b}+\Gamma_{b}\Gamma_{a} =
2\,g_{a b}\cdot \mathbf{I}
\end{equation}
where $\mathbf{I}$ is the unit operator and $g_{ab}$ is the metric for $\mathbb{R}^{p+1,q+1}$:
$$
 g_{ab} = {\rm diag}(\underbrace{1,\dots,1}_p , \underbrace{-1,\dots,-1}_q , 1,-1) \; .
$$
Operators $\Gamma_{a}$ are
generators of the Clifford algebra which, as a vector
space, has dimension $2^{n+2}$ by integer $n$. The standard basis in this
space is formed by anti-symmetrized products
$$
\Gamma_{A_0} = \mathbf{I}\  ,\  \Gamma_{A_1} = \Gamma_{a}\ ,
\ \Gamma_{A_2} = \Gamma_{a_1 a_2} = \frac{1}{2!}
[\Gamma_{a_1}\Gamma_{a_2}-\Gamma_{a_2}\Gamma_{a_1}]\,,
$$
$$
\Gamma_{A_k} = \Gamma_{a_1\ldots a_k} =
\mathrm{Asym}(\Gamma_{a_1}\cdots\Gamma_{a_k}) =
\frac{1}{k!}
\sum_{s}(-1)^{\mathrm{p}(s)} \Gamma_{s(a_1)} \cdots\Gamma_{s(a_k)}\,,
$$
where $A_k$ is multi-index $a_1\ldots a_k$, summation is over all permutations of k indices $\{a_1, \dots, a_k\}$ and
$\mathrm{p}(s)$ -- is a parity of the permutation $s$.

When the number of dimensions is integer,
then $k\leq n+2$ and the total number of these matrices is $2^{n+2}$.
However, when $n$ is non integer (dimensional regularization)
the Clifford algebra is infinite-dimensional.
Let us introduce the generating
function for the matrices $\Gamma_{A_k}$
$$
\sum_{k=0}^{\infty} \frac{1}{k!}\,u^{a_k}\cdots u^{a_1} \cdot\mathrm{Asym}(\Gamma_{a_1}\cdots\Gamma_{a_k}) = \sum_{k=0}^{\infty} \frac{1}{k!}\,\left(u^{a}\Gamma_{a}\right)^k = \mathrm{exp} (u\Gamma)\,.
$$
Inside the sym-product gamma-matrices behave
like anti-commuting variables so that the auxiliary vector variables $u^{a}$
have to be anti-commuting: $u^{a}\,u^{b} = - u^{b}\,u^{a}$ but it is not the end of the story.
The simplest exponential form of the generating function is obtained by the condition that $u^{a}\,\Gamma_{b} = - \Gamma_{b}\,u^{a}$ so that we fix these rules of the game and have
$$
\Gamma_{a_1\ldots a_k} =
\left.\frac{\delta^k}
{\delta u^{a_1}\cdots\delta u^{a_k}}\, \exp(u\Gamma)\right|_{u=0}\ \ \ ;\ \ \
\mathrm{Asym}\left[ \exp(u\Gamma)\right] = \exp(u\Gamma)\,.
$$
There exists the general formula which allows to transform the product of generating functions to the anti-symmetrized product
\begin{equation}\label{base}
e^{u_1\Gamma}\cdots e^{u_k\Gamma} =
e^{-\sum_{i<j} u_i u_j}\cdot e^{(u_1+\cdots+u_k)\Gamma}\,.
\end{equation}
It is the consequence of the simplest relation
$$
e^{u\Gamma}\,e^{v\Gamma} =
e^{-u v}\cdot e^{(u+v)\Gamma}\,,
$$
obtained from the Backer-Hausdorff formula
$$
e^A\,e^B = e^{A+B +\frac{1}{2}[A,B]} \,,
$$
where $A=u\Gamma$ , $B=v\Gamma$ and $[A,B] = - u^{a} v^{b}\left(\Gamma_{a}\Gamma_{b}+\Gamma_{b}\Gamma_{a}\right) = -2u\,v$.

The second very useful formula is
\begin{equation}\label{use}
\left. \exp\left(\lambda \frac{\delta^2}{\delta u \delta v}\right)
\,\exp(ux+vy+uvz)\right|_{u=v=0} = (1-z\lambda)^{n+2}
\,\exp\left(\frac{\lambda}{1-z\lambda}xy\right)\,,
\end{equation}
where $u^{a}\,,\,v^{a}\,,\,x^{a}\,,\,y^{a} $ are anti-commuting variables,
$\lambda$ and $z$ are commuting scalar variables and as usual $\frac{\delta^2}{\delta u \delta v} \equiv
\frac{\delta^2}{\delta u^{a} \delta v_{a}}$.
The expression from the right hand side is derived by using the
standard representation of the operation
$\mathrm{exp}\left(\lambda \frac{\delta^2}{\delta u \delta v}\right)$
through gaussian integral over anti-commuting variables.
In fact all calculations are based on two formulae:~(\ref{base}) and ~(\ref{use}).

\subsection{Tensor product}

In the present Subsection we collect some formulae
needed for the calculations related to the Yang-Baxter equation.
Let us introduce the generating functions $\exp(u\Gamma_k)$
for the gamma-matrices acting in different spaces $\mathbb{V}_k$.
In fact we shall use the gamma-matrices acting in the tensor product of two spaces and we assume that these operators are anti-commuting
$$
\Gamma_{a}\otimes \mathbf{I} = (\Gamma_1)_{a}\ ;\
\mathrm{I}\otimes\Gamma_{a} = (\Gamma_2)_{a}\ \ ;\ \ \Gamma_1\,\Gamma_2 = - \Gamma_2\, \Gamma_1\,.
$$
This assumption of anti-commutativity leads to unusual signs in
comparison to the usual tensor product convention but all needed
formulae have the simplest form.
It is very similar to our convention that auxiliary variables $u_{\mu}$
anti-commute with gamma-matrices $\gamma_{\mu}$.

Let us represent the tensor product $\Gamma_{A_k} \otimes \Gamma^{A_k}$ in
the following form
$$
\Gamma_{A_k} \otimes \Gamma^{A_k} =\left. s_k
\left(\frac{\delta^2}{\delta u^{a}\delta v_{a}}\right)^k
\,e^{u\Gamma_1}\,e^{v\Gamma_2}\right|_{u=v=0} =
\left. s_k\, \partial_{\lambda}^k\,
e^{\lambda\frac{\delta^2}{\delta u\delta v}}
\,e^{u\Gamma_1+v\Gamma_2}\right|_{\lambda=u=v=0}\,,
$$
where $s_k\equiv(-)^{\frac{k(k-1)}{2}}$.
Note that at this stage we cannot use the simplest variant of the formula~(\ref{use}) for $z=0$ but it is possible to do under the sign of anti-symmetrization
$$
\left.
e^{\lambda\frac{\delta^2}{\delta u\delta v}}
\,e^{u\Gamma_1+v\Gamma_2}\right|_{u=v=0} =
\left.
e^{\lambda\frac{\delta^2}{\delta u\delta v}}
\,\mathrm{Asym}\left[e^{u\Gamma_1+v\Gamma_2}\right]
\right|_{u=v=0} =
\mathrm{Asym}\left[e^{\lambda\Gamma_1\Gamma_2}\right]\,,
$$
so that one obtains the compact expression
$$
\Gamma_{A_k} \otimes \Gamma^{A_k} = s_k \partial_{\lambda}^k \left.
\, e^{\lambda\frac{\delta^2}{\delta u\delta v}}
\, e^{u\Gamma_1+v\Gamma_2}\right|_{\lambda=u=v=0}
= s_k \partial_{\lambda}^k
\,\left.\mathrm{Asym}\left[e^{\lambda\Gamma_1\Gamma_2}\right]\right|_{\lambda=0} =
s_k\cdot\mathrm{Asym}\left[(\Gamma_1\Gamma_2)^k\right]\,.
$$
In detailed notations everything looks as follows
$$
\mathrm{Asym}\left[\Gamma_{1a_1}\cdots\Gamma_{1a_k}\right]\cdot
\mathrm{Asym}\left[\Gamma_{2}^{a_1}\cdots\Gamma_{2}^{a_k}\right] = \mathrm{Asym}\left[\Gamma_{1a_1}\cdots\Gamma_{1a_k}\right]\cdot
\Gamma_{2}^{a_1}\cdots\Gamma_{2}^{a_k} =
$$
$$
= \mathrm{Asym}\left[\Gamma_{1a_1}\cdots\Gamma_{1a_k}\cdot
\Gamma_{2}^{a_1}\cdots\Gamma_{2}^{a_k}\right] =  s_k\cdot\mathrm{Asym}\left[(\Gamma_1\Gamma_2)^k\right]\,.
$$
At first step one can forget about one of the signs $\mathrm{Asym}$ because there is convolution of two antisymmetric tensors. Next it is possible to rearrange everything in the rest product using our rule of the game
$\Gamma_1\,\Gamma_2 = - \Gamma_2\,\Gamma_1$. In any case the compact expression in the right hand side means exactly the expression from the left hand side so that decoding procedure is simple.

Using this expression it is possible to represent
operator acting in $\mathbb{V}_1\otimes \mathbb{V}_2$
in any of the forms
$$
\mathrm{R} = \sum_{k=0}^{\infty} \frac{\mathrm{R}_k}{k!}
\cdot \Gamma_{A_k} \otimes \Gamma^{A_k} =
\sum_{n=0}^{\infty} \frac{\mathrm{R}_k\,s_k}{k!}\, \partial_{\lambda}^k
\,\left.\mathrm{Asym}\left(e^{\lambda \Gamma_1\Gamma_2}\right)\right|_{\lambda=0} =
\mathrm{R}(\lambda)\ast\mathrm{Asym}\left(e^{\lambda \Gamma_1\Gamma_2}\right)\,,
$$
where for simplicity we introduce the compact notation
for the used operation of convolution
$$
\mathrm{R}(\lambda)\ast \mathrm{F}(\lambda) \equiv
\left.\mathrm{R}(\partial_{\lambda})\, \mathrm{F}(\lambda)\right|_{\lambda=0}
\,.
$$
Note that all information about operator $\mathrm{R}$ is encoded in
the coefficient function $\mathrm{R}(x)$
$$
\mathrm{R}(x) = \sum_{k=0}^{\infty} \frac{\mathrm{R}_k\,s_k}{k!}\, x^k \longleftrightarrow \mathrm{R} = \sum_{k=0}^{\infty}
\frac{\mathrm{R}_k}{k!}\cdot \Gamma_{A_k} \otimes \Gamma^{A_k}\,.
$$
Let us consider as example the permutation operator
 which is defined by the equation
$$
\mathrm{P}\cdot\mathbf{I}\otimes\Gamma_{a} =
\Gamma_{a}\otimes\mathbf{I}\cdot\mathrm{P}\,.
$$
First we rewrite this equation with the help of generating functions
$$
\left.\mathrm{P}(\lambda)\ast\,\frac{\delta}{\delta s^a}\,
e^{s\Gamma_1}\cdot\mathrm{Asym}\left(e^{\lambda \Gamma_1\Gamma_2}\right) \right|_{s=0} = \left.\mathrm{P}(\lambda)\ast\,\frac{\delta}{\delta t^a}\,\mathrm{Asym}
\left(e^{\lambda \Gamma_1\Gamma_2}\right)\cdot e^{t\Gamma_2}
\right|_{t=0}
$$
The products of generating functions can be transformed
to the anti-symmetrized product in a following way
$$
e^{s\Gamma_1}\cdot\mathrm{Asym}\left(e^{\lambda \Gamma_1\Gamma_2}\right) = \mathrm{Asym}\left(e^{\lambda \Gamma_1\Gamma_2+s(\Gamma_1+\lambda\Gamma_{2})}\right)\ \ ;\ \
\mathrm{Asym}
\left(e^{\lambda \Gamma_1\Gamma_2}\right)\cdot e^{t\Gamma_2} =
\mathrm{Asym}\left(e^{\lambda \Gamma_1\Gamma_2+t(\Gamma_2+\lambda\Gamma_{1})}\right)
$$
Derivation is very simple and we perform it
step by step for the first product
$$
e^{s\Gamma_1}\cdot\mathrm{Asym}\left(e^{\lambda \Gamma_1\Gamma_2}\right) =
\left.
e^{\lambda\frac{\delta^2}{\delta u\delta v}}
\,e^{s\Gamma_1}\cdot e^{u\Gamma_1+v\Gamma_2}
\right|_{u=v=0} = \left.
e^{\lambda\frac{\delta^2}{\delta u\delta v}}
\,e^{u s+u\Gamma_1+v\Gamma_2+s\Gamma_1}
\right|_{u=v=0} =
$$
$$
= \left.
e^{\lambda\frac{\delta^2}{\delta u\delta v}}
\,\mathrm{Asym}
\left(e^{us+u\Gamma_1+v\Gamma_2+s\Gamma_1}\right)
\right|_{u=v=0} =
\mathrm{Asym}
\left(e^{\lambda(\Gamma_1+s)\Gamma_2+s\Gamma_1}\right)\,.
$$
In fact this calculation is prototype of all similar
manipulations with generating functions. In the following we shall omit all intermediate steps and state the final identities.
Next we calculate the derivatives with respect $s^a$ and $t^a$
$$
\mathrm{P}(\lambda)\ast\,
\mathrm{Asym}\left[(\Gamma_{1a}+\lambda\Gamma_{2a})\,e^{\lambda \Gamma_1\Gamma_2}\right] =
\mathrm{P}(\lambda)\ast\,
\mathrm{Asym}
\left[(\Gamma_{2a}+\lambda\Gamma_{1a})\,e^{\lambda \Gamma_1\Gamma_2}\right]
\,,
$$
or equivalently
\begin{equation}\label{eq1}
\left[\mathrm{P}(\lambda)-
\mathrm{P}^{\prime}(\lambda)\right]\ast\,
\mathrm{Sym}\left(\Gamma_{1a}\,e^{\lambda \Gamma_1\Gamma_2}\right) = \left[\mathrm{P}(\lambda)-
\mathrm{P}^{\prime}(\lambda)\right]\ast\,
\mathrm{Sym}
\left(\Gamma_{2a}\,e^{\lambda \Gamma_1\Gamma_2}\right)\,,
\end{equation}
where in the last transformation we use the
simplest variant of the general formula
$$
\mathrm{P}(\lambda)\ast\,\lambda^n\,\mathrm{F}(\lambda) = \mathrm{P}^{(n)}(\lambda)\ast\,\mathrm{F}(\lambda)\,
$$
and $\mathrm{P}^{(n)}(\lambda)$
is the $n$-th derivative of the function $\mathrm{P}(\lambda)$.
As evident consequence of~(\ref{eq1}) we have
$$
\mathrm{P}^{\prime}(\lambda)=\mathrm{P}(\lambda) \longrightarrow
\mathrm{P}(\lambda) = e^{\lambda}\,,
$$
so that the permutation operator can be represented
in one of the following forms
$$
\mathrm{P} = e^{\lambda}\ast\, \mathrm{Asym}\left(e^{\lambda
\Gamma_1\Gamma_2}\right) =
\mathrm{Asym}\left(e^{\Gamma_1\Gamma_2}\right) = \sum_{k=0}^{\infty}
\frac{s_k}{k!}\cdot \Gamma_{A_k} \otimes \Gamma^{A_k}  =
 $$
 $$
 = \sum_{k=0}^\infty  \left( \sum_{a_1 < a_2 < \dots < a_k}
 \Gamma_{a_1} \Gamma_{a_2} \cdots \Gamma_{a_k} \otimes
 \Gamma^{a_k} \Gamma^{a_{k-1}}\cdots \Gamma^{a_1} \right) \,.
 $$

\subsection{Yang-Baxter equation}
Consider Lie algebra $so(p+1,q+1)$ with generators $M_{ab}$
$(a,b = 0,1,\dots, p+q+1)$ subject relations
$$
[ M_{ab} , M_{dc} ] = i( g_{bd} M_{ac}
+ g_{ac} M_{bd} - g_{ad} M_{bc} -
g_{bc} M_{ad} )  \; ,
$$
where $g_{ab}$ is the metric for $\mathbb{R}^{p+1,q+1}$.
The $\mathrm{L}$-operator (\ref{LKaz}) for $so(p+1,q+1)$ can be written as
 \begin{equation}
 \label{LKaz-so'}
 \mathrm{L}(u) = u \, I_n \otimes \mathbf{1} + \frac{1}{2}\cdot T(M^{a b})\otimes T'(M_{a b})  =
 u \, I_n \otimes \mathbf{1} - \frac{1}{4}\cdot\, \Gamma^{ab}\otimes\, g (e_{ab}  - e_{ba})\,,
 \end{equation}
where we choose $T$ and $T'$ to be spinor and defining
representations of $so(p+1,q+1)$, respectively:
$$
T(M^{a b}) = \frac{i}{2}\, \Gamma^{ab}
\ \ ;\ \ T'(M_{a b}) = i \, g (e_{ab}  - e_{ba})\,.
$$
Then by direct calculation one can prove that operator $\mathrm{L}(u)$ (\ref{LKaz-so'}) satisfies intertwining relation
$$
 \mathrm{R}'_{23}(u - v) \, \mathrm{L}_{12}(u) \,  \mathrm{L}_{13}(v) =
  \mathrm{L}_{12}(v) \, \mathrm{L}_{13}(u)  \,  \mathrm{R}'_{23}(u - v)
   \in  {\rm End}(V \otimes V' \otimes V')
$$
with $so(p+1,q+1)$-type Yangian $\mathrm{R}$-matrix \cite{Zam}
$$
\mathrm{R}'_{23}(u) = u \, \mathrm{P}_{23} + \mathrm{I}_{23} - \frac{u}{u+\frac{n}{2}} \mathrm{K}_{23}\,,
$$
where matrices $\mathrm{I}_{23}$, $\mathrm{P}_{23}$ were described in (\ref{IdPe}) and operator $\mathrm{K}_{23}$ is
$$
\mathrm{K}_{23} (\vec{e}_a \otimes \vec{e}_b) =
\left(\, \vec{e}_c \otimes \vec{e}_d\, g^{cd}\, \right) \cdot g_{ab}  \; .
$$
where $\vec{e}_a$ are basis vectors in the space $V'$
of the defining representation $T'$.

In \cite{Witten} it was also shown that there exists a spinorial
Yang-Baxter $\mathrm{R}$-matrix $\mathrm{R}(u) \in {\rm End}(V \otimes
V)$
 which satisfies the Yang-Baxter equation
 in the braid form (\ref{Y-B1}) and
intertwines $\mathrm{L}$-operators (\ref{LKaz-so'})
 in spinorial spaces
 \begin{equation}
 \label{RLLf}
\mathrm{R}_{12}(u - v) \, \mathrm{L}_{13}(u) \,  \mathrm{L}_{23}(v) =
\mathrm{L}_{13}(v) \, \mathrm{L}_{23}(u)  \,  \mathrm{R}_{12}(u - v)
   \;\; \in  \;\; {\rm End}(V \otimes V \otimes V') \; .
 \end{equation}
 There is a natural question about generality of the
representation (\ref{LKaz-so'}): what happens when we choose $T$ and
$T'$ to be spinor and arbitrary representations of $so(p+1,q+1)$,
respectively? So we are going to find the $\mathrm{R}$-matrix, acting
in the tensor product of two spinor representations
\begin{equation}
 \label{RLLf-1}
\mathrm{R}(u) = \sum_{k=0}^{\infty} \frac{\mathrm{R}_k(u)}{k!} \cdot
\Gamma_{A_k} \otimes \Gamma^{A_k}
 = \mathrm{R}_{even}(u) + \mathrm{R}_{odd}(u) \,,
 \end{equation}
and obeying the intertwining relation (\ref{RLLf}),
 %$$ \mathrm{R}_{12}(u-v)\,\mathrm{L}_{13}(u)\,\mathrm{L}_{23}(v) =
 %\mathrm{L}_{13}(v)\,\mathrm{L}_{23}(u)\,\mathrm{R}_{12}(u-v)\,, $$
where $\mathrm{L}$-operator is universal:
 \begin{equation}
 \label{RLLf-2}
\mathrm{L}(u) = u+\frac{i}{4}\,\Gamma_{a b}\otimes M^{a b}\,.
 \end{equation}
 In (\ref{RLLf-1}) we have used notations
 $$
 \mathrm{R}_{even}(u) =  \sum_{k=0}^{\infty} \frac{\mathrm{R}_{2k}(u)}{(2k)!}
\cdot \Gamma_{A_{2k}} \otimes \Gamma^{A_{2k}} \; , \;\;\;
\mathrm{R}_{odd}(u) =  \sum_{k=0}^{\infty}
\frac{\mathrm{R}_{2k+1}(u)}{(2k+1)!} \cdot \Gamma_{A_{2k+1}}
 \otimes \Gamma^{A_{2k+1}} \; ,
 $$
 and in view of (\ref{Y-B1}) we obviously have $\mathrm{R}_{odd}(u) = 0$,
 i.e. $\mathrm{R}_{2k+1}(u)=0$ for all $k$. The substitution of
 (\ref{RLLf-1}) and (\ref{RLLf-2}) in (\ref{RLLf}) gives
 %In details we have
$$
\sum_{k=0}^{\infty} \frac{\mathrm{R}_k(u-v)}{k!}\cdot
\Gamma_{A_k} \otimes \Gamma^{A_k}\cdot \left(u+\frac{i}{4}\,\Gamma_{a b}\otimes\mathbf{I}\cdot M^{a b}\right)\,
\left(v+\frac{i}{4}\,\mathbf{I}\otimes\Gamma_{c d}\cdot M^{c d}\right) =
$$
$$
=\sum_{k=0}^{\infty} \frac{\mathrm{R}_k(u-v)}{k!}\cdot\left(v+\frac{i}{4}\,
\Gamma_{a b}\otimes\mathbf{I}\cdot M^{a b}\right)\,
\left(u+\frac{i}{4}\,\mathbf{I}\otimes\Gamma_{c d}\cdot M^{c d}\right)
\cdot\Gamma_{A_k} \otimes \Gamma^{A_k}\,.
$$
This relation contains terms linear and quadratic in generators
$M_{a b}$.
We transform the product of two generators using commutation relations
$$
M_{a b}\,M_{c d} =
\frac{1}{2}\,\left[M_{a b}\,,M_{c d}\right]+
\frac{1}{2}\,\left\{M_{a b}\,,M_{c d}\right\}
= \frac{i}{2}\,\left[g_{b c}M_{a d}-g_{a d}M_{c b}-
g_{a c}M_{b d}+
g_{b d}M_{c a}\right]+
\frac{1}{2}\,
\left\{M_{a b}\,,M_{c d}\right\}
$$
so that
$$
\Gamma_{a b}\otimes\Gamma_{c d}\cdot
M^{a b}\,M^{c d} =
-2 i\,\Gamma_{a}^{\,\,\, c}\otimes\Gamma_{b c}\cdot M^{a b}
+\frac{1}{2}\,\Gamma_{a b}\otimes\Gamma_{c d}\cdot
\left\{M^{a b}\,,M^{c d}\right\}\,.
$$
The linear on the spectral parameters contributions
are combined in the one term $\sim(u-v)$ due to relation
$$
\Gamma_{A_k}\otimes \Gamma^{A_k}\,\Gamma_{a b} -
\Gamma_{a b}\,\Gamma_{A_k}\otimes \Gamma^{A_k} =
\Gamma_{A_k}\otimes\Gamma_{a b}\,\Gamma^{A_k} -
\Gamma_{A_k}\,\Gamma_{a b}\otimes\Gamma^{A_K}\,,
$$
which is consequence of the invariance condition
$$
\left[\,
\Gamma_{a b}\otimes\mathbf{I} + \mathbf{I}\otimes\Gamma_{a b}\,,
\Gamma_{A_k}\otimes \Gamma^{A_k}\,\right] = 0\,.
$$
After all these transformations our main
equation is reduced to the form
$$
\sum_{k=0}^{\infty} \frac{\mathrm{R}_k(u)}{k!}\cdot u\cdot M^{a b}\,\biggl(\Gamma_{A_k}\otimes \Gamma^{A_k}
\,\Gamma_{a b} -
\Gamma_{a b}\,\Gamma_{A_k}\otimes\Gamma^{A_k}\biggl) -
$$
$$
+\frac{1}{2}\,\sum_{k=0}^{\infty} \frac{\mathrm{R}_k(u)}{k!}\cdot M^{a b}\cdot
\biggl(\Gamma_{A_k}\,\Gamma_{a}^{\,\,\, c}\otimes\Gamma^{A_k}\,\Gamma_{b c}-
\Gamma_{a}^{\,\,\, c}\,\Gamma_{A_k}\otimes\Gamma_{b c}\,\Gamma^{A_k}
\biggl) +
$$
$$
+\frac{i}{8}\,\sum_{k=0}^{\infty} \frac{\mathrm{R}_k(u)}{k!}\cdot
\biggl(\Gamma_{A_k}\,\Gamma_{a b}\otimes\Gamma^{A_k}\,\Gamma_{c d}-
\Gamma_{c d}\,\Gamma_{A_k}\otimes\Gamma_{a b}\,\Gamma^{A_k}\biggl)
\cdot
\left\{M^{a b}\,,M^{c d}\right\} = 0\,.
$$
Using the rules of the game with generating functions
$$
\mathrm{Asym}\left(e^{\lambda \Gamma_1\Gamma_2}\right)
\cdot e^{s\Gamma_1} = \mathrm{Asym}\left(e^{\lambda \Gamma_1\Gamma_2+s(\Gamma_1-\lambda\Gamma_{2})}\right)\ \ ;\ \
e^{s\Gamma_1}\cdot\mathrm{Asym}\left(e^{\lambda \Gamma_1\Gamma_2}\right) = \mathrm{Asym}\left(e^{\lambda \Gamma_1\Gamma_2+s(\Gamma_1+\lambda\Gamma_{2})}\right)\,,
$$
$$
\mathrm{Asym}
\left(e^{\lambda \Gamma_1\Gamma_2}\right)\cdot e^{t\Gamma_2} =
\mathrm{Asym}\left(e^{\lambda \Gamma_1\Gamma_2+t(\Gamma_2+\lambda\Gamma_{1})}\right)\ \ ;\ \ e^{t\Gamma_2}\cdot\mathrm{Asym}
\left(e^{\lambda \Gamma_1\Gamma_2}\right) =
\mathrm{Asym}\left(e^{\lambda \Gamma_1\Gamma_2+t(\Gamma_2-\lambda\Gamma_{1})}\right)\,,
$$
it is easy to derive the compact expression for the first contribution
$$
\sum_{k=0}^{\infty} \frac{\mathrm{R}_k}{k!}\,M^{a b}\,
\left[\Gamma_{A_k}\otimes \Gamma^{A_k}\,\Gamma_{a b} -
\Gamma_{a b}\,\Gamma_{A_k}\otimes\Gamma^{A_k}\right] =
\mathrm{R}(\lambda)\ast
M^{a b}\,\frac{\delta^2}{\delta s^{a}\delta s^{b}}
\,\mathrm{Asym}\,e^{\lambda \Gamma_1\Gamma_2}\,\left[e^{s(\Gamma_2+\lambda\Gamma_{1})} - e^{s(\Gamma_1+\lambda\Gamma_{2})}\right] =
$$
$$
= \mathrm{R}(\lambda)\ast\,
M^{a b}
\,\mathrm{Asym}\,e^{\lambda \Gamma_1\Gamma_2}\,
\biggl[(\Gamma_{2a}+\lambda\Gamma_{1b})(\Gamma_{2b}+\lambda\Gamma_{1a}) - (\Gamma_{1a}+\lambda\Gamma_{2a})
(\Gamma_{1b}+\lambda\Gamma_{2b})\biggl] =
$$
$$
= \mathrm{R}(\lambda)\ast\,\left(\lambda^2-1\right)\,
M^{a b}\,
\,\mathrm{Asym}\,e^{\lambda \Gamma_1\Gamma_2}
\left[\Gamma_{1a}\Gamma_{1b}-
\Gamma_{2a}\Gamma_{2b}\right]\,,
$$
so that finally we have
$$
\sum_{k=0}^{\infty} \frac{\mathrm{R}_k}{k!}\,M^{a b}\,
\left[\Gamma_{A_k}\otimes \Gamma^{A_k}\,\Gamma_{a b} -
\Gamma_{a b}\,\Gamma_{A_k}\otimes\Gamma^{A_k}\right]
= \biggl(\mathrm{R}^{\prime\prime}(\lambda) - \mathrm{R}(\lambda)\biggl)\ast\,
M^{a b}\,
\,\mathrm{Asym}\,
e^{\lambda \Gamma_1\Gamma_2}\,\left[
\Gamma_{1a}\Gamma_{1b}-\Gamma_{2a}\Gamma_{2b}\right]\,.
$$
In a similar way using
$$
\mathrm{Asym}\left(e^{\lambda \Gamma_1\Gamma_2}\right) \cdot e^{s\Gamma_1+t\Gamma_2} = \mathrm{Asym}\left(e^{\lambda (\Gamma_1-s)(\Gamma_2-t)+s\Gamma_1+t\Gamma_{2}}\right)\,,
$$
$$
e^{s\Gamma_1+t\Gamma_2}\cdot\mathrm{Asym}\left(e^{\lambda \Gamma_1\Gamma_2}\right) = \mathrm{Asym}\left(e^{\lambda (\Gamma_1+s)(\Gamma_2+t)+s\Gamma_1+t\Gamma_{2}}\right)\,,
$$
we obtain
$$
\sum_{k=0}^{\infty} \frac{\mathrm{R}_k}{k!}\,M^{a b}\,
\left(\Gamma_{A_k}\,\Gamma_{a}^{\,\,\, c}\otimes\Gamma^{A_k}\,\Gamma_{b c}-
\Gamma_{a}^{\,\,\, c}\,\Gamma_{A_k}\otimes\Gamma_{b c}\,\Gamma^{A_k}\right) =
$$
$$
= -2\,\mathrm{R}(\lambda)\ast\,
M^{a b}\,
\,\mathrm{Asym}\,e^{\lambda \Gamma_1\Gamma_2}\,
\left[\Gamma_{1a}\Gamma_{1b}-
\Gamma_{2a}\Gamma_{2b}\right]
\biggl[(\lambda^3+\lambda)\,\Gamma_{1c}\Gamma_2^{c}-n\,\lambda^2
\biggl] =
$$
$$
= -2\biggl[\lambda\,\mathrm{R}^{\prime\prime\prime}
(\lambda) + \lambda\,\mathrm{R}^{\prime}
(\lambda)-n\,\mathrm{R}^{\prime\prime}
(\lambda)\biggl]\,
M^{a b}\,
\,\mathrm{Asym}\,
e^{\lambda \Gamma_1\Gamma_2}\,\left[\Gamma_{1a}\Gamma_{1b}-
\Gamma_{2a}\Gamma_{2b}\right]\,,
$$
and there is expression for the last contribution
$$
\sum_{k=0}^{\infty} \frac{\mathrm{R}_k}{k!}\,
\biggl(\Gamma_{A_k}\,\Gamma_{a b}\otimes\Gamma^{A_k}\,\Gamma_{c d}-
\Gamma_{c d}\,\Gamma_{A_k}\otimes\Gamma_{a b}\,\Gamma^{A_k}\biggl)
\cdot
\left\{M^{a b}\,,M^{c d}\right\} =
$$
$$
= 4\,\mathrm{R}(\lambda)\ast\,(\lambda^3-\lambda)\,
\left\{M^{a b}\,,M^{c d}\right\}\,
\,\mathrm{Asym}\,e^{\lambda \Gamma_1\Gamma_2}\,
\biggl[\Gamma_{1a}\Gamma_{1b}\Gamma_{1c}\Gamma_{2d} - \Gamma_{2a}\Gamma_{2b}\Gamma_{2c}\Gamma_{1d}
\biggl] =
$$
$$
= 4\,\biggl[\mathrm{R}^{\prime\prime\prime}
(\lambda)-\mathrm{R}^{\prime}
(\lambda)\biggl]\ast\,
\left\{M^{a b}\,,M^{c d}\right\}\,
\,\mathrm{Asym}\,e^{\lambda \Gamma_1\Gamma_2}\,
\biggl[\Gamma_{1a}\Gamma_{1b}\Gamma_{1c}\Gamma_{2d} - \Gamma_{2a}\Gamma_{2b}\Gamma_{2c}\Gamma_{1d}
\biggl]\,.
$$
Collecting everything we obtain that intertwining
relation for the $\mathrm{R}$-matrix is equivalent to the relation
$$
\biggl[\lambda\,\mathrm{R}^{\prime\prime\prime}
(\lambda) + \lambda\,\mathrm{R}^{\prime}
(\lambda) - n\,\mathrm{R}^{\prime\prime}
(\lambda)-u\cdot\left(
\mathrm{R}^{\prime\prime}
(\lambda)-\mathrm{R}(\lambda)\right)\biggl]\ast\,
M^{a b}\,
\,\mathrm{Asym}\,
e^{\lambda \Gamma_1\Gamma_2}\,\left[\Gamma_{1a}\Gamma_{1b}-
\Gamma_{2a}\Gamma_{2b}\right] -
$$
$$
-\frac{i}{2}\,\biggl[\mathrm{R}^{\prime\prime\prime}
(\lambda)-\mathrm{R}^{\prime}
(\lambda)\biggl]\ast\,
\left\{M^{a b}\,,M^{c d}\right\}
\,\mathrm{Asym}\,e^{\lambda \Gamma_1\Gamma_2}\,
\biggl[\Gamma_{1a}\Gamma_{1b}\Gamma_{1c}\Gamma_{2d} - \Gamma_{2a}\Gamma_{2b}\Gamma_{2c}\Gamma_{1d}
\biggl] = 0\,.
$$
There are two independent gamma-matrix structures so that we have
differential equation for the coefficient function $\mathrm{R}(x)$
$$
x\cdot\biggl[\mathrm{R}^{\prime\prime\prime}
(x) + \mathrm{R}^{\prime}(x)\biggl] - n\,\mathrm{R}^{\prime\prime}(x)
   -u\cdot\biggl[\mathrm{R}^{\prime\prime}
(x)-\mathrm{R}
(x)\biggl] = 0
$$
and requirement
$$
\left\{M_{[ a b}\,,M_{c ] d}\right\} = 0\,.
$$
The differential equation gives the recurrence
 relation for the
coefficients $\mathrm{R}_k(u)$ for even $k$:
$$
\mathrm{R}(x) = \sum_{k=0}^{\infty}\frac{s_k\,\mathrm{R}_k(u)}{k!}\cdot
x^k\ \longrightarrow \mathrm{R}_{k+2}(u) = - \frac{u+k}{u+n-k}
\cdot\mathrm{R}_{k}(u)\, ,
$$
and for odd $k$ we fix $\mathrm{R}_k(u)=0$.

%%%%%%%%%%%%%%%%%%%%%%%%%%%%%%%%%%%%%%%%%%%%%%%%%%%%%%%%%%%%%%%%%%%%%%%%%%%%%%%%%%%%%%%%%%%%%%%%%%%%%%%%%%%%%%%%%
%%%%%%%%%%%%%%%%%%%%%%%%%%%%%%%%%%%%%%%%%%%%%%%%%%%%%%%%%%%%%%%%%%%%%%%%%%%%%%%%%%%%%%%%%%%%%%%%%%%%%%%%%%%%%%%%%
%%%%%%%%%%%%%%%%%%%%%%%%%%%%%%%%%%%%%%%%%%%%%%%%%%%%%%%%%%%%%%%%%%%%%%%%%%%%%%%%%%%%%%%%%%%%%%%%%%%%%%%%%%%%%%%%%


\begin{thebibliography}{99}

%%%%%%%%%%%%%%%%%%conformal%%%%%%%%%%%%%%%%%%%%

\bibitem{Dirac} P. A. M. Dirac,
{\it Wave equations in conformal space.}
Annals Math. 37 (1936) 429-442

\bibitem{MackSalam} G.~Mack and A.~Salam,
{\it Finite-Component Field Representations of Conformal Group,}
Ann. Phys. 53 (1969) 255.

\bibitem{Koller} K.~Koller,
{\it The Significance of Conformal Inversion in Quantum Field Theory.}
Commun.Math.Phys. 40 (1975) 15-35

\bibitem{Mack} G.~Mack,
{\it All Unitary Ray Representations of the Conformal
Group SU(2,2) with Positive Energy.}
Commun.Math.Phys. 55 (1977) 1

\bibitem{DMPPT} V.K. Dobrev, G. Mack, V.B. Petkova, S.G. Petrova, I.T. Todorov,
{\it Harmonic Analysis on the n-Dimensional Lorentz Group and Its Application to Conformal Quantum Field Theory.}
Lecture Notes in Physics 63:1-280,1977

\bibitem{TMP}
I.T. Todorov, M.C. Mintchev, V.B. Petkova,
{\it Conformal Invariance in Quantum Field Theory.}
Pisa, Italy: Sc. Norm. Sup. ( 1978) 273p

\bibitem{DP} V.K. Dobrev, V.B. Petkova,
{\it Elementary Representations and Intertwining Operators
for the Group SU*(4).}
Rept.Math.Phys. 13 (1978) 233-277\\
V.K. Dobrev,
{\it Elementary Representations And Intertwining Operators For Su(2,2).}
J.Math.Phys. 26 (1985) 235

\bibitem{FGG} S. Ferrara, A.F. Grillo, R. Gatto,
{\it Tensor representations of conformal algebra and conformally covariant operator product expansion.}
Annals Phys. 76 (1973) 161-188

\bibitem{Siegel} W. Siegel, {\it Embedding versus 6D twistors.}
e-Print: arXiv:1204.5679 [hep-th]

\bibitem{Weinberg} S.~Weinberg,
{\it Six-dimensional Methods for Four-dimensional
Conformal Field Theories.}
Phys.Rev. D82 (2010) 045031\,,
e-Print: arXiv:1006.3480 [hep-th]

\bibitem{Shadow} D.~Simmons-Duffin,
{\it Projectors, Shadows, and Conformal Blocks.}
e-Print: arXiv:1204.3894 [hep-th]

\bibitem{CPPR} M.S.~Costa, J.~Penedones, D.~Poland, V.~Rychkov,
{\it Spinning Conformal Correlators.}
JHEP 1111 (2011) 071\,,
e-Print: arXiv:1107.3554 [hep-th]

\bibitem{Kuz} S.M.~Kuzenko, {\it Conformally compactified Minkowski superspaces revisited},
preprint (2012), arXiv:1206.3940 [hep-th].

%%%%%%%%%%%L-operators%%%%%%%%%%%%%%%%%%%
\bibitem{KS}P.P. Kulish and E.K.Sklyanin ,
{\it On the solutions of the Yang-Baxter equation}
Zap. Nauchn. Sem. LOMI {\bf 95} (1980) 129\\
{\it Quantum spectral transform method. Recent developments},
Lect. Notes in Physics, {\bf v 151}, (1982) , 61-119

\bibitem{KRS}
  P.~P.~Kulish, N.~Y.~Reshetikhin and E.~K.~Sklyanin,
  {\it Yang-Baxter Equation And Representation Theory. 1,}
  Lett.\ Math.\ Phys.\  {\bf 5} (1981) 393.

\bibitem{KR}
P. P. Kulish\ and\ N. Yu. Reshetikhin,
{\it On ${\rm GL}\sb{3}$-invariant solutions of the Yang-Baxter
equation and associated quantum systems,}
Zap. Nauchn. Sem. LOMI {\bf 120} (1982), 92--121.


\bibitem{FRT}   L.~D.~Faddeev, N.~Yu.~Reshetikhin and L.~A.~Takhtajan,
{\it Quantization Of Lie Groups And Lie Algebras,}
Lengingrad Math.\ J.\  1 (1990) 193
[Alg.\ Anal.\  1 (1989) 178].

\bibitem{Fad}
L.D. Faddeev,~{\it How Algebraic Bethe Anstz works for integrable
model}, In: Quantum Symmetries/Symetries Qantiques,
Proc.Les-Houches summer school, LXIV. Eds. A.Connes,K.Kawedzki,
J.Zinn-Justin. North-Holland, 1998, 149-211, hep-th/9605187,

\bibitem{DM}
S.~E.~Derkachov and A.~N.~Manashov,
{\it R-Matrix and Baxter Q-Operators for the Noncompact SL(N,C) Invarianit Spin
Chain,} SIGMA {\bf 2} (2006) 084.\\
S.~E.~Derkachov and A.~N.~Manashov,
{\it Factorization of R-matrix and Baxter Q-operators for generic sl(N) spin
chains,}
J.\ Phys.\ A  {\bf 42} (2009) 075204.\\
S.~Derkachov and A.~Manashov, \textit{General solution of the Yang-Baxter
equation with the symmetry group $\mathrm{SL}(n,\mathbb{C})$},
Algebra i Analiz {\bf 21} (4) (2009), 1--94
(St. Petersburg Math. J. {\bf 21} (2010), 513--577).

\bibitem{Resh} N.Yu.~Reshetikhin, {\it Algebraic Bethe-Ansatz for $SO(N)$
invariant transfer-matrices}, Zap. Nauch. Sem. LOMI, vol. 169 (1988) 122
(Journal of Math. Sciences, Vol.54, No. 3 (1991) 940-951).

\bibitem{Okad} M.~Okado, {\it Quantum $R$ matrices related to the spin representations of
$B_n$ and $D_n$}, Comm. Math. Phys., Vol.134, No. 3 (1990), 467-486.

\bibitem{KuSu} A.~Kuniba and J.~Suzuki,
{\it Analytic Bethe ansatz for fundamental representations of Yangians},
Comm. Math. Phys., Vol.173, No. 2 (1995) 225-264; hep-th/9406180.

\bibitem{Drin} V.G.~Drinfeld, {\it Hopf algebras and quantum Yang-Baxter equation}, Sov.Math.Dokl.
Vol. 32 (1985) 254--258.

\bibitem{Mol} A.I.~Molev, M.~Nazarov and G.~Ol'shanskii,
{\it Yangians and Classical Lie Algebras,} Russ. Math. Surv. 51 (1996) 205;
hep-th/9409025.

\bibitem{IsOM} A.P.~Isaev, A.I.~Molev and O.V.~Ogievetsky,
{\it A new fusion procedure for the Brauer algebra and evaluation homomorphisms},
Int. Math. Research Notices, doi:10.1093/imrn/rnr126, (2011);
arXiv:1101.1336 (math.RT).

 \bibitem{Bax} R.J.~Baxter,
 {\it Exactly solved models in statistical mechanics,}
 (Academic Press, London, 1982).

\bibitem{BazSer}
V.V.~Bazhanov and S.M.~Sergeev,
{\it A master solution of the quantum Yang-Baxter
 equation and classical discrete integrable equations},
(2010), arXiv:1006.0651 [math-ph]; {\it Elliptic gamma-function and
multi-spin solutions of the Yang-Baxter equation}, Nucl.Phys. B856
(2012) 475-496, arXiv:1106.5874 [math-ph].

\bibitem{SLN}
V. V. Bazhanov, R.Frassek, T.Lukowski, C. Meneghelli, M.Staudacher,
{\it Baxter Q-Operators and Representations of Yangians.}
Nucl.Phys. B850 (2011) 148-174
e-Print: arXiv:1010.3699 [math-ph]\\
A. Rej, F. Spill,
{\it The Yangian of sl(n|m) and the universal R-matrix.}
JHEP 1105 (2011) 012
e-Print: arXiv:1008.0872 [hep-th]\\
R.Frassek, T.Lukowski, C.Meneghelli, M.Staudacher,
{\it Baxter Operators and Hamiltonians for 'nearly all' Integrable Closed gl(n) Spin Chains.}
e-Print: arXiv:1112.3600 [math-ph]\\
Zengo Tsuboi,
{\it Wronskian solutions of the T, Q and Y-systems related to infinite dimensional unitarizable modules of the general linear superalgebra $gl(M|N)$.}
e-Print: arXiv:1109.5524 [hep-th]\\
A. Alexandrov, V. Kazakov, S. Leurent, Z. Tsuboi, A. Zabrodin,
{\it Classical tau-function for quantum spin chains.}
e-Print: arXiv:1112.3310 [math-ph]

\bibitem{Witten}
R.Shankar and E.Witten,~{\it The $S$-matrix of the kinks of the $(\bar{\psi}\psi)^2$ model},
Nucl.Phys. B141 (1978) 349-363.

\bibitem{KarT} M.Karowski and H.J.Thun,~{\it Complete S-matrix of the $O(2N)$ Gross-Neveu model},
 Nucl.Phys. B190[FS3] (1981) 61-92.

 \bibitem{ZamL} Al.B.~Zamolodchikov, "Factorizable Scattering in Assimptotically
 Free 2-dimensional Models of Quantum Field Theory", PhD Thesis, Dubna (1979), unpublished.

 \bibitem{Lipat} L.N. Lipatov,
High-energy asymptotics of multicolor QCD and exactly solvable lattice
models, JETP Lett.\  59 (1994) 596 [Pisma Zh.\ Eksp.\ Teor.\ Fiz.\  59
(1994) 571], hep-th/9311037;
%%CITATION = HEP-TH 9311037;%%
High-energy asymptotics of multicolor QCD and two-dimensional conformal
field theories, Phys.\ Lett.\ B 309 (1993) 394.
%%CITATION = PHLTA,B309,394;%%


\bibitem{Zam} A.B. Zamolodchikov and Al.B. Zamolodchikov,
{\it Factorized S-Matrices In Two Dimensions As The Exact Solutions Of
Certain Relativistic Quantum Field Models},
Ann. Phys. (N. Y.) 120 (1979) 253;
{\it Relativistic Factorized S Matrix In Two-Dimensions Having O(N) Isotopic Symmetry},
Nucl. Phys. B 133 (1978) 525.

\bibitem{Zam1} A.B. Zamolodchikov,
{\it "Fishnet"\,  Diagrams As A Completely Integrable System.}
Phys.Lett. B97 (1980) 63-66


\bibitem{GN}I. M. Gelfand , M. A. Naimark,
{\it Unitary representations of the classical groups,} Trudy Mat.
Inst. Steklov., vol. 36, Izdat. Nauk SSSR, Moscow - Leningrad, 1950; Gernman transl.:
Academie - Verlag, Berlin, 1957.

\bibitem{Knapp}
A. Knapp and E. Stein, {\it Intertwining operators
for semi-simple Lie groups,} Ann. of Math. (2) 93 (1971), 489-578.\\
Knapp A.W. {\it Representation theory of
semisimple groups: an overview based on examples.}
Princeton,N.J.:Princeton Univ.Press,1986.

\bibitem{Zaikov} G.M. Sotkov, R.P. Zaikov,
{\it Conformal Invariant Two Point and Three Point
Functions for Fields with Arbitrary Spin.}
Rept.Math.Phys. 12 (1977) 375

\bibitem{FrP}  E.S. Fradkin and M.Ya. Palchik, {\it Recent Development in Conformal Invariant Field Theory},
Phys. Rep. {\bf 44} No. 5, (1978) 249.

\bibitem{Vas} A.~N.~Vasilev, Y.~M.~Pismak and Y.~R.~Khonkonen,
{\it $1/N$ Expansion: Calculation of the Exponents $\eta$ and $\nu$
in the Order $1/N^2$ for Arbitrary Number Of Dimensions},
Theor.\ Math.\ Phys.\  {\bf 47} (1981) 465.

\bibitem{Vas1}
A. N. Vasil'ev, {\it The Field Theoretic Renormalization Group in Critical Behavior Theory and Stochastic Dynamics} (Routledge Chapman  Hall) 2004

\bibitem{Lip} L.N. Lipatov,
~{\it Duality symmetry of reggeon interactions in
multicolor QCD,}
Nucl.Phys. {\bf B 548}, (1999) 328.

\bibitem{Der} S.E. Derkachov,
{\it Baxter's Q-operator for the homogeneous XXX spin chain.}
J.Phys.A A32 (1999) 5299-5316
e-Print: solv-int/9902015

\bibitem{Br}    D. J. Broadhurst,
{\it Summation of an infinite series of ladder diagrams,}
Phys. Lett. B 307 (1993) 132.

\bibitem{dual0}
J.M. Drummond, J. Henn, V.A. Smirnov, E. Sokatchev,
{\it Magic identities for conformal four-point integrals.}
JHEP 0701 (2007) 064
e-Print: hep-th/0607160

\bibitem{dual1} G.P. Korchemsky, J.M. Drummond, E. Sokatchev,
{\it Conformal properties of four-gluon planar amplitudes and Wilson loops.}
Nucl.Phys. B795 (2008) 385-408
e-Print: arXiv:0707.0243 [hep-th]

\bibitem{dual2}
J.M. Drummond, J. Henn, G.P. Korchemsky, E. Sokatchev,
{\it Dual superconformal symmetry of scattering
amplitudes in N=4 super-Yang-Mills theory.}
Nucl.Phys. B828 (2010) 317-374
e-Print: arXiv:0807.1095 [hep-th]

\bibitem{DKM}
S.~E.~Derkachov, G.~P.~Korchemsky and A.~N.~Manashov,
{\it Noncompact Heisenberg spin magnets from high-energy QCD. I: Baxter
Q-operator and separation of variables,}
Nucl.\ Phys.\ B {\bf 617} (2001) 375
[arXiv:hep-th/0107193].

%\bibitem{LipVeg} H.J.~de~Vega and L.N.~Lipatov,
%{\it Interaction of Reggeized gluons in the Baxter-Sklyanin
%representation}, Phys. Rev. D {\bf 64} 114019.

\bibitem{Isaev1}
A.P. Isaev, {\it Quantum groups and Yang-Baxter equations}, Sov.J.Part.Nucl. 26
(1995) 501-526; (see also extended version: A.P. Isaev, {\it Quantum groups and
Yang-Baxter equations}, preprint MPIM (Bonn), MPI 2004-132 (2004),
http://www.mpim-bonn.mpg.de/html/preprints/preprints.html).

\bibitem{Isaev2}
A.~P.~Isaev,
  \textit{Multi-loop Feynman integrals and conformal quantum mechanics,}
  Nucl.\ Phys.\ B {\bf 662} (2003) 461
  [arXiv:hep-th/0303056].
\bibitem{Isaev2a}  
A. P. Isaev, {\it Operator approach to analytical evaluation of Feynman diagrams},
Phys.Atom.Nucl.71:914-924,2008,
arXiv:0709.0419.

\bibitem{Mitr} Indrajit Mitra, {\it External leg amputation in conformal invariant three-point function},
Eur.Phys.J. {\bf C71} (2011) 1621; arXiv:0907.1769.

\bibitem{VDK}   A.N. Vasiliev, S.E. Derkachov, N.A. Kivel,
{\it A Technique for calculating the gamma matrix structures of the diagrams of a total four fermion interaction with infinite number of vertices in d = (2+epsilon)-dimensional regularization.}
Theor.Math.Phys. 103 (1995) 487-495, Teor.Mat.Fiz. 103 (1995) 179-191

\end{thebibliography}
\end{document}